\documentclass[11pt,preprint]{aastex}
\usepackage{color}
\usepackage{graphicx}
 
\newcommand{\ly}{Lyman}
\newcommand{\Lya}{\mbox{Ly$\alpha$}}
\newcommand{\lya}{\mbox{Ly$\alpha$}}
\newcommand{\lyb}{\mbox{Ly$\beta$}}

\newcommand{\kms}{\mbox{km s$^{-1}$}}
\newcommand{\cmm}{\mbox{cm$^{-2}$}}
\newcommand{\cmmm}{\mbox{cm$^{-3}$}}

\newcommand{\het}{\mbox{$^3$He}}
\newcommand{\hef}{\mbox{$^4$He}}
\newcommand{\lisv}{\mbox{$^7$Li}}

\newcommand{\yp}{\mbox{Y$_p$}}

\newcommand{\qone}{PKS~1937--1009}
\newcommand{\qtwo}{Q1009+2956}
\newcommand{\qthree}{Q0130--4021}
\newcommand{\qfour}{HS~0105+1619}
\newcommand{\qfive}{Q1243+3047}
\newcommand{\object}{Q1243+3047}
\newcommand{\qhst}{PG~1718+4807}
\newcommand{\qpettini}{Q2206--199}
\newcommand{\qdodorico}{Q0347--3819}

\newcommand{\etal}{{\it et al.}}
\newcommand{\zabs}{\mbox{$z_{\rm abs}$}}
\newcommand{\chisq}{\mbox{$\chi^2$}}
\newcommand{\lyaf} {\lya\ forest}
\newcommand{\nhi}{\mbox{N$_{\rm H I}$}}
\newcommand{\lnhi}{\mbox{log \nhi}}
\newcommand{\ndi}{\mbox{N$_{\rm D I}$}}
\newcommand{\lndi}{\mbox{log \ndi}}

\newcommand{\zdh}{\mbox{$z = 2.526$}}

\newcommand{\ETA}{\mbox{$\eta $}}
\newcommand{\ob}{\mbox{$\Omega_b$}}
\newcommand{\obh}{\mbox{$\Omega_bh^{2}$}}
\newcommand{\om}{\mbox{$\Omega_m$}}
\newcommand{\mf}{\mbox{$10^{-5}$}}

\newif\ifdraftmodep
\draftmodepfalse

\newcommand{\NOTE}[1]{\ifdraftmodep {\color {red} [{\it #1}]} \fi}

\newif\ifapjp
\apjpfalse

\draftmodepfalse
\apjptrue

\ifapjp
  \usepackage{emulateapj5}
\fi

\begin{document}

\title{THE COSMOLOGICAL BARYON DENSITY FROM THE DEUTERIUM TO HYDROGEN RATIO 
TOWARDS QSO ABSORPTION SYSTEMS: D/H TOWARDS Q1243+3047}

\author{ David Kirkman\altaffilmark{1,2,3}, 
         David Tytler\altaffilmark{1,2},        
         Nao Suzuki\altaffilmark{1,2},  
         John M. O'Meara\altaffilmark{1,2}, 
         Dan Lubin\altaffilmark{1,2}  
\altaffiltext{1} {Visiting Astronomer, W.M. Keck Observatory which
is a joint facility of the University of California, the California
Institute of Technology and NASA.}
\altaffiltext{2} {Center for Astrophysics and Space Sciences,
University of California, San Diego,
MS 0424; La Jolla; CA 92093-0424}
\altaffiltext{3} {E-mail: dkirkman@ucsd.edu}
}

\begin{abstract}
We report the detection of Deuterium absorption at redshift 2.525659
towards Q1243+3047. We describe improved methods to estimate the
Deuterium to Hydrogen abundance ratio (D/H) in absorption systems,
including improved modelling of the continuum level, the \lyaf\ and the velocity
structure of the absorption. Together with improved relative flux calibration,
these methods give D/H
$=2.42^{+0.35}_{-0.25}\times $\mf\ \cmm\ from our Keck-I HIRES spectra of
Q1243+3047, where the error is from the uncertainty in the shape of
the continuum level and the amount of D absorption in a minor second
component. The measured D/H is likely the
primordial value because the [O/H] $= -2.79 \pm 0.05$. This absorption
system has a neutral Hydrogen column density
\lnhi\ $= 19.73 \pm 0.04$ \cmm , it shows five D lines and is mostly ionized.
The best estimate of the primordial D/H is $2.78 ^{+0.44}_{-0.38}
\times$\mf , from the log D/H values towards five
QSOs.  The dispersion in the five values is larger than we expect from
their individual measurement errors and we suspect this is because
some of these errors were underestimated.  
We observe a trend in D/H with \lnhi\ that we also suspect is spurious.
The best value for D/H is
$0.6\sigma $ smaller than we quoted in O'Meara
\etal\ (2001) from three QSOs, and
although we have more values, the error is similar because the
dispersion is larger. 
In standard big bang nucleosynthesis (SBBN),
the best D/H corresponds to a baryon-to-photon ratio $\eta = 5.9 \pm
0.5 \times 10^{-10}$ and gives precise predictions for the primordial
abundances of the other light nuclei.  We predict more $^4$He than is
reported in most measurements, although not more than allowed
by some estimates of the systematic errors.  
We predict a $^3$He abundance very similar to that reported by Bania
\etal\ (2002), and we predict 3 -- 4 times more $^7$Li than is seen in
halo stars. It is unclear if those stars could have destroyed this much of 
their $^7$Li.  The $\eta $ value from D/H corresponds to a cosmological
baryon density $\Omega _b h^2 = 0.0214 \pm 0.0020$ (9.3\%) that 
agrees with values from the anisotropy of the Cosmic Microwave
Background: $\Omega _b h^2 = 0.021 \pm 0.003$ from the 
Netterfield \etal\ (2002) analysis of BOOMERANG data and $\Omega _b h^2 = 0.022
^{+0.004}_{-0.003}$ from the Pryke \etal\ (2002) analysis of the DASI
results.
\end{abstract}


\section{Introduction \NOTE{section\_a.tex}}
\label{introduction}

It is well established that the light nuclei hydrogen (H), deuterium
(D), \het , \hef\ and \lisv\ are all made during big bang
nucleosynthesis.  The relative primordial abundances created in the Standard 
theory of big bang nucleosynthesis (SBBN) for these five nuclei depend
on one parameter, the cosmological baryon-to-photon ratio, \ETA $=
{n_{b}}/{n_{\gamma}}$ (Kolb \& Turner 1990; Walker \etal\ 1991;
Schramm \& Turner 1998; Nollett \& Burles 2000; Olive, Steigman \&
Walker 2000).  A measurement of the ratio of any
two primordial abundances gives $\eta$.  The three other primordial
abundances are predicted once \ETA\ is known and measurements of them
test the theory.

SBBN has now been validated in two main ways. First, it successfully
accounts for measurements of the approximate relative primordial
abundances of all these light nuclei  (Boesgaard \& Steigman 1985;
Walker \etal\ 1991;
Copi, Schramm \& Turner 1995;
Schramm 1998;
Schramm \& Turner 1998;
Tytler \etal\ 2000 and references there in).
Second, the baryon density
required is roughly consistent with that measured in other ways, including the
\lyaf\ at high redshifts 
(Rauch \etal\ 1997;
Weinberg \etal\ 1997;
Zhang \etal\ 1998;
Hui \etal\ 2002), and the baryon fraction in
clusters of galaxies 
(Babul \& Katz 1993;
Boute \& Canizares 1996;
Bludman 1998;
Rines \etal\ 1999;
Arnaud \& Evrard 1999;
Wu \& Xue 2000;
Sadat \& Blanchard 2001). 
Recently, the cosmic microwave
background (CMB) has given the same baryon density to within 10\%.

We use the D~I/H~I ratio in QSO absorption line systems
to estimate $\eta$.  This method is attractive because 
the deuterium abundance is more sensitive to \ETA\ than other light nuclei,
the D/H ratio in QSO absorption systems is apparently the primordial 
value, and we do not apply any corrections for unobserved ions.  

We measure D/H by dividing the column densities that we measure for
absorbing D~I and H~I atoms. Since D/H is low, we could see D in only
absorption systems that have high H~I column densities, at most one
per QSO.  We would like to measure the column densities to better than
10\% accuracy and ideally $<1$\% to better test early universe
physics.  Echelle spectrographs give ample spectral resolution, and
integration times approaching 1 day give the required signal-to-noise
ratio (S/N). Once D has been found in a well calibrated QSO spectrum, there
are four main factors that limit the measurement accuracy: the
continuum level, the \lyaf, the velocity structure of the absorber,
and contamination by absorption other than D.

Each of the factors that complicate the measurement of D/H are hard to
model and quantify.  We need to know the level of the unabsorbed
continuum to measure the amount of absorption. The \Lya\ and other
weak emission lines make it hard to estimate the precise continuum
shape, and stochastic \lyaf\ absorption only makes matters worse.  The
\lyaf\ H~I absorption is random in wavelength and opacity, and is
ubiquitous over the redshift intervals in which we can measure D/H
from ground based telescopes.  All the H and D lines are in the \lyaf\
portion of QSO spectra, which introduces confusion and in the worse
case can mimic D absorption.  The hardest problem is to ensure that
we explore all possible velocity structures that might explain a given
spectrum.  D/H values and their errors are often dictated by the
choice of velocity structure.

It appears likely that the measurement errors have been underestimated
in at least one of the published D/H values, because these errors are
hard to calculate, and the dispersion in D/H measurements is larger
than expected.  In O'Meara \etal\ (2001) we found that the dispersion
between three D/H values that we had measured towards different QSOs
was larger than expected from the quoted errors. We concluded that
this dispersion was probably not real, and instead a result of our
underestimation of the errors.  Since then, Pettini \& Bowen (2001)
have found D in a fourth QSO, \qpettini\ and quote a D/H value which
is much less than the previous measurements, while D'Odorico \etal\ (2001)
and Levshakov \etal\ (2002) have presented two very different measurements of
D/H towards a 5th QSO, \qdodorico .  These new measurements further
increase the dispersion in the reported D/H values, and again we
suspect that this is because the errors have been underestimated.

In this paper we announce the detection of D in an absorption system
at \zdh\ in the spectrum of \qfive.  We describe how we have improved
our exploration of some factors which determine the measurement
errors.  In particular we describe improved calibration of relative
flux levels, our estimates of the continuum level, our modelling of the
velocity structure, and our modelling of the \lyaf\ and blended lines.

We present material in the following order: observations and
reductions (\S \ref{observations}), overview and velocity structure of
the absorption system that shows D (\S \ref{llssec}), measurement of
the D column density (\S \ref{ndi}), evidence that we are measuring D
(\S \ref{itisd}) measurement of the H column
density (\S \ref{nhi}), the metal abundance and ionization in the
absorbing gas(\S \ref{metallicitysec}), the D/H from \object\ (\S
\ref{best1243dh}), and from all QSOs (\S \ref{pridh}), and the
deduced cosmological parameters (\S \ref{cospar}).  The summary ends
with a discussion of the improvements in methods that we use here, and
a list of the issues that remain. In the Appendices we discuss how
we model the continuum, our optimizing code, error estimates from the 
covariance  matrix and D/H values from other QSOs.

\section{OBSERVATIONS AND DATA REDUCTION \NOTE{section\_b.tex}}
\label{observations}

We report the detection of D absorption in the QSO Case Stellar Object
CSO~0167, which we call by its B1950 coordinate name: \qfive.  This
QSO was reported as a V = 17 blue object at B1950 RA 12h 43m 44.9s
DEC +30d 47.9', with a 15" error, (equivalent to J2000 12h 46m 10.9s
+30d 31m 31.2s) in an objective prism photograph by Sanduleak \& Pesch
(1984), and identified as a QSO at emission line redshift $z_{em} =
2.56$ by Everett \& Wagoner (1995).  This QSO is not well known. The
NASA extra-galactic database (NED) gives no other primary references
except for a 2MASS QSO search.

We have spectra with three different spectral resolutions, from three
spectrographs, summarized in Tables \ref{table_a} and
\ref{resolution}, and shown in Figures 1, 2 and 3.  

We obtained five spectra of \object\ from the Kast double spectrograph
on the Lick Observatory 3.1m Shane telescope.  These low resolution
spectra are used to flux calibrate HIRES spectra.
We compare these spectra to check the
flux calibration, and we can sum them to improve the S/N.  All integrations 
were obtained using the d46 dichroic
that splits the spectrum near 4600~\AA, the 830 line/mm grism
blazed at 3460~\AA\ for the blue side, and the 1200 line/mm grating
blazed at 5000~\AA\ for the red side. 

We have one intermediate resolution spectrum from the ESI spectrograph
(Epps \& Miller 1998; Bigelow \& Nelson 1998; Sheinis \etal\ 2000),
mounted at the Cassegrain focus of Keck-II.  This echellette covers
from 3900 -- 11,000~\AA\ in a single setting in ten overlapping
orders.  We also use this spectrum to flux calibrate the HIRES
spectra, and to look for metal line absorption at wavelengths larger 
than those covered in our HIRES spectra.

We have eight high resolution spectra from the HIRES spectrograph
(Vogt \etal\ 1994) on the Keck-I 10m telescope. These were obtained
using the C5 dekker that has a 1.14" slit width and gives a FWHM of
8.0 \kms\ sampled with 2.1 \kms\ per pixel on the original HIRES
Tektronix 2048x2048 CCD.  To improve the relative flux levels along
each spectrum, all observations were taken with the spectrograph slits
aligned with the local vertical.  All but one of these spectra cover
down to wavelengths below the Lyman limit break in the absorption
system that shows D, while the exception extends to larger
wavelengths to cover some metal lines. The S/N per pixel in the summed
HIRES spectra (Table \ref{table_b})
increases linearly with wavelength up to 4200~\AA, and
then rises faster in the \lya\ emission line, reaching 105 on either side of
the \lya\ absorber of the D/H system. The S/N in the center of an
echelle order is approximately 1.4 times greater than that at the
ends.

We describe the data reduction and flux calibration in detail in
Suzuki \etal\ (2003).  The error in the wavelength solution for the
HIRES spectra is at least 0.05 -- 0.1 pixels, or 0.1 -- 0.2~\kms, and
may be as large as 1 -- 2 \kms. Our analysis of other similar spectra
(Levshakov, Tytler \& Burles 2000) revealed errors of order 1 \kms.
We placed all the HIRES integrations on the same logarithmic
wavelength scale, which has a constant velocity of 2.1~\kms\ per
pixel, similar to the original pixel size, and we shifted the Kast and
ESI spectra to put them on a similar wavelength scale.  All spectra
were also converted to vacuum wavelengths and shifted to the solar
rest frame.  We applied a relative flux calibration to the HIRES
spectra using the Kast and ESI spectra to transfer the flux
information from standard stars to the QSO spectra.

\section{THE \zdh\ SYSTEM THAT SHOWS D \NOTE{section\_c.tex}}
\label{llssec}

In this section, we discuss the spectrum of \object , we introduce the
absorption system that shows the D, and we discuss its velocity structure.

\subsection{Overview of the \qfive\ spectrum}
\label{overview}

We observe D in the conspicuous absorption system at \zdh\ that
produces both a clear Lyman limit break and has the strongest \lya\
absorption line in the spectra, near 4285~\AA.  We see no flux below
the limit that is near 3210 \AA\ in the Kast and HIRES spectra. 
%(Figure
%\ref{lyseries}).  
No flux is expected because the H~I column density
\lnhi\ is clearly near $\simeq 19$~\cmm\ from the shape of the \lya\
line.  The presence of a strong O~I line at this redshift also implies
that \lnhi $> 19$~\cmm .  In addition to five D absorption lines, this
system shows many Lyman series H~I transitions, but only a few metal
lines, because the metal abundances are approximately 0.001 solar in
the main component (see \S \ref{metallicitysec}).  In comparison,
HS~0105+1619 has half as much H~I, but shows 21 metal lines, as its
metal abundance is approximately 14 times larger (O'Meara et
al. 2001).

The system at \zdh\ shares some of the properties of both the Lyman
limit systems (LLSs) and the Damped \lya\ systems (DLAs).  It is a LLS
because it is optically thick in the Lyman continuum, and it has 25\%
of the minimum \lnhi\ to qualify as a DLA (Turnshek \& Rao 2002).  It
shows O~I like DLAs, but unlike DLAs (Wolfe \& Prochaska 2000), it is
mostly ionized and the different low ionization lines have different
velocity profiles.  
There is little published work on systems with \lnhi $= 18-20$~\cmm ,
(e.g. O'Meara \etal\ 2001; Peroux \etal\ 2002)

The spectrum shows metal lines from unrelated redshifts, none of which
effect our D/H measurement.  A weaker \lya\ absorption line near
3710~\AA ($z \sim 2.052$) appears to have weak damping wings that
imply a high \nhi.

\subsection{How we Measure D/H}

The D/H value is the ratio of two column densities, \ndi\ for the
neutral Deuterium (D~I) and \nhi .  We must estimate the errors on
each of these quantities and show that both measurements apply to the
same gas.

%
% I'm pretty sure we don't need the following paragraph.  --dk
%
%We use Voigt profiles to parameterize the distribution of absorption
%in velocity. Most absorbers are too complex to fit with a single Voigt
%profile, and hence we model lines with a sum of Voigt profiles, one
%for each component.  These components are typically centered at
%different velocities, measured in the rest frame of the absorption
%system.  Each Voigt component is described by three parameters: the
%column density $N$, the central redshift $z$ and the parameter $b =
%\sqrt{2}\sigma $, which measures the velocity distribution of the ion.

We measure the \nhi\ to high accuracy from the \lya\ line alone,
because the shape of this line is dominated by damping wings that are
sensitive to \nhi, and less sensitive to the velocity information.
However, the \nhi\ value is more accurate and reliable when we use
accurate velocity information, and hence we measure the \nhi\ last,
after we have examined the velocity information.  Unlike the \nhi,
the \ndi\ that we measure depends strongly on the velocity
information.

We obtain the velocity information from the D~I lines themselves,
taking guidance from the high order H Lyman lines and the metal lines,
especially O~I.

\subsection{Velocity Structure of the \zdh\ absorption system}
\label{velstructure}

We now discuss the clues that metal line absorption give
for the velocity structure.
The system at \zdh\ shows many 
components, with different structure in different
ions, but the bulk of the H~I is in just two components that we see in
absorption by O~I.

We expect that the H~I and the D~I should have identical velocity
distributions, except for the effects of thermal broadening.  We
expect the D/H ratio to be constant from component to component, 
because the metal abundances are
too low for significant destruction of D.  We also expect that \ndi
/\nhi = D/H because the D and H should have nearly identical
ionization (Savin 2002).

We would like to measure the velocity structure of the H and D
separately, allowing for a direct comparison.  However, the H lines
are all saturated, broad and blended, and give little information.
Instead, we use the unsaturated metal lines, especially the O~I lines,
as guides.

We do not know how closely the metal lines will trace the velocity
structure of the H~I and D~I.  We expect to find some H~I wherever we
find metals, but the metal lines can have different velocity
distributions in detail because the ionization and metal abundance can
vary from component to component, and perhaps with velocity inside a
component (\S \ref{velagree}).

In Figure \ref{p5}, we present the regions of the spectrum where we
expect metal line absorption.  We observe strong absorption in only a
few metal ions: O~I, C~II, and Si~II. C~III and Si~III also show
absorption, but are poorly constrained since their lines appear
saturated and may be highly contaminated by \lyaf\ absorption. We see
weak C~IV and Si~IV absorption that has very different
velocity structure from the low ionization metals.  Al~II is seen in the ESI
spectrum, but is not covered by our HIRES data.  Fe~II 1608 is not
detected at our S/N, and although some absorption 
appears near the
expected positions of N~I, N~II, N~III, and Fe~III, it is very weak,
and can readily be accounted for as \lyaf.  We also measure $\log
N_{C~II*}< 12.3$~\cmm\ in all components.

The O~I line suggests that two components will be needed to model the
velocity distribution of the gas that shows the D.  O~I provides the
best indication of the velocity distribution of the H~I and D~I
absorption, because O~I/H~I is similar to O/H in gas of low ionization
(O'Meara et al. 2001).  The O~I 1302 transition is in a high S/N
region of the spectrum well separated from other lines. This line is
asymmetric, with extra absorption at larger wavelengths. We fit the
O~I with the two components that we list in Table \ref{dhlinetab}.

In Figures \ref{lowions1} \& \ref{highions1} we show fits to the O~I,
C~II, Si~II, C~IV, and Si~IV lines.  Voigt profile fitting usually
does not produce unique results, and this is especially true for C~II
and Si~II, which can be synthesized with a variety of components at
different velocities that are all heavily blended.  The C~II and
Si~II fits listed in Table \ref{dhlinetab} and shown in Figures
\ref{lowions1} \& \ref{highions1}. Irrespective of the exact model
used, the C~II and Si~II require gas at $v\simeq -40$, $\simeq 0$, and
$ \simeq 95$ \kms, blended with additional absorption between 16 and
35 \kms.

The C~IV and Si~IV lines have a different velocity structure from the
low ionization transitions O~I, C~II and Si~II, which means they
primarily come from different gas.  Both C~IV and Si~IV show
absorption at $v \simeq $40, 100, and 200 \kms. There is also a
component centered near $v \simeq 5$~\kms\ for C~IV and $v \simeq -10
$~\kms\ for Si~IV.

In summary, the metals indicate that H~I may be found in components
near: $v = -40$, 0, 13, 20, 40, 95 and 200~\kms .  We are confident
that the O~I marks the velocity of most of the H~I.  The low
ionization lines of C~II and Si~II have most of their column density
at these velocities, and O~I is not seen at other velocities.

There is no sign of any metal components near $v \simeq -81.6$~\kms,
the velocity that would place H~I lines at the position expected for
the D~I lines. This means that there does not need to be any H~I that
could contaminate the D lines, although there could be if the metal
abundances were very low.

\subsection{Velocity Structure of the Gas that Gives D/H}
\label{velofdh}

While the system shows many components, the D/H measurement will
depend on just three.  We expect to see D~I in the two components
that we see in the O~I, C~II and Si~II, while the third component is
an unfortunate blend that adds uncertainty but does not show D.
\begin{itemize}
   \item Component 1, near $v=0$~\kms , is the strongest component in
         O~I, C~II, Si~II and D~I. We will refer to the D~I absorption
         in this component as D-1, with column density \ndi (D-1) and
         velocity dispersion $b$(D-1) and similarly for the other
         ions.  The parameters of the O-1 and D-1 components are well
         determined and they contain approximately 90\% of the column
         density.  We use the redshift of the O-1 component, $z$
         (O-1)$ = 2.5256916$, to define a reference frame with radial
         velocity $v=0$~\kms.  In this frame, we measure the velocity
         of the O-1 as $v=0.00 \pm 0.14$~\kms , where the error is
         from the fit alone and ignores the poorly known wavelength
         scale error.
   \item Component 2, near $v=13$~\kms , is seen as an asymmetric
         extension of the O~I that we call component O-2. There is
         also some C~II, Si~II near this velocity. We call other
         associated lines D-2 and H-2.  Neither the velocity nor the
         $b$-value of the O-1 component are well known, and they are
         correlated. Smaller $b$-values are needed as the line center
         moves to larger wavelengths. If we have some additional
         velocity or $b$-value information, then the second component
         is well determined.  For example, if both O~I components have
         the same $b$-value, this value is 6.78~\kms , component 2 is
         at $13.3$~\kms , and its column density at 13\% of the total.
   \item Component 3, near $v=-40$~\kms , is an even weaker component
         in C~II and Si~II. We will refer to the H~I absorption from
         this component as H-3, which is at $v(H-3) = -44.3 \pm
         7.0$~\kms , with \lnhi (3) $= 15.90 \pm 0.03$~\cmm\ from the
         higher order Lyman lines.  This \lnhi\ value is low enough
         that we would not expect to see O~I or D~I and neither is
         seen.  H-3 effects D/H because the short wavelength side of
         the Lyman lines from H-3 accidentally blend with the long
         wavelength side of the D-2 lines, and hence some of the
         absorption near --70~\kms\ can be explained by either H-3 or
         D-2. In Figure \ref{fig_d3}, we see that D-2 is strongly
         blended with D-1 and H-3.
\end{itemize} 

There is a large amount of H~I absorption, \lnhi $\simeq 19.7$~\cmm,
near components 1 and 2, but we have little information on the
velocity distribution of this H~I.  In Figs. \ref{p4} \& \ref{p15} we
show the Lyman series lines.  The spectra require H~I Lyman lines
corresponding to the metal line components near $v = -40$, 0, 95 and
200~\kms.  The spectra are also consistent with, but do not need,
components at 13, 20 and $40 \kms$.  The parameters that we find for
these components are in Table \ref{dhlinetab}.  If there is H~I
absorption at $40\ \kms$ with $b > 10\ \kms$, the blue edge of the
higher order Lyman lines show that it has $\lnhi < 16.4\ \cmm$.
Figure \ref{p15} shows there is no flux in the range $-40$ to
$40$~\kms\ in lines up to Ly-14, and in higher order lines the S/N is
rather low, and the Lyman lines start to overlap.  We can measure the
shape and width of the higher order lines, but unfortunately this
tells us little about the gas at $v= 0 - 13.3$~\kms.  The absorption
on the negative velocity side of these lines is determined by the H~I
at --40~\kms\ with $\lnhi \sim 16\ \cmm$, and the positive velocity
side of these lines may be influenced by $\lnhi \sim 16\ \cmm$
absorption at $40 $~\kms.

A single component at $v=0$~\kms\ with a $b \simeq 17.7$~\kms \ gives
a fair fit to all of the H lines without separate lines at $\pm
40~\kms$ .  This $b$-value is an upper limit on the $b$-value for the
H~I in component 1, since we must use a lower $b$-value when we
include separate lines at $\pm 40~\kms$ .  We will see below that we
can also obtain an excellent fit to the H lines using components H-1
and H-2.

\section{Measurement of the D column density \NOTE{section\_k.tex}}
\label{ndi}

In this section we explore a variety of models that can account for the
spectrum. We estimate the \lndi\ from the models that fit the spectrum with 
the lowest \chisq\ values (\S \ref{grid}) and 
in \S \ref{derrors}
we discuss the many factors that effect the accuracy and reliability of 
the \lndi\ value and our estimate of the error.

To make a reliable estimate of the D/H ratio in this system, we must
explore the full range of models that can explain the H and D
absorption in the spectrum.  Different models employ different
assumptions about issues such as the velocity structure of the
absorber, and whether D or H is making parts of the absorption.

A large part of the uncertainty over the D/H comes from the \lndi\ in the D-2
component.  We can not predict the amount of D in D-2, because we do
not know the fraction of the H~I in this component, and the O/H need
not be the same in the two components.  The spectrum requires some
absorption near --70~\kms\ that can be either D-2 or H-3. We must now
determine how much of this absorption is from D-2.

Most of the information on the D is in transitions Ly-2 to 8, which we
show in Figure \ref{p4}.  In each transition we see absorption that
has all the expected characteristics of component D-1 associated with
H-1 and O-1.  In Ly-4 to 8 we see a resolved D line that is well separated
from the H absorption, while in Ly-2 and 3 the D absorption is fully
blended with the H lines, as we expected because the H line is wider.
The velocity structure, central velocity, $b$-value and column density
are all well determined from Ly-4 to 8, especially since the last few
lines are not saturated. The absorption is coming from D in gas with
velocity near zero.  The D-1 lines are all very narrow, and they can all
be fit by a single component.

However, we have less information on the D-2 component, because it is
fully blended with the D-1 on its short wavelength side, and H-3 on
the other side.

In Figure \ref{p8} we show the $b$ and \lndi\ values that we obtain
when we fit each of the D lines individually with a single component.
While there is general agreement, the dispersion is clearly larger
than expected for the errors that we show.  These errors come from the
covariance matrix of the \chisq\ optimization and we explain in the
Appendix \S \ref{apperror} why such errors are often too small.  

Two of the D-1 lines are contaminated by other lines. The
Ly-5 D line contains a second narrow Ly-3 line at $z=2.3987515$ that
has \lnhi \ 14.459~\cmm\ and $b = 28.81$~\kms .  This contaminant is
well determined by its other Ly lines and we fit it when we fit the
Ly-5 D line.  The D Ly-3 line includes a slight contribution from a
\lyb\ line with $z=2.342177$, $b=35.6$~\kms\ and $\lnhi=13.81$~\cmm .

\subsection{Grid Search to Find \lndi}
\label{grid}

In this section we estimate the \lndi\ value and its error. 
We have explored various models that might explain the absorption at
all wavelengths relevant to deuterium.  When we calculated the \chisq\ 
difference between the simulated spectra from the models and the data 
we found that there was a well defined set of parameters that give 
excellent \chisq\ values.  We obtained \lndi\ and its error from the set 
of models that had \chisq $< \chi^2_{min} +1$.

We made a six dimensional grid of models, one model for each combination of the
allowed parameter values. We specified these models using
six parameters that we vary, because they might change the D/H, and
other parameters that we hold constant. We made simulated spectra for
15,750,000 different models, one for each possible combination
of the parameter values. We allowed each of the six parameters to have every 
discrete value given by the (minimum value, maximum value, and the number of 
uniformly spaced steps), as follows:
\begin{itemize}
   \item \lndi (total) (15.07~\cmm , 15.2,  15).
   \item \ndi (D-2) / \ndi (total) (0.01, 0.35, 35).
   \item $v_{sep}$, the velocity separation of the two D
         components, (5~\kms, 20, 20).  The maximum value lets the D
         absorption cover the full range of the O~I absorption, while
         we found that \lndi (D-2) was insignificant whenever the
         separation was $<5$~\kms.
   \item $b$(D-1) (9~\kms, 9.48, 10). 
   \item $b$(D-2) (4~\kms, 15, 10). The upper limit is guided by the
         largest value for $b$(H-2) and we discuss the lower limit in 
         \S \ref{lowbd2}.
   \item $b$(H-3) (15.7~\kms, 19, 15).
\end{itemize}
We found that the \lndi (D-1) varies very little, as expected, because
its lines are strong and clear. The main variation in \lndi (total)
comes from \lndi (D-2), and hence the first two parameters are
strongly correlated. All except the first and last parameters in the
list describe the velocity structure of the D, while the last
parameters varies the amount of H absorption near the D-2 lines.  In
most cases the ranges and the step sizes were chosen by trial and
error to sample all models that can give fits with low $\chi ^2$.
Exploring the range of parameters in this grid search was a
computationally intensive task that took over a month and required
tens of runs, many of which took over 24 hours on a 1.5Ghz personal
computer. We used a similar grid search, with only three parameters,
in Kirkman \etal\ 2001.

Several other parameters of the model were not varied, because we
believed they would have less effect on D/H, and the computations take
approximately ten times longer for each additional parameter.  The
fixed parameters include:
\begin{itemize}
   \item The wavelength ranges included in the fit: --40 to
         --120~\kms\ for Ly-2 to Ly-8.
   \item $v$(D-1) = $-2.8$~\kms , which is well constrained by
         the spectrum and is also used as the value for $v$(H-1).
         We discuss why this differs from the O-1 velocity
         in \S \ref{velagree}.
   \item The specification of the main H~I, including its \lnhi (total)
         and its two components, H-1 and H-2. We use the values given
         in Table \ref{dhlinetab}, some which come from the fits that we will
describe in \S \ref{nhi}. We can fix the main H~I because the
         high order Lyman lines of the H~I are all strongly saturated
         and they do not absorb near the D lines.
   \item \lnhi (H-3) is constrained because H-3 does not absorb in the
         higher order Ly-lines (Figs. \ref{fig_d3} \&
         \ref{p15}) and its precise \nhi\ value does not effect the
         \lndi.
   \item $v$(H-3) $= -44$~\kms.  We found that $b$(H-3) is effectively
         degenerate with v(H-3), and hence either parameter suffices
         to explore the range of acceptable models.  We changed
         $v$(H-3) from --50~\kms\ to --30~\kms\ and found that the
         $b$(H-3) was larger in compensation, but there was no change
         in the \lndi (total) that gave acceptable $\chi ^2$ values.
         We believe that this is a sufficient range to explore for
         v(H-3) because a metal component is centered near --41~\kms.
   \item The continuum shape and height. In Figure \ref{fig_d4} we
         show the continuum that we used near the Ly-limit, while in
         Figs. \ref{fig_d3}, \ref{p4}, \ref{p15} \& \ref{fig_d1} we
         show the fit to the spectrum using this continuum and
         absorption by obvious lines.  We found that the bumps in the
         continuum are required to fit the spectrum.  We do not claim
         that the QSO continuum itself is this bumpy, rather that the
         combination of the QSO continuum, emission lines, broadly
         distributed and weak blended absorption, and flux calibration
         errors combine to make this the effective continuum. Had we
         chosen a smooth or flat continuum across Fig. \ref{fig_d4}
         then we would have also required extra broad absorption
         distributed like the dips in the continuum in Fig.
         \ref{fig_d4}, or emission near 923~\AA\ in the rest frame of
         the QSO, to obtain fits similar to those seen in
         Fig. \ref{fig_d1}.
   \item The absorption in the vicinity of D that is H~I at different
         redshifts.
\end{itemize}
In \S \ref{derrors} we show that the last two factors are related and
have little effect on \lndi.

In Figure \ref{dgrid} we show contours of constant \chisq\ as a
function of the total \ndi\ against each of the other five parameters
that we adjust.  These plots show projections of the minimum \chisq\
values onto the planes specified by the axes, and hence the values of
the four other parameters may vary slightly as we move along a
particular contour. We found these minimum \chisq\ values by
stepping each adjustable parameter over its full range, to explore all
combinations of all parameters.

Our best estimates for the parameters are the values that together
give the minimum \chisq\ value, $\chi ^2 _{min}$. These values are all
contained within the innermost contour on each panel.  The $m \sigma $
confidence intervals for the parameters are the ranges allowed by the
successive contours on the plots, that we have placed at values of
$\chi ^2_{min} +m^2$.  Hence the $1\sigma$ ranges are the smallest
and largest values of the parameters that are allowed by the innermost
contour, at $\chi ^2_{min} +1$.  These critical \chisq\ values apply
because we have varied all the parameters to find the minimum $\chi
^2$ value for each \lndi (total).  We have only one effective degree
of freedom, the \lndi (total).  The distribution of the $ \chi ^2$
statistic as a function of \lndi (total) is then expected to be
approximately the $\chi ^2_1$ distribution function, with one degree
of freedom.  We shall refer to this method of estimating the errors on
a parameter as the $\delta $\chisq\ method.  

In Figure \ref{dchi} we show the minimum \chisq\ that we found for
each \lndi.  The critical \chisq\ values are the same as in Figure
\ref{dgrid}, for the same reason.  Our best estimate for the total D
in components 1 and 2, the column density that we will use to get D/H,
is
\begin{equation}
\lndi (total) = 15.113 ^{+0.042}_{-0.026} ~\cmm ,
\end{equation}
which is a $1\sigma $ range of $+$10\% and $-$6\%. This is the \lndi
(total) value that gives the lowest \chisq\ when combined with the
following values for the other parameters, all of which are collected
in Table \ref{table_c}): $b$(D-1) $= 9.2 \pm 0.2$~\kms , $b$(D-2) $=
4$~\kms , $v_{sep} = 12^{+2}_{-3.5}$ \kms , \ndi (D-2) / \ndi (total)
$ = 0.12^{+0.16}_{-0.05}$, and hence \lndi (D-2) $=14.191$~\cmm. 

The $\chi _{min}^2 = 271.4$ is reasonable for 274 pixels.  We have
used 3 degrees of freedom fitting the parameters of each of the two D
lines and 2 parameters for the H-3 lines. The continuum, other
contaminating lines, and \nhi (H-3) are well constrained by pixels
outside the region used in the grid search, while $v$(H-3) is
degenerate with $b$(H-3). There are then approximately 274-8 degrees
of freedom.  The probability $\chi ^2_{266} > 271.3 = 0.40$.

We have a simple explanation for the range of \ndi(total).  The
\chisq\ distribution is asymmetric, rising more slowly as \lndi(total)
increases from the best value, because of the contribution from D-2.
The \lndi (D-1) is well determined and varies little for all models
that give a small \chisq\, while the \lndi (D-2) varies from zero to a
well determined maximum.  Figure \ref{dgrid} shows an anti-correlation
between \lndi\ and $b$(H-3), which arises because the absorption near
--70~\kms\ can be explained by various combinations of D-2 and H-3. As
\lndi (D-2) decreases, the $b$(H-3) can increase to give models with a
similar shape and $\chi ^2$.  The lower bound on \lndi\ comes from our
assumption that the absorption near $v=0$ is D, which we discuss in
section \ref{itisd}.  The lower bound on \lndi\ applies when \lndi
(D-2) is insignificant and all the absorption in D-1 is D.  The upper
bound on \lndi\ is met when D-2 absorbs photons that we see at
velocities $\simeq 70$~\kms.  We find a continuous range of acceptable
models between these two bounds.

The fit shown in Figures \ref{p15}, \ref{fig_d3}, \ref{fig_d1}, \&
\ref{p4} is that with the minimum \chisq .  Other fits with \chisq\
values larger by several look indistinguishable in such plots.

We found that we could also find good fits to the D lines using the
optimization methods that we describe in \S \ref{nhi}. Before we
conducted the grid search, our best fit had a reasonable \chisq\
= $\chi ^2_{min} + 0.7$.  After we completed this grid search, we
confirmed that the optimization methods can find the same best fit,
but only when we limit the parameter to narrow ranges centered on 
the best fit.

\subsection{Discussion of the \lndi\ Value and its Error}
\label{derrors} 

We now discuss a few of the many factors that can effect the \lndi\
value and its error. In each case we have made some reasonable choice
for the relevant parameters when we conducted the grid search. Here we
discuss some of these other options that we did not quantitatively
explored in the grid search. In all cases we suspect that these
effects are smaller than the error that we quote on the \lndi, except
for missing D components that might be comparable to the error.  Some
of these factors could appear as random or systematic errors,
depending on details such as potential bias. An error is random if its
effect on \lndi\ is such that the mean \lndi\ from many spectra and
model fits would converge on the true value. 

\subsubsection{Flux Calibration}

Flux calibration is a possible cause of the bumpy continuum in
Fig. \ref{fig_d4}. Four different HIRES orders contribute at $\lambda
< 3400$~\AA . The wavelength scale of the bumps are consistent with
the 10~\AA\ correlation length that we expect for flux calibration
errors in this part of the spectrum.  The amplitude is also
consistent, except for the largest bump near 3290~\AA\ that is 6 --
12\% in amplitude, two -- three times larger than the expected
calibration errors. There are two components to the flux calibration
error, the S/N of the HIRES integrations, and the S/N of the Kast
spectra. We find that the total error in the calibration of the summed
HIRES spectrum is approximately a 4\% near 3290~\AA , and 8\% near
3250~\AA\ where the S/N is lower.

The method we used to fit the continuum near the D line should have
corrected for the errors in the flux calibration without adding to the
error on the \lndi . The continuum fitting will also have corrected
for any other effects that are correlated over $> 10$~\AA .

\subsubsection{Continuum Level}

We find that the continuum level is not a significant part of the
error in the \lndi\ value. If we multiply the adopted continuum by
1.05, the result is clearly too high for all the D line, while if we
multiply by 0.95 the result is too low, except perhaps for Ly-3. This
means that the error in the local continuum is $<5$\%, and since we
fit the continuum independently for each of the 5 main D lines, the
corresponding error on the \lndi\ is $< 5$\%/$ \sqrt{ 5}$, which we
will ignore.

\subsubsection{Velocity Error}

We took $v(D-1)=-2.8$~\kms\ and we did not vary this value, to account
for the error of 0.6~\kms\ from the fit to D-1 or the wavelength scale
error. If the D-1 were not centered at this velocity, it would have a
smaller \lndi\ and some additional narrow D or H absorption would be
needed to account for all the absorption. This does not seem a likely
way to change the \lndi\ value.

\subsubsection{Could $b$(D-2) be very Low? }
\label{lowbd2}

The grid search hints that $b$(D-2) could be very small.
The model with $\chi ^2_{min}$ had $b$(D-2) $= 4$~\kms , but with 
a $1\sigma $ range extending to  12~\kms .
The minimum $b$ value that we considered in the grid search 
was 4~\kms\ 
because the instrumental resolution is similar, and hence $b < 4$~\kms\
has little effect on the spectrum.

Small values for $b$(D-2) do not allow very large \lndi (D-2).
The spectrum places an upper limit on the opacity near --70~\kms , and a weaker
limit on the equivalent width. We can then trade \ndi (D-2) against $b$(D-2) 
to find acceptable fits for a wide range of $b$(D-2). 
Solutions with low $b$(D-2) tend to have low \ndi(D-2), the 
opposite of the situation for lines with constant equivalent width.
Fortunately the optical depth prohibits large \lndi (D-2),
and hence the uncertainty in the $b$(D-2) value 
has little effect on \lndi (D-2), as seen in the third panel of 
Figure \ref{dgrid} .

Low $b$(D-2) values require temperatures that are unlikely.
If $b$(D-2) $< 4$~\kms, the absorbing gas would have $T < 1900$ K, and it 
should be neutral if ionized by intergalactic radiation.
However, the C~II/O~I and Si~II/O~I ratios appear to be higher near to 
component 2 than in component 1, which implies that component 2 is more ionized,
and less likely to be neutral (\S \ref{ionization12}).

\subsubsection{Velocity Structure}

When we calculated the maximum \lndi\ in D-2, we assumed that both D-2
and H-3 were Voigt profiles. More complex velocity distributions could
give different column densities, a topic that has been explored by
Sergei Levshakov and collaborators (Levshakov \etal\ 2003).  This is
unlikely to be a major effect for H-3, which is wide enough
(Fig. \ref{dgrid}) that its profile is dominated by thermal motions,
and hence should be insensitive to the detailed velocity
distribution. The same argument wil apply to D-2 if its $b$-value is
close to the $1\sigma $ upper limit of 12~\kms , but not if $b$(D-2) is
near 4~\kms .

\subsubsection{Missing D Components}
\label{missingd}

The D/H would be systematically low if we have included velocity
components in \lnhi (total) that are not detected in D~I because the
D~I is hidden by H~I absorption.

We see no evidence that we have missed D components, and we do not
know how to estimate the chance that they exist, and hence we will not
include them in the error on \lndi . However, they could exist and
have a major effect on \lndi, even doubling the D/H if their metal
abundances are very low.

We need to examine the D and H lines in detail to establish the
velocities that could contain missing D components.  Fig. \ref{p15}
shows that the H~I included in the total \lnhi\ all lies well within
the range --40 to 40~\kms . At other velocities, \lnhi $< 17$~\cmm\
because see flux at the expected positions of one or more of the
higher order Lyman lines.  Figs.  \ref{p4v0} \& \ref{fig_d1} show that
there are no significant D lines from components at --40~\kms\ $< v <
0$~\kms , except for D-1.  At positive velocities there are
opportunities to hide D.  Component 2 was hard to detect, some of
component H-3 could be D, and there might also be significant D
between the two, at $13 < v < 40$~\kms , where we see C~II, Si~II,
C~IV and Si~IV.

The H~I lines alone allow that a major portion of the total \lnhi\ is
at velocities approximately 16 -- 25~\kms . H~I lines alone do not
give the $v$ of the main H~I to within approximately 25~\kms\ because
the core of the \lya\ contains many other components, and the higher
order Lyman lines are blended on both sides.  However, the C~II, S~II
and especially the O~I all indicate that most of the H~I is in
component H-1. The H~I in components other than 1 and 2 is probably
$<10$\% of the total because we do not see O~I at other velocities.
If D/H is a constant, we expect that any missing D is also less than
10\% of the total. We would have seen the O~I from any component with
more than 10\% of the total \lnhi\ provided it had [O/H] $> -3$.

We do not know whether the absence of O~I and other metals is
sufficient to rule out missing components that contain $>10$\% of the
total column density.  Abundances $< -3$ are very rare for absorption
systems as a whole. Not one of 34 DLAs listed by Pettini \etal\ (1997)
had [Zn/H] $< -2$, although several were upper limits and some systems
with lower abundances are known (e.g. Fan \& Tytler 1994).  
However, we know little about the abundances in components of systems.

Although H-3 might contain as much optical depth in D as D-1, H-3 is
predominantly H and not D because a distinct component of C~II and
Si~II is centered near $v = -40$~\kms\ (Table \ref{table_c},
Figure \ref{lowions}).

\subsubsection{Contamination of the D-1 component by H}

In \S \ref{itisd} we explain why it is unlikely that the D-1 lines are
contaminated by H, other than the blends that we mentioned above with
D Ly-3 and 5.  Several of the D lines seem relatively free of close
contamination and hence we do not expect that the way in which we
fitted the contaminants to D Ly-3 and 5 is a significant source of
error.

\subsubsection{Validity of the Model and the \chisq\ Constraint}
\label{deltachisq}

The error we quote for \lndi\ comes from the range of fits with \chisq
$< \chisq _{min} +1$.  The \lndi\ and the error will be unreliable if:
(1) the model or the parameters with the lowest \chisq\ are
unrealistic, or physically unacceptable; (2) we did not fully explore
the parameter space of the model; (2) we did not explore a wide enough
range of models, for example allowing D in enough components or at
enough velocities.

We believe that the \lndi\ value and its error are reasonable.  We
explored a relatively large range of possible models using coarse
grids before we settled on the fine grid presented above. However, we
may have misunderstood the velocity structure because the S/N is low
near the Lyman limit and we did not explore more complex models
because they would have taken too long to compute.

\section{THE ABSORPTION NEAR THE D-1 POSITION IS MOSTLY D}
\label{itisd}

Several lines of argument imply that the absorption that we identify
as D-1 is mostly D. The evidence is convincing, but not as strong as for
some other QSOs.

Ideally, we would compare the velocity structures of the D~I and H~I
lines, which would be identical if we have correctly identified the
transitions and there is minimal contamination. This comparison is not
possible for \object\ because the velocity structure of the H~I is not
observed, and hence we are less certain that we have seen D than we
were in other QSOs.

The only reasonable identifications for D-1 are D and H, because D-1
shows a Lyman series.

If the D-1 absorption is H, $b$(D-1) implies a very low temperature of
$< 5130$~K. This temperature is much too low for low density gas
photoionization by the intergalactic ionizing UV radiation. It is
certainly too low for the \lyaf\ and perhaps for components with \lnhi
$= 15$~\cmm\ in DLAs and LLS, unless they are shielded from the
radiation by H~I.  There are no
metals at the redshift of H at the position of D-1, but this is not
significant, since metals would not be seen if their abundances were
low.

There are three main reasons why D-1 is D: its velocity, line width,
and column density are all close to the values that we expect.

The column density of the absorption near D-1 could have been orders
of magnitude different from the measured value. D-1 is likely to be D
because it has a \lndi\ that gives D/H similar to the values found
toward four other QSOs. The converse is not true: we would not reject a
D line that gives a different D/H value if it otherwise appears to be
D.

\subsection{Velocity Agreement}
\label{velagree}

The wavelengths of the centers of the D, O and H lines indicate that the
D-1 absorption is D, but they also suggest that O/H varies with velocity.

In the  frame of the main  component of the  O-1, D-1 is at  $-2.8 \pm
0.6$~\kms , where the error is from the line fit alone.  This is close
enough  to show that  D-1 is  mostly D,  since the  D-1 and  O-1 lines
overlap in velocity.

The O-1, D-1 velocity difference of 1.3 pixels is larger than we
expect from measurement errors, but we have been surprised in the past
by wavelength scale errors in HIRES spectra.  The error might be with
the single O~I line, rather than the D lines, since the several D
lines appear at the same velocity (Figs. \ref{fig_d3} \& \ref{p4}).

The H~I velocity is consistent with that of the D~I, but it is not
known well enough to show that the D-1 line is D.

\subsubsection{Does O/H Increase with Velocity?}

The O~I might have a different velocity structure from the H~I and the
D~I if O/H varies and is correlated with velocity. For example,
we can imagine that all of the H~I, D~I and O~I come from a single
component. This component might have a Gaussian distribution of
velocities.  The O-1 component would be centered at $+2.8$~\kms\ in
the frame of the D-1 because O/H is larger at larger velocities, and
the O-2 component might arise from the O/H gradient alone.  In this
case we should model the H and D with a single component.  We do not
favor this model, because C~II and Si~II show components 1 \& 2. 
However, we can fit to the H~I lines with a single component 
(\S \ref{velofdh}), and this can have a $b$-value that
is consistent with the D and O lines (\S \ref{linewidth}).
If we have a single component the \lnhi\ is unchanged (\S
\ref{nhi}), while the D/H is lower because \lndi\ tends to be lower
when we loose D-2 (\S \ref{grid}).

\subsubsection{Metal Abundance Gradients in Other Absorbers}
\label{gradients}

There is almost no published information on the variation of metal
abundances or metal line kinematics in components of absorption
systems with \lnhi $\simeq 19$~\cmm , especially on velocity scales of
a few \kms . The closest analogue is \qfour\ where we found that ions
C~II, N~II, Si~II and Fe~II did not come from the same gas as the H~I,
D~I and O~I, because the line centers differed by 1 -- 2 ~\kms\ and
the lines widths were different (O'Meara \etal\ 2001). Toward
absorption systems with much lower \nhi\ we have found that the
abundances can vary greatly between components with similar velocities
and relatively high ionization.  Toward \qone\ Tytler, Fan \& Burles
(1998) found that two components separated by 15~\kms\ had very
different metal abundances: [Si/H] $= -2.7$ and $-1.9$. Toward \qtwo\
Burles \& Tytler (1998) found that two components also separated by a
similar amount had abundances of [Si/H] $=-2.7$ and $-2.4$.  Hence for
these two QSOs the metal lines do not give a sufficient description of
the H~I and D~I.  Toward DLAs with \lnhi $> 20.3$~\cmm, Wolfe \&
Prochaska (2000) find that different low ionization ions show similar
velocity distributions, which implies that both the ionizations and
relative abundances are similar, but does not constrain the absolute
abundances.  In those cases the differences were in different
components, but for \object\ the velocity offset would require
structure within component 1, for which we have no other evidence.

\subsubsection{Velocity Agreement for D-2}

We suspect that D-2 is D for three reasons, but the case is weak, because 
D-2 is not well determined.
First, the absorption that we fit with D-2 is very narrow, narrower then 
typical H~I lines. 
Second, the separation of D-2 from D-1, 12~\kms , with a $1\sigma $ range of 
8.5 -- 14~\kms , matches the separation of the O~I components, that is $\simeq
13.3$~\kms . In the grid search we examined D separations from 4 -- 20
~\kms , and the D separation could have been different from that of
the O~I. However, this argument is weak, in part because the D-1 and
O-1 are at slightly different velocities.
Third, the fraction of the D in D-2 ($0.12^{+0.16}_{-0.05}$)
matches the fraction of O in O-2, 0.13, when we force $b$(O-1) $= b$(O-2).

\subsection{Line Width Agreement}
\label{linewidth}

The widths of the D~I, H~I and O~I lines indicate that the D-1 absorption is D.
If the D~I, H~I and O~I arise in the same homogeneous gas, and the
velocity distribution in this gas is described by a single temperature,
and a Gaussian distribution of turbulent velocities, then we can
predict the $b$-value of the D from the H and O $b$ values.  Following
O'Meara \etal\ (2001), the intrinsic $b$ value of the lines, $b_{int}$
is given by $b_{int}^2 = b_{temp}^2 + b_{turb}^2$, where the
temperature term is $b^2_{temp} = 2kT/m = 166.41(\kms
)^2(T/10^4~K)/$mass (amu), where T is the gas temperature and
$b_{turb}$ represents the bulk turbulent motions, and $m$ is the mass
in atomic mass units. The $b$ values in this paper are all intrinsic
values, since we have convolved the intrinsic line profiles by the
instrumental broadening before we fit to the spectrum.

The observed width of the D-1 component agrees with the prediction
from the widths of the O-1 and H-1 components, given in Table
\ref{dhlinetab}, and shown in Fig. \ref{p7}.  The straight line that
fits $b^2$ against $1/m$ gives $T=1.1 \pm 0.6 \times 10^4$~K and
$b_{turb} = 5.8 \pm 0.6$~\kms , both reasonable values, similar to
those that we found in other D/H systems.  This fit predicts $
b$(D-1)$ = 11.3 \pm 1.8$~\kms , $1.2\sigma $ larger than the observed
$b$(D-1)$ = 9.2 \pm 0.2$~\kms . This agreement suggests that D-1 is D
and not H. However the evidence is weak because the $b$(H-1) value and
its error, from the covariance matrix, are both poorly constrained,
and because the velocity difference of the D-1 and O-1 components may
mean that the model we used to predict $b$(D-1) is too simplistic.

When we accept that the D-1 component is D, we can use D-1 together
with the H-1 and O-1 to improve our estimates of the gas properties.
We find $T=0.55 \pm 0.04 \times 10^4$~K, which is low but reasonable
for the large \lnhi , and $b_{turb} = 6.3 \pm 0.2$~\kms .  The data
are also fully consistent with this fit, which we show as the dashed
line in Fig. \ref{p7}.  Compared to this line, the measurements give
\chisq $=1.38$ for one degree of freedom, where 
Prob$(\chi ^2_1 > 1.38) = 0.24$.

\section{Measurement of the H column density \NOTE{section\_l.tex}}
\label{nhi}

In this section, we describe the measurement of the Hydrogen column
density in the system that shows D, along with new methods that we
developed to improve the accuracy and reliability of \nhi\ measurements.  
We first find the approximate \lnhi\ value (\S \ref{approxnhi}), then
a detailed model that fits the spectrum (\S \ref{initialnhi}),
and then we examine other models that give acceptable fits
(\S \ref{explorationsection}). We find the \lnhi\ value and its error in
\S \ref{besth}, and we discuss these values in \S \ref{herror}.

Most information on the \nhi\ comes from the shape of the damped \lya\ line
at 4285~\AA\ that we will call the DLA because it shows damping wings,
although its \lnhi\ value is less than the usual definition for a DLA
system.  Compared to the methods we used to get \lndi, for the \lnhi\
we use different ways to fit the continuum and \lyaf, to explore the
parameter values and to assign an error to \lnhi.

\subsection{The Approximate \lnhi\ Value}
\label{approxnhi}

We can quickly establish that the DLA has \lnhi $\simeq 19.7$ \cmm.
If we fit the HIRES spectrum by a \lya\ line with a lower column
density, there remains unexplained absorption which looks like damping
wings. This is very clear when \lnhi $\leq 19.5$~\cmm , and can still
be seen at larger \lnhi .  On the other hand, when we fit with \lnhi =
19.8~\cmm\ we absorb flux which is seen, which is un-physical. To allow
this extra absorption, the QSO continuum, including any emission line
flux, would have to bend upward, on either side of the damped line,
near 4288~\AA\ and 4293~\AA , which is best seen when we divide the
spectrum by the line profile.  We then know that $19.5 <$ \lnhi ~\cmm\
$< 19.8$, where the larger limit is twice the smaller, a range that is
20 times the $1 \sigma $ range on the \lndi (total).

\subsection{An Initial Model for the Spectrum Near the DLA}
\label{initialnhi}

We now describe how we made an initial model of the spectrum. In \S
\ref{explorationsection} we will use this model as a starting point
when we explore alternative models.

A model of the flux in the spectrum is determined by three main
factors: the flux emitted by the QSO, the \lyaf\ and the DLA.  We
developed our own software to control these coupled factors.  We can
change the model manually, or automatically, optimizing over
parameters to give models that have the smallest \chisq\ difference
from the spectrum. These tools are described in Appendices I and II.

To construct our model, we first placed a DLA at \zdh\ with $\lnhi =
19.7 \ \cmm$.  We then set a preliminary continuum level from 4200\AA\
to 4350\AA , such that the continuum plus DLA touched most of the
peaks in the spectrum.  This continuum was defined by 10 control
points: six between 4320\AA\ and 4340\AA\ to define the top of the
\lya\ emission line, two at $\lambda < 4250$ \AA\ to define the
continuum far from the emission line, and two between 4260~\AA\ and
4325\AA\ nearest to the DLA.  This continuum has enough freedom to
take on a wide variety of shapes in the crucial region where the DLA
is strongly absorbing.  In addition, we add \lyaf\ absorbers where we
see lines.  We then made made small modifications to the continuum,
the \lnhi\ of the DLA and the \lyaf , until we arrive at a model that
appeared to accurately reproduce the spectrum. This model used $\lnhi
= 19.695 \ \cmm$.  Finally, we used this model as the starting point
for an automatic optimization that returned $\lnhi = 19.73$~\cmm .
This model, shown in Figure \ref{f1008_v1}, has a reasonable \chisq\ ,
a smooth looking continuum, the \lyaf\ absorbers are not unusual, and
there is little discernible structure in the residuals.

The error on the \lnhi\ from the optimization covariance matrix,
$\sigma (\lnhi ) = 0.0002$~\cmm , is too small for at least two
reasons: (1) the final covariance matrix has significant off-diagonal
term involving the \nhi , and (2) the \chisq\ manifold may have
multiple minima. We believe that errors derived from the covariance
matrix will be too small for all similar models that contain many
correlated parameters. We discuss this issue in \S \ref{apperror}.

The initial model has two problems. First, we do not know the error on
the \lnhi\ value, and second, we have no reason to believe that it is
unique.

\subsection{Restarting the Optimization to Find the \nhi\ Range}
\label{explorationsection}

We now try to determine the robustness and uniqueness of our
fits by seeing if, for a variety of input values for the parameters,
we return to our best value of \lnhi\ $ = 19.73$ \cmm .  
A fit is robust if it can be found when we start from a wide
variety of input parameters. A fit is unique if there are no other
fits which have significantly lower $\chi ^2$. Although we can readily
find local minima in the manifold of parameters, it is well known that
there are no simple ways of showing that we have found the unique, or
global minimum.  On the contrary, we expect that there are other fits,
which may have different \lnhi\ values, and similar or lower $\chi ^2$
values.

We used the optimizing code to searched for fits with low \chisq\
values.  We performed thousands of optimizations, each beginning with
different parameter values.  We added random numbers, selected from
normal distributions (NDs), to the parameter values of our initial
model (\S \ref{initialnhi}).  The ND for the continuum control points
had $\sigma = 1$ unit of flux. Since the continuum near the damped
line is 12 units of flux, these input continua usually have
significant bumps and dips.  The ND for the \lnhi, $b$, and $z$ of
each \lyaf\ line had a $\sigma $ equal to the prior measurement error.

We allowed the optimizer to move the two continuum points between
4260~\AA\ and 4325~\AA\ (shown as squares in Figure \ref{f1008_v1}) in
both wavelength and flux.  The optimizer also varied the flux, but not
the wavelengths of all the other continuum points. Some examples of
the starting continua are shown in Fig. \ref{f1009_v2}.

We found that the velocity structure of the H~I has little effect on
the \lnhi\ value. For all optimizations, we distributed the total
\lnhi\ of the DLA between the two velocity components seen in the O~I
line, and we put 13\% of the total \lnhi\ in the second component at
$v = 13.3$~\kms . This proportion is similar to that seen in O~I and
D~I.  However, we measured the same total \lnhi\ if all of the H~I is
in one component.  Our fits to the DLA give \lnhi (total) that
excludes the H~I from components 3, 4 and 5. We fit these other
components separately with the results listed in Table \ref{dhlinetab}.
We did not fit separate components near 20 and 40~\kms\ because they
are poorly constrained, none are needed near 40~\kms , and gas in this
range should have \lnhi $<17$~\cmm .

The total \lnhi\ was not allowed to vary in any of the optimizations.
However, the input \lnhi\ was different in different optimization
re-starts.

We conducted thousands of re-starts for each chosen \nhi\ value.  The
\chisq\ at the starts were typically $10^5$, while at the end,
depending on the \nhi\ value, 5700-6700, similar to the degrees of
freedom.  We found reasonable solutions with low \chisq\ values for a
wide variety of \nhi\ values.  In Figure \ref{f1014_v1} we show the
\chisq\ values for these re-starts.

Some of the final solutions found by the re-start procedure had \lyaf\
absorbers that are not typical of those seen in the \lyaf\ of other
QSOs (Pettini \etal\ 1990; Hu \etal\ 1995; Kirkman \& Tytler 1997).
In particular, many of the solutions with $\lnhi < 19.68\ \cmm$ had
very wide \lyaf\ absorbers near to 4295\AA.  We do not believe that
these solutions indicate that there may be wide \lyaf\ absorption near
to the LLS, but rather that the optimizer can, if it needs to, use
\lyaf\ absorption with large $b$ values to compensate for low values
of \nhi\ in the LLS.  For this reason we have rejected,
from the bottom panel of Figure \ref{f1014_v1}, all fits that
included \lyaf\ lines with $b > 150$~\kms .
In Figure \ref{f1012_v3} we show the effect of
one broad \lyaf\ line on a portion of the spectrum.

\subsection{The Best Estimate for \lnhi\ and its Error}
\label{besth}

We now discuss our best estimate for the \lnhi\ value and its error.

We found that the re-starts with \lnhi = 19.73~\cmm\ consistently gave
the lowest \chisq\ values.
If our re-starts had fully explored the parameter space, we would have
concluded \lnhi $= 19.73 \pm 0.005$~\cmm , where the error is the
range of \lnhi\ values have fits with \chisq $< \chi ^2_{min} +1$, the
$\delta $\chisq\ method that we used for the \lndi\ in \S \ref{ndi}.
For no other values of \lnhi , including 19.72 and 19.74~\cmm\ did we
find fits with \chisq\ within 10 of the minimum, let alone 1.

We do not use this method to estimate the \lnhi\ error because our
optimization process is not efficiently finding fits with the lowest
\chisq\ values.  For each \lnhi\ value, Figure \ref{f1014_v1} shows a
range of \chisq\ values of approximately 200, and the lowest \chisq\
is often 10 lower than the second lowest.  To define the error on
\lnhi\ using $\delta \chisq = 1$, we would require many fits with 
\chisq\ values within 1 of the minimum, and we would also have to prove that 
we had adequately explored all relevant models and parameter ranges
(\S \ref{deltachisq}).

We determine the error on the \lnhi\ value by testing the hypothesis
that there is at least one acceptable fit for a given \lnhi\ value.
We accept an \lnhi\ value if the restarts include one or more models
with an acceptable \chisq\ value. We accept a \chisq\ if there is
$>5$\% probability of a larger value when the hypothesis, that the
data came from the model, is true. We fit 5048 pixels near \lya\ with
289 parameters, leaving 4759 degrees of freedom and hence the maximum
acceptable \chisq\ value is 4920.6.  We increased this maximum
acceptable \chisq\ by a factor of 1.23, to 6052.3, for reasons
discussed in \S \ref{chisqscale}.

We found acceptable fits for the re-starts for \lnhi $=19.68$ --
19.78~\cmm\ when we reject fits with \lyaf\ lines with $b > 150$~\kms
, and 19.70 -- 19.78~\cmm\ when we reject $b > 100$~\kms .  Therefore,
we choose 19.68 -- 19.78~\cmm\ as the range of acceptable \lnhi\
values.

It is helpful to assign a probability distribution function to
summarize the range of likely \lnhi\ values. We made the following
choices: a normal function, centered at 19.73~\cmm , with an 80\%
probability that the true \lnhi\ is in the range 19.68 -- 19.78~\cmm . 
We choose 80\% because we guess that there is of order a 20\%
chance that the true \lnhi\ is not in the range 19.68 -- 19.78~\cmm ,
and because we do not want the function to be too sharply centered on
19.73~\cmm .

Our best estimate for the column of the H~I in components 1 and 2,
that is associated with \lndi\ in the same components, is then
\begin{equation}
\label{besthequation}
 \lnhi (total) = 19.73 \pm 0.04~\cmm ,
\end{equation}
which is a $1\sigma $ error of 9\%, similar to the error on the \lndi.

\subsection{Discussion of \nhi\ Value and its Error}
\label{herror}

We now discuss our \lnhi\ value and its
error.  We discussed a similar set of issues for the \lndi\ value in
\S \ref{derrors}.

\subsubsection{The Coupling of the Continuum, \lyaf\ and the \lnhi\ }

The flux emitted by the QSO varies smoothly over scales of thousands
of \kms .  Its precise shape is unknown {\it a priori} and is hard to
gage from spectra because of the absorption and emission lines. 

The \lyaf\ lines are narrow relative to the uncertainties in the shapes of the
continuum and the DLA. We
can identify and fit these lines individually, and we can distinguish
their effects from the shape of the emitted flux and the DLA.  It is
unlikely that such lines would have the velocities and column
densities which make absorption that varies smoothly over hundreds or
thousands of ~\kms , however the optimizer will attempt to do this to
accommodate errors in the \lnhi\ and the continuum.

The damped \lya\ line has a very well determined shape that gives the
\lnhi .  The information on the \nhi\ comes from the whole profile of
the damped \lya\ line.  While the line is most conspicuous over the
central 20~\AA , the damping wings extend much further, and continue
to absorb about 1\% of the flux from approximately 4233~\AA\ to
4340~\AA .  This range covers approximately $\pm 3700$~\kms\ in either
direction from the line center, or 3500 pixels in total, and extends
over the peak of the QSO \lya\ emission line at 4331~\AA .

In Fig. \ref{deltaflux} we show the change in the flux corresponding
to a change in \lnhi\ of 0.04~\cmm , the $1\sigma $ error that we
quote.  This is the derivative of the flux with respect to the column
density, in different units.  The change in the flux is largest $\pm
5.4$~\AA\ or 380~\kms\ from the line center, where the line absorbs
60\% of the flux.  The change in the flux decreases nearer to the line
core because little flux remains there. The fractional change in the
flux keeps increasing toward the line center, but we can not detect
this.  The information in the spectrum that gives \lnhi\ comes from
the whole of the \lya\ line. The change in the flux is significant in
the core where there is almost no flux, and also far from the core
where we can integrate over a wide wavelength range.

The error on the \lnhi\ value for the DLA is larger than would be
thought from considering just the change in the spectrum produced by
changing the DLA line alone, because the \lyaf\ and continuum can be
adjusted to accommodate some of the effect of changing the \lnhi.

Figure \ref{deltaflux} shows the change in flux required of a fit to
the spectrum when the \lnhi\ changes slightly. If we start with a
model that gives a good fit to the spectrum and then increase \lnhi, we
could maintain a good fit if we add bumps to the continuum
approximately 5.4~\AA\ from the center of the DLA. For small changes
in \lnhi\ the optimizer is able to avoid adding bumps, and instead it
can find good fits using different smooth continua shapes, and
different amounts of \lyaf\ absorption; less where the dips were
expected and more on either side of them.
The optimizer adjusts all
parameters of the model so that the residuals appear uniformly
distributed.  This is one type of behavior that explains why the
errors on \lnhi\ are greater than we expect by varying \lnhi\ alone.

We now discuss in more detail the connection between the error on
\lnhi\ and how we fit the \lyaf\ and the continuum.

\subsubsection{Fitting the \lyaf}
\label{fitting}

The models accommodate different \lnhi\ values by changing both the
continuum and the \lyaf\ together.  We must be careful to not allow
the optimizer too much freedom to adjust the \lyaf\ parameters to
simulate smoothly varying absorption, to accommodate an excessive
range of \lnhi.  For example, when we divide the spectrum by a DLA
with \lnhi\ = 19.8~\cmm\ we make a spike in the flux near 4293~\AA\
that is unlike any feature seen in QSO spectra. Hence, we should not
allow the optimizer to obtain acceptable fits for this \lnhi\ value by
raising the continuum over a wide wavelength range and using smooth
\lyaf\ absorption to remove this extra flux everywhere, except at the
spike.  This is reason that we require all of the \lyaf\ $b$-values
to be less than 150 \kms.

We may have over-estimated the error on the \lnhi\ if we have allowed
the \lyaf\ lines to have more variety than is typical.  The \lyaf\ can
accommodate changes in the \lnhi\ in two ways: by making a few lines
unusually broad, and by using more lines than normal.  We rejected the
fits that used lines with $b > 150$~\kms, a value that is not well
defined. If we instead reject fits with $b < 100$~\kms , we reduce the
allowed range of \lnhi.

Alternatively, we may have underestimated the error on \lnhi\ because
we did not adequately explore the \lyaf\ absorption. We began all
restarts using the parameter values that deviated about the values of
the initial fit.  Perhaps the deviations that we gave to the \lyaf\
lines were too small to allow them to adequately fit the continum
required for differing \lnhi . The deviations that we gave to the
\lyaf\ parameters were uncorrelated, so that increased absorption by
one line would tend to cancel decreased absorption by its
neighbors. We might have explored the \lyaf\ more thoroughly if we had
used correlated deviations.

We have not performed any quantitative checks of whether we have used
an accurate \lyaf\ absorption model, but the excessive \chisq\ value
of our model (\S \ref{chisqscale}) suggest that we may have
systematically used too few components in our \lyaf\ model.  We
suspect that the effect this has \nhi\ is systematic, though we do not
know in which direction.  We also suspect the effect is small, in part
because two of us developed independent models of the \lyaf\
absorption, both of which gave the same $\lnhi = 19.73\ \cmm$.  It's
hard to proceed quantitatively with this issue because it is extremely
difficult to develop alternative models of the \lyaf\ absorption
(number of lines, their approximate positions, etc) which are equally
compelling to the one we used.

\subsubsection{Continuum Shape}
\label{contshape}

Because the DLA absorbs over a very large wavelength range, we suspect
that the largest class of systematic errors which might affect the
\nhi\ are those where the flux emitted by the QSO can not be
adequately fit by our continuum model.  This might occur, for example,
if there are weak emission lines or high frequency errors in the
relative flux calibration, in the vicinity of the DLA.

The continuum shape can be effected by errors in the flux
calibration. These errors are approximately 3\% near the DLA, from
comparison of HIRES spectra that we calibrated in different ways.  We
find the same \lnhi $= 19.73$~\cmm\ gives the lowest \chisq\ values
when we fit the DLA in HIRES spectra that we calibrated in different
ways.

The models described above all used only two continuum control points
that were free to move in both flux and wavelength near the DLA. We
also performed re-starts with three such continuum control points in
the region between 4260~\AA\ and 4325~\AA.  The results of these
re-starts are shown in Figure \ref{f1013_v2}.  The top panel contains
a number of fits that are unreasonable, having either very unusual
\lyaf\ absorption (like some fits in the top panel of
Fig. \ref{f1014_v1}), or very bumpy continua that we could not make
with only two continuum points.

We filter the results using 3 continuum control points to exclude the
unreasonable fits, leaving the results shown in the bottom panel of
Figure \ref{f1013_v2}.  As with the two continuum point re-starts, we
first removed all fits containing \lyaf\ absorbers with $b > 150$~\kms .  
In addition, we removed models having continuum bumps larger
than we expect to be present due to either weak emission lines, broad 
shallow absorption or errors from the relative flux calibration. 

To help us decide what size of bumps and dips were
reasonable in QSO continua, we fitted the \lyaf\ of three other QSOs
in a similar manner to \object .  Our HIRES spectra of these QSOs had
similar, or in two cases a bit lower S/N. Near the position of the DLA
in \object, approximately 3000~\kms\ from the peak of the \lya\
emission line, we did not require any bumps or dips in the continuum
that exceeded approximately 2\% in amplitude, similar to the error in
the relative flux calibration for \object.

The filter that we used to reject bumpy continua was guided by this
result, but it was imperfect in its treatment of the \lya\ emission
line.  Examples of continua that it ejected are shown in Figure
\ref{f1002ab_v1}, and those that it accepted in Figure
\ref{f1002bb_v1}.  

We find that the results from the models with two-
and three-control point in the continuum are similar. This suggests
that we have explored an adequate range of continuum shapes to
accommodate the emitted flux, weak emission lines, broad shallow
absorption and errors in the relative flux calibration.

The first thing that we note is that the best model with
three-continuum points has $\lnhi = 19.73$~\cmm, the same value as
with two control points.  Second, the range of solutions is also
approximately the same; $\lnhi = 19.71$ to just above $\lnhi = 19.78$,
which was unfortunately the highest \lnhi\ we explored with three
points.

The third finding is that most of the filtered 3-point continua with
acceptable \chisq\ had shapes similar to those from the 2-point fits.
In Figure \ref{f1022} we compare the continua from the two and
three-point continua, for the models with that gave the lowest \chisq\
values for \lnhi = 19.69~\cmm , the value $1\sigma $ below our
estimate for \lnhi .  They differ by at most $< 1$\% in flux, which is
less than the error from the flux calibration.  In Figure
\ref{f1023} we show the same for \lnhi = 19.783~\cmm , our best
estimate for \lnhi . The differences are at most $\simeq 0.3$\%.  For
\lnhi = 19.783~\cmm , shown in Figure \ref{f1024}, the differences
are $<3$\%.

We use the 2-point rather than the 3-point continua to define the
\lnhi\ because we have explored the former in more detail, and the
latter give a large number of fits with unacceptably bumpy continua,
that we can not readily filter.

Although the 2 and 3-point continua give very similar results, we are
not certain that we have explored an appropriate range of continua. We
might have allowed either too much or too little freedom to represent
the continuum, and we do not know how such errors might effect the
\lnhi\ value and its error.

\subsubsection{The Largest Acceptable \chisq\ Value}
\label{chisqscale}

As we stated in \S \ref{besth}, we increased the maximum acceptable
\chisq\ by a factor of 1.23. We did this scaling because we found that
when we fit a similar region of the \lyaf\ without unusually strong
lines, we obtained a \chisq\ value of 6111 over 5290 pixels using 336
parameters. This is \chisq\ per degree of freedom is much larger than we
would expect.

One possible explanation is that we are under-fitting the
\lyaf. This is reasonable because we fit lines only where they are
clearly required.  We did not fit lines that changed the residuals by
less that approximately $1\sigma $, and hence we have underfit the
\lyaf\ because the need for such lines would be apparent where the S/N
higher.  We did not fit such lines because the spectrum gives little
guidance in how to obtain a lower \chisq . The residuals, like those
shown in Figure \ref{f1008_v1}, do not show the need for distinct
lines.

A less likely explanation is that the error array associated with the
spectrum is systematically too small.  When we fit 5882 pixels to the
red of the \lya\ emission lines, using just 32 parameters, we found a
\chisq\ value of 5264, or 0.90 per degree of freedom, which suggests
that the errors are on average too large.

Several factors will determine the \chisq\ value we obtain when we fit
the \lyaf. The \chisq\ will decrease when we add lines.  As the S/N
increases, we need more lines to reach a given \chisq\ per pixel, and
the higher S/N helps show that these lines are needed.  The S/N
decreases systematically with decreasing wavelength in our HIRES
spectrum, and hence we do not know if the 1.23 factor applies at all
wavelengths.  In addition, the \lyaf\ evolves, with more absorption at
higher redshifts, until we near the QSO, where the proximity effect
reverses the trend, further complicating the issue.

Although we can not readily determine the accuracy of the correction
factor of 1.23, Fig. \ref{f1014_v1} shows that the \lnhi\ range is
rather insensitive to the precise maximum acceptable \chisq\ value.

\subsubsection{Speculation on the Errors Associated with the \chisq\ Range}
\label{random}

The portion of the error on the \lnhi\ that comes from only the
quality of the fit and the S/N of the spectrum, might be up to an
order of magnitude smaller than 0.04~\cmm . We speculate this because
the 0.04~\cmm\ error corresponds to a range of \chisq\ of 250, while
we could have used a range of \chisq\ of 1 had we adequately explored
both the parameter and model space.  Because we have not done the
required exploration, we do not know the shape of the $\chi ^2_{min}$
as a function of \lnhi\ to the required accuracy.  If we make the
guess that the shape remained similar to that suggested in
Fig. \ref{f1014_v1}, then the range of \lnhi\ corresponding to $\delta
\chi^2 =1$ would be much less than 0.04~\cmm.  Of course, the central
value could end up anywhere in the range indicated by Equation
\ref{besthequation}.

\section{Abundances and Ionization of the Heavy Elements in the Absorbing Gas
         \NOTE{section\_d.tex}}
\label{metallicitysec}

In this section we discuss the metal abundance, ionization, and
physical conditions in the components (1 and 2) where we measure D/H.
In \S \ref{otherabund} we discuss the components at $-40$ and
100~\kms.

We modelled the level of photoionization using CLOUDY version
94.00, developed by G. Ferland (Ferland, 1991).  We used the solar
Oxygen abundance log O/H $= - 3.31$ from Allende, Prieto, Lambert \&
Asplund (2001).  For other elements we used the CLOUDY defined solar
abundance ratios.  We assumed a plane-parallel geometry and we
approximated an isotropic background by placing a point source at a
very large distance. We used the Haardt-Madau (Haardt \& Madau,
1996) ionizing spectrum at $z=2.526$.  The ionization is given by the
parameter $\log U$, where $U \equiv \phi/{c n_{H}} = J_{912}/{4\pi h c
n_H}$, 
$\phi $ (\cmm ~s$^{-1}$) is the surface flux of ionizing photons,
and $J_{912}$ is the intensity of
the incident radiation at 1 Rydberg.  We set $J_{912} = 10^{-21}$ ergs
\cmm\ s$^{-1}$ Hz$^{-1}$ sr$^{-1}$ , a typical, if slightly high value
for the intensity of the intergalactic radiation field (Scott \etal\
2000; Hui \etal\ 2002; but see Prochaska 1999). 
The gas density, $n_H$~\cmmm , is then related
to the ionization by log~$U = -4.34 - log~n_H$ (\cmmm ).

\subsection{Metal Abundance in Components 1 \& 2}

Photoionization models show that the gas in components 1 \& 2 
has relatively low ionization and 
hence we used O~I/H~I to find [O/H] $\simeq -2.79$.

We ran a set of CLOUDY models for \lnhi $=19.73$~\cmm\ and metal
abundance [X/H]$ = -2.77$, similar to the value we expected to find
based upon the observed O~I/H~I ratio.  We considered all the gas in
components 1 and 2 together because there are large errors on the
fraction of each ion in component 2.  Figure \ref{cloudy1975} shows
the predicted column densities for ions of interest as a function of
gas density.  

The absence of strong C~IV and Si~IV shows that the density is high
enough that the metal abundance can be taken from the O~I/H~I ratio.
The components that we fit to C~IV and Si~IV nearest to $v=0$~\kms\
have $b \simeq 38$~\kms\ , very different from the $b \simeq 7$~\kms\
that we expect for components 1 \& 2.  In Table \ref{table_c} we list
upper limits on the column densities that could be associated with
components 1 \& 2, obtained when we fit lines with fixed $b=7$~\kms\ and
$v=0$~\kms .  The spectrum allows these column densities, but not
much larger values.
We also list upper limits that we found for Si~III and
Al~II.  These limits all require relatively high densities: 
$\log n_H $ ~(\cmmm ) $> $ --2.3
(Al~II), --1.7 (Si~IV), --1.6 (Si~III) and --1.5 (C~IV).

We obtain the [O/H] from the observed O~I/H~I, making a slight
adjustment for ionization.  The O~I column density in components 1 \&
2, $13.63 \pm 0.02$~\cmm , equals the model prediction for [O/H]
$=-2.77$ and log~$n_H = -2.24$~\cmmm .  Since the density is relatively
high, the
[O/H] is insensitive to the precise density. The other ions prefer a
higher density of $\log n_H \simeq -1.5$~\cmmm, where the predicted
O~I column density is 0.02~\cmm\ higher than we observe.  Hence we
conclude [O/H] $=-2.79 \pm 0.05$, which includes the error on \lnhi .
Had we chosen a much higher density of $\log n_H =
+1$~\cmmm , the [O/H] would decrease by only 0.01.  We checked that
that the [O/H] value is insensitive to the shape of the ionizing
spectrum.

\subsection{Ionization, Size and Mass of Components 1 \& 2}
\label{ionization12}

Overall, the column densities of the ions indicate log~$n_H \simeq
-1.5$~\cmmm , or $\log U \simeq -2.84$.  Although a lower density of
$\simeq -2.1$~\cmmm\ is preferred by Si~II and is allowed by Al~II,
the C~IV and Si~IV rule this out, and the C~II prefers a higher
density, $\log~n_H \simeq -0.44$~\cmmm .  
We used the column 
densities in the components at the velocities that we footnote in
Table \ref{table_c}.
The fit is improved if we add C~II at $v=-5.6$~\kms , or if C is 
slightly under-abundant and Si
is over-abundant, but none of these differences are significant
because the column densities are not well known. 

The level of ionization indicated by the ions implies that the gas is
mostly ionized. Figure \ref{abs_size} shows the neutral fraction is of
order 21\% for $\log U=-2.84$.

The $\log n_H$ value implies an absorbing region of typical size and
mass.  The bottom half of the Fig. \ref{abs_size} shows the size for a
given ionization, for either constant $n_H$ or constant $J_{912}$.
For a given $\log U$, $n_H$ is proportional to $J_{912}$ by
definition, the size is proportional to $n_{H}^{-1}$, and the mass to
$n_{H}^{-2}$.  For an ionization of $\log U = -2.84$ and $J_{912} =
10^{-21}$ ergs \cmm\ s$^{-1}$ Hz$^{-1}$ sr$^{-1}$ , the size is 2700
pc along the line of sight, and the mass of H in a sphere 
with this diameter, and the density $\log n_H = -1.5$~\cmmm\ is $\rm
8.0 \times 10^6 M_{\odot}$.

Both the size and mass of the absorbing region are uncertain by at
least an order of magnitude because $J{912}$ and especially
the gas density are uncertain. Consider three types of change we can make to
lower the sizes and masses: we can increase the density, we can
decrease the ionization, or we can increase the density and the $J_{912}$
in proportion to maintain the ionization, which we prefer.  
For example, if we increase both the $J_{912}$ and the $n_H$ by 10 times, the
ionization is unchanged, the size drops by 10 times, and mass by 100 times.  

If we consider low ionization solutions, we can have high gas
densities that give small sizes and masses.  We view this
alternative as unlikely, because at lower ionization we expect less
Si~II than we see, and less Al~II than allowed by the upper
limit. This is possible if most of the absorption from these ions is
from different gas.  However, this would also require C~II to come
from different gas, yet Fig. \ref{lowions1} suggests that much of the
O~I, C~II and Si~II do some from the same gas, and the ionization
model is consistent with this.  But if we do consider lower ionization
solutions and we keep the density constant at $\log n_H = -1.5$ \cmmm , 
the size comes down by at most a factor of 5, and the mass by a factor of 125.
This is because the gas becomes neutral, at which point the size
depends only on the density.

The properties of the gas that shows D towards \object\ are typical.
Table \ref{dhinferred} summarizes the inferred total gas column
densities, sizes and gas masses of six D/H absorbers.  For consistency
we scaled the gas densities, and hence sizes and masses so that the
observed ionization was obtained for $J_{912} = 10^{-21}$ ergs \cmm\
s$^{-1}$ Hz$^{-1}$ sr$^{-1}$. We see that the total gas column
densities, log~H, may all be very similar.  The sizes are of order 1
kpc, but with a huge range covering over a factor of 1000, while the
masses cover an even larger range. 

Finally, CLOUDY provides estimates for the gas temperature that we
show in Fig. \ref{gas_temp}.  For $\log U \simeq -2.84$ we expect an
equilibrium gas temperature near 10,000~K. At this temperature we
expect $b$(D-1) $> 9.1$~\kms , similar to the observed value. We saw
in \S \ref{linewidth} that the widths of the O-1, D-1, and H-1
components together indicate $T=0.55 \pm 0.04 \times 10^4$~K, which is
lower, perhaps because the gas is, after all,  neutral or the ionizing
spectrum is softer than the one we used.

\subsection{Abundances in other components}
\label{otherabund}

Photoionization calculations suggest, with large uncertainty, that components
3 near $v \simeq -40$, and 4 near $v \simeq +100$~\kms , have different
metal abundances from each other and from components 1 \& 2.

We performed additional photoionization calculations for a metal abundance
$[X/H] = -1.5$ and $n_H = 0.001$ \cmmm\ and
\lnhi $=16.00$ \cmm\ (component 3) and 
\lnhi $=16.50$ \cmm\ (component 4).  
The $n_H$ was determined using the relation between column
density and physical density as given by Hellsten et al. (1998).
Figure \ref{cloudylowcol} illustrates the results.

For component 3 we estimate [C/H] $= -0.5$ to $-1.5$, from Si~II/C~II
and the lack of Si~IV and C~IV. The ionization is in the range
$-5 < \log U < -2.75$, and the Si is not enhanced.
The equilibrium temperature requires $ 13.36 < b_{H} < 17.50 $~\kms, when 
we assume a turbulent velocity width of 3 \kms, as implied by the widths
of the Si~II and C~II absorption. This range is consistent with the
grid search value $b(H-3) = 16.3 \pm 0.7$~\kms (\S \ref{grid}).

For component 4 we estimate [Si/H] $\simeq $ [C/H] $\simeq -1.9 $ with a
large uncertainty, because different ions suggest different ionization.
If Si is enhanced by 0.3 dex, C~II/Si~II suggests $\log U \simeq -1.8$
and [Si/H] $\simeq $ [C/H] $\simeq -1.7$. However, C~IV/C~II 
and C~IV/Si~IV
indicate higher ionization, $\log U \simeq -2.6$, and [Si/H] $\simeq -2.1$.

Three of the components of the absorption system at $z \simeq 2.526$
appear to have very different abundances: $-2.79$ for the gas in
components 1 \& 2, $-2$ for component 4 and $-1$ for component
3. We note that the component with the most H~I and the
lowest ionization also has the lowest metal abundance.
\section{Best Fit Values For \qfive \NOTE{section\_n.tex}}
\label{best1243dh}

We have made the following estimates for the parameters that
describe the gas in which we measure D/H:
\begin{itemize}
\item The velocity structure of the gas has two main components that 
      we see in the asymmetry of the O~I line.
\item Our grid search of models that fit the D lines showed that the
      components are separated by 12~\kms , with a $1 \sigma$ range of
      8.5 -- 14~\kms , consistent with the asymmetry of the O~I line.
\item The main D component, D-1, is at $z = 2.525659$ and has $b = 9.2
      \pm 0.2$~\kms .
\item The wavelength, $b$-value and column density show that D-1 is D
      absorption. We believe that the D-2 component is D because it is
      very narrow, the separation of the D-1 and D-2 components is
      the same as that of the O-1 and O-2 components, and the fraction 
      of the D in D-1 is similar to the fraction of the O in O-1 
      (\S \ref{itisd}). However, these arguments are weak because D-2
      is not well defined.
\item The grid search also indicated that component D-2, at the larger
      redshift, contains 12\% of the total D column density, with a
      range of 7\% to 28\% depending on how much of the absorption
      near component 2 is H in another component, H-3.
\item Total \lndi = $15.113 ^{+0.042}_{-0.026}$ \cmm , (i.e. $+$10\%,
      $-$6\%) where the error range includes fits with $\chi^2 <
      \chi^2_{min} +1$.  The errors are larger for higher values of
      \lndi\ because of the second component can make a significant
      contribution.
\item Total \lnhi $= 19.73 \pm 0.04$~\cmm\ (9\%), where the error is a
      Gaussian distribution centered on the \lnhi\ value that
      consistently gave the best fits. We set the width of the
      Gaussian to give an 80\% chance that the true \lnhi\ lies in the
      range 19.68 -- 19.78~\cmm\ where we found acceptable fits to the
      \lya\ line, the continuum and \lyaf.
\item log~(D/H) $=-4.617 ^{+0.058}_{-0.048}$ (+14\%, --10\%).
\item D/H $=2.42^{+0.35}_{-0.25}\times $\mf ~\cmm .
\item{ [O/H] = $-2.79 \pm 0.05$.}
\item The absorber is probably mostly ionized, with of order 21\% of H atoms
      neutral (\S \ref{metallicitysec}).
\end{itemize}

\section{THE PRIMORDIAL D/H RATIO \NOTE{section\_e.tex}}
\label{pridh}

In this section we compare D/H measurements from different QSOs, we
discuss the dispersion in these value, and whether D/H might correlate
with metal abundance or \nhi, and we give our estimate for the
primordial D/H value.

\subsection{The Weighted Mean D/H from Five QSOs and the Dispersion of the
Values}
\label{weighted}

We find the weighted mean of the D/H values from five QSOs and we show that the
individual values show more dispersion than we expect.
In Tables \ref{table_f} \& \ref{table_g} we list all reported D/H
measurements that remain viable.  We have previously measured D/H in
three QSOs (Tytler, Fan \& Burles 1996; Tytler \&\ Burles 1997; Burles
\&\ Tytler 1998a; Burles \&\ Tytler 1998b; O'Meara \etal\ 2001), and
placed a strong upper limit on D/H in a fourth (Kirkman \etal\ 2000).
We will also use the D/H measurement by Pettini \& Bowen (2001)
towards Q2206--199, although this measurement is less secure because
the HST spectra are of much lower S/N and resolution than those from
the ground.  We discuss Q2206--199 and QSOs that we will not use
in Appendix III (\S \ref{otherqsos}).

The weighted mean of the first five log~D/H values from Table \ref{table_g}
is,
\begin{equation}
\label{meandh}
log~D/H =-4.556, 
\end{equation}
where the weights we use are the $1\sigma $ errors on the quantity 
$Y_i = log (D/H)_i$.  We use log values because they were used to find all
but one of the individual D/H values and errors.  We obtain a slightly
smaller mean D/H if we instead work with the linear D/H values.

The D/H measurements towards the five QSOs are more dispersed than we
expect.  In O'Meara \etal\ (2001) we noted that the dispersion in the
first three measurements was larger than expected, with a 3\% chance
of a larger $\chi ^2$ value. We interpreted this to mean that we had
underestimated one or more of those errors.  With the addition
\object\ the dispersion of log D/H remains approximately 0.10, but
adding the low D/H from Q2206--199 increased the dispersion to 0.14,
the $\chi ^2$ value for all five measurements increases to 12.35 for 4
degrees of freedom, and the probability that we would have obtained a
larger $\chi ^2$ value by chance drops to 1.5\%.

\subsection{Factors that Determine Measurement Errors on D/H Values}
\label{measerrors}

The accuracy and reliability of a D/H value will depend on many
factors, that fall under three main headings:
\begin{itemize}
\item The details of the absorption system: the \nhi\ value, the
   number of velocity components, their velocities, column densities
   and $b$-values, the chance placement of \lyaf\ lines and other
   contamination, the redshift and the flatness of the continuum near
   the key lines.
\item The quality of the spectrum: the spectral resolution, the S/N,
   the ions observed, and the accuracy of the wavelengths and relative
   flux calibration.
\item The adequacy of the model: 
   Have all possible velocity structures been explored?  
   Have alternative line identifications been considered?
   Where contaminating and blended lines fit?
   Were the hidden components explored?  
   Is the continuum over- or under-fit?  Have the continuum, \lyaf\ and
   the lines that give D/H been fit simultaneously?
\end{itemize}

The dominant error with any particular D/H value might depend on any
of these factors, or some combination of them.  Many of these factors
are different between the D/H values, and hence each value can have a
different dominant error.  We do not expect the measurement errors, or
any uncorrected systematic errors to correlate with a single
parameter, such as the \zabs , \nhi\ or metal abundance. Although
unlikely, we do not know enough to rule this out.

The errors that are associated with any D/H value could be too small
for any of the factors in the list above, such as: poor flux
calibration, especially near the Lyman limits, inadequate fitting of
the \lyaf , and inadequate modelling of the continuum and velocity
structure, and especially, inadequate exploration of all possible
models.  The error on a D/H value could be larger than that deduced
from the \chisq\ values for any of these reasons.

\subsection{The Dispersion in the D/H Values May Come from 
Measurement Errors}
\label{dispersioniserrors}

We suspect that the dispersion in the D/H values arises from
measurement errors and is not real. If the measurement errors have
been underestimated for at least one QSO, then we can explain the
excess \chisq\ value.

The dispersion of the D/H values is not much larger than we expect.
In Table \ref{table_g} we list $X_i = (Y_i - \rm mean)/\sigma (Y_i)$, the
deviation of each measurement from the weighted mean, in units of the
individual measurement errors.  The D/H from Q1009+2956 is $1.95
\sigma $ above the weighted mean, while the D/H from Q2206--199 is
$2.17 \sigma $ below.  A \chisq $<9.5$ would have been expected if all
five log D/H values were consistent with the weighted mean, since
Prob$(\chi ^2_4 > 9.5) = 0.05$.  We would obtain \chisq $< 9.5$ if
either \qtwo\ or Q2206--199 were within $1\sigma $ of the mean, or if
the measurement errors on all five QSOs were increased by 1.14, both
small changes. However, to obtain a typical \chisq = 3.36, where 
Prob$(\chi ^2_4 > 3.36) = 0.5$, 3 or more D/H values, or their errors,
would need to change, or the measurement errors on all five QSOs would
need to increase by a factor of 1.92. These are a large, but still
credible changes.

The methods that we introduced in this paper have allowed us to
explore some of the issues that effect a D/H value more thoroughly
that in past work.  We see many ways in which the errors might have
been underestimated, both here, and in past work. In \S \ref{derrors}
and \S \ref{herror} we discuss factors that might change the $1\sigma
$ error that we give for the D/H value from \object\ by a factor of
two. We further believe that the D/H value for \object\ has benefited
from the most thorough exploration of these error related issues. We
would not be surprised if the errors on some of the other D/H values
were too small by a factor of two.  

We are now more confident that measurement errors explain the
dispersion in D/H values than we were in O'Meara \etal . Our work on
\object\ better illustrates the many ways in which errors can be
underestimated. We see that the three D/H values that we suspect are
the least reliable are also the values farthest from the mean.  For
\qone\ and \qtwo\ some of the data reduction and analysis methods were
less reliable than those that we used for HS0105+1619 and Q1243+3047,
and for Q2206--199 there is much less data available.

\subsection{How D/H Depend on Metal Abundance: D/H Chemical Evolution}
\label{astration}

The mean D/H from QSOs is similar to the primordial D/H
value that we would predict from
the D/H and metal abundance in the local interstellar medium (LISM) using 
standard Galactic chemical evolution. However, chemical evolution can not 
account for the dispersion in the D/H values from QSOs.

 With the two latest D/H measurements, there is no
longer a hint of a correlation between D/H and metal abundance in the
QSO absorbers that we noted in O'Meara \etal\ (2001).  Prantzos \&
Ishimaru (2001) showed that standard chemical evolution could not
reproduce the correlation, while Fields \etal\ (2001) discussed an
unconventional scenario that could.
In Fig. \ref{p10} we show D/H values against metal abundance, mostly
measured with Si when the ionization is high, and O otherwise (Table
\ref{table_g}). 

With the elimination of the correlation, Galactic chemical evolution
more clearly supports the idea that our D/H measurement towards
\object, and the mean D/H from five QSOs are very close to the
primordial D/H value.  We include two curves in Fig. \ref{p10} that
show the expected decrease of D/H against metal abundance in a simple
closed box model, without in-fall, and with the instantaneous recycling
approximation (Tinsley 1974; Tinsley 1980; Ostriker \& Tinsley
1975). We normalized the two curves in different ways. The solid curve
uses the mean D/H from Eqn. \ref{meandh} as the primordial D/H and
predicts D/H as a function of metal abundances, including that in the
LISM.  The dotted curve is normalized to
give the D/H abundance in the LISM (Moos \etal\ 2002; Oliveira \etal\
2002) and it then predicts the primordial D/H abundance, (D/H)$_p$.  
To draw these curves, we use the equation
\begin{equation}
\label{box}
{\rm D/H}={\rm (D/H)_{p}} \times {\rm exp} \left(
-\frac{Z}{y}\frac{R}{1-R} \right),
\end{equation}
and the values for two parameters from Prantzos \& Ishimaru(2001); the
returned mass fraction $R=0.31$, and the yield $y=0.6$, where $Z$ is
the metallicity, or metal abundance in solar units, on a linear scale.
These parameters were derived from a Salpeter IMF and the yields of
Woosley \& Weaver (1995).  The choice of parameters and D/H chemical
evolution are discussed in Steigman \& Tosi(1992), Vangioni-Flam,
Olive \& Prantzos (1994), Galli \& Palla (1995), Prantzos (1996) and
Prantzos \& Silk (1998).

Both curves connect the D/H values towards \object\ and \qfour\ to the
D/H in the LISM, over three orders of magnitudes in metallicity.
The curves show that the decline in D/H, as stars eject gas that lacks D, is
insignificant when metal abundances are low; e.g. [O/H] $<-1$.  The
simple closed box model (Eqn. \ref{box}) predicts that by the metal
abundance of \object , (D/H) $\rm = 0.9987(D/H)_p$, and for \qfour ,
0.986.  We have not applied these corrections because they are much
smaller than our measurement errors.

We predict the primordial D/H when we normalize the simple model to
give the D/H in the LISM. Ostriker \& Tinsley (1975) predicted the
primordial D/H would be 1.5 -- 2 times that in the LISM.  With the
modern parameters given above, the same model predicts 1.62 times,
again without using any D/H measurements.  Using the D/H and $Z$
measurements in the LISM from FUSE and HST spectra, $ {\rm D/H} _{\rm
LISM}$ $= 1.52 \pm 0.07 \times $\mf , and [O/H] $=-0.189$ or $Z=0.647$
(Oliveira \etal\ 2002, and using the solar $\log \rm O/H = -3.310 $ from 
Allende Prieto, Lambert \& Asplund 2001), the predicted
primordial abundance is D/H $= 2.47 \pm 0.13 \times$\mf , very similar
to the value from \object\ and from Eqn. \ref{meandh}: $2.78 \times
$\mf . The errors on the prediction are substantial.
When we allow the two parameters to simultaneously take values
that maximize the change in D/H, from $0.26 < $ R $ < 0.36$ and
$0.5Z_{\odot} < {\rm yield} < 0.9Z_{\odot}$ (Prantzos, private
communication), we find D/H $= 1.94$ -- $3.16 \times$\mf .  The
range is further increased when we consider in-fall of gas to the
Galactic disk (Lubowich \etal\ 2000) and dispersion of D/H in the LISM
(Vidal-Madjar \etal\ 1998; Sonneborn \etal\ 2000).
Hence, although the D/H values from both \object\ and \qfour\ are closest to 
the predicted D/H, we can not use the chemical evolution model to rule out 
primordial D/H values that are suggested by the measurement to the other QSOs.

\subsection{Does D/H Depend on \nhi ?}
\label{dhvnhi}

The D/H values appear to decline with increasing \lnhi . We do not believe that
this correlation is real, because we suspect that the dispersion in the D/H 
values is not real (\S \ref{dispersioniserrors}). 
We will discuss potential systematic errors, and then potential cosmological 
and astrophysical origins for a correlation.

In Figure \ref{p9v1} we see a clear trend of declining D/H with \nhi .
This trend was apparent with just the three D/H values discussed by
O'Meara \etal\ (2001) and it was accentuated by Q2206-199 from Pettini
\& Bowen (2001). The trend rests upon the relatively high D/H for the
two LLS (\qone\ and \qtwo ), and the relatively low D/H for Q2206--199
and \object\ that we will call DLAs.

\subsubsection{Systematic Measurement Errors that Depend on \lnhi }

We are not aware of any systematic error in the measurements of \qone,
\qtwo, \qfour\ and \object\ that could readily account all of the
trend with \nhi.  Rather, we expect that potential errors are complex,
and specific to each spectrum (\S \ref{measerrors}). However, we can
think of three types of systematic error that might depend on \nhi.

One possibility is that we have included absorption by H in the D
measurements, making D/H too large. Hydrogen absorption is more likely
to appear like D in the LLS compared to the DLAs for three reasons.
First, the LLS have weaker D lines, and hence small amounts of H can
have a larger effect than with the DLAs.  Second, in the LLS we see
only the D \lya\ line, and hence we have less information on the D
velocity structure than we have for the DLAs. Third, the $b$ values
for the H and D lines in the LLS are larger than those in the systems
with higher \nhi\ values, which makes the D lines less different from
common H lines. A strong argument against this possibility is that the
D lines in both \qone\ and \qtwo\ had $b$-values that were smaller
than \lyaf\ lines, and that agreed with predictions from their H and
metal lines that we collect in Table \ref{table_f}.

A second possibility is that we have underestimated the \nhi\ for the
two LLS. The method of measuring \nhi\ is different for the LLS than
for the systems with higher \nhi , hence this could be a systematic
effect.

A third possibility is that we have missed some of the components of
D, making D/H too low in some QSOs (\S \ref{missingd}).  D components
at positive velocities can hide in the parts of the spectrum where the
main H absorbs.  This can happen in any system where the $v$
distribution of the H~I is poorly known, and the H absorption extends
over nearly 82~\kms , enough that the H with large positive velocities
has D that lies among the absorption of the H with negative
velocities.

We do not know whether missing components are more likely at low or
high \lnhi .  It would be header for this effect to go undetected in
systems with the lowest column densities and unsaturated Ly series
lines, like \qtwo .  We might expect this effect to be worst in
systems with the highest \lnhi\ because the H lines are then
widest. However two other factors make it easier to find components of
the systems with higher \lnhi\ values: the components then have enough
gas to show metal lines, and the $b$ values for the H~I are smaller
than for the LLS.

\subsubsection{Both LLS and DLAs can give Reliable D/H}

In general, both DLAs and LLS can give accurate D/H values, and both
can have significant advantages. On the other hand, both LLS and DLAs
can, in specific cases, give unreliable D/H values.  If we had many
more D/H measurements, we could determine observationally the fraction
of DLAs and LLS that had given errors, but we can not estimate this
today.

The DLAs have the advantages that we see several D lines, the $b$
values can be much lower than in the LLS or the \lyaf , the D lines
are strong enough that they are less effected by the \lyaf\ alone, and
we can see many metal lines that can help us understand the velocity
structure.

The LLS with the lowest \lnhi\ also have major advantages. They have
unsaturated H lines that give the velocity distribution of the H.  We
can also measure the \lnhi\ independently the drop in flux across the
Lyman break, as we did for \qone\ (Burles \& Tytler 1997) and \qtwo\
(Burles \& Tytler 1998b). These two advantages make the \lnhi\ much
easier to measure than in other systems.  We can also directly compare
the velocity distributions of the H and D, proving that the absorption
is D, and guaranteeing that we are measuring the D and H in the same
gas, and not missing components of either the D or the H.  These
advantages are offset by the greater similarity between the D and H
lines.

\subsubsection{Cosmological and Astrophysical Reasons why D/H Might 
Depend on \lnhi }
\label{astrophys}

If D/H were correlated with \lnhi , the explanation might involve
inhomogeneous BBN, or the creation, removal or destruction of D.  No
such plausible mechanisms are known.

If we ignore measurement errors, the correlation in Fig. \ref{p9}
suggests that the D/H ratio has a range of approximately a factor of
2.4.

Although the correlation is seen against \lnhi , this might not be the
most readily interpreted parameter. In Table \ref{dhinferred} we list
approximate, order of magnitude estimates for the physical conditions
where we measure D/H.  The \lnhi\ values correlate with the ionization
of the gas, and with the approximate density and size and mass of the
absorbing region. The highest \lnhi\ values (low D/H values) tend to
correspond to low ionization, higher gas density and smaller, less
massive absorbing regions.  \object\ stands apart from the trend
because it has both a high \lnhi\ and it is ionized, giving a large
size and mass, perhaps the largest of any of the D/H regions.  Most of
the masses are large, $10^5$ -- $10^9$~M$_{\odot}$, which makes it
hard to make or destroy the D in local events, unless they effect a
large fraction of the higher density portions of the universe.

In SBBN the range in D/H values would correspond to a range of
approximately 2 in the \ETA\ value. This range might arrive is the
universe is has inhomogeneous \ETA\ on large scales at the time of BBN
(Kainulainen, Kurki-Suonio, \& Sihvola 1999).  The scale of the
inhomogeneity would need to be $>1$~kpc, and perhaps, for the \lnhi\
values of the D/H absorbers, $>100$~kpc (Mike Norman, private
communication), to avoid mixing before the time of observation. The
scale is also but is limited to $<1$~Mpc by the near isotropy of the
CMB (Jedamzik \& Fuller 1995; Copi \etal\ 1998; Jedamzik 2002).  The
1~Mpc scale corresponds to a baryon mass of $5.9 \times 10^9$
M$_{\odot}$ that is larger than the typical masses of baryon in the
gas showing D/H.  Inhomogeneity can produce varying D/H, but the power
spectrum of the fluctuations should be cut off at both small and large
scales to prevent the overproduction of \hef\ and \lisv , and to avoid
CMB constraints.  The lack of variation of the \lisv / H in stars in
the halo of our galaxy is a weak argument against inhomogeneities on
these scales, since we expect that the gas that we see in QSO
absorption systems is the type of gas that makes the halo stars. The
argument is weak because the \lisv\ in the halo stars may not be the
primordial value (\S \ref{nuclei}).

Dolgov \& Pagel (1999) present a different scheme (discussed by
Kurki-Suonio 2000; Whitmire \& Scherrer 2000 and Dolgov 2002) that is
inhomogeneous on scales of 100 -- 1000 Mpc, small enough to give
variation in D/H to QSOs, but large enough to avoid large variations
in existing \hef\ and \lisv\ measurements that are all in local
objects.  They employ inhomogeneities of the different neutrino
flavors that add up to give a constant total energy density, thus
avoiding the CMB constraint.

We can speculate that significant D has been destroyed in those parts
of the universe that, by the time of observation, had the largest H~I
column densities. These are regions with larger over-densities
relative to the mean density.  Fields \etal\ (2001) discuss a highly
unconventional chemical evolution models and find three constraints on
the conditions required to destroy significant D while keeping metal
abundances very low.  Most observed baryons must have been inside an
early generation of stars, the early stars must all have had
intermediate initial masses in the range $2 - 8 M_{\odot }$, and they
must not have ejected much C or N.  

Other astrophysical explanations seem equally unlikely (Epstein \etal\
1976; Jedamzik \& Fuller 1997; Fuller \& Shi 1997; Famiano, Boyd \&
Kajino 2001; Pruet, Guiles \& Fuller 2002; Jedamzik 2002).  We can not
hide significant D in dust or molecules because neither are abundant
enough in typical absorption systems. We would require $>10$\% of all
the gas to be molecular if HD/H$_2 \simeq 10^{-4}$, or a large
proportion of all heavy elements to be in dust, neither of which are
likely, except in molecular clouds.

\subsection{Our Estimate for the Primordial D/H from all QSOs }
\label{best5qsodh}

We believe that the best value for primordial D/H is the weighted mean
(\S \ref{weighted}) of the log D/H values for the first five QSOs
listed in Table \ref{fourqsos}:
\begin{equation}
log D/H =-4.556 \pm 0.064,
\end{equation}
($1\sigma $ error of 15\%), which is equivalent to
D/H$=2.78^{+0.44}_{-0.38} \times$ \mf , where the errors are the $1
\sigma $ errors on the mean, given by the standard deviation of the
five log D/H values divided by $\sqrt 5$.  We use this error on the
mean instead of the usual error on the weighted mean because the
individual D/H values show more dispersion than we expect.  The error
on the mean depends on just the dispersion of the D/H values, and the
number of measurements, but not on the errors on the values.  Had the
D/H values been consistent with a single value (\S \ref{weighted}) we
would have used the error on the weighted mean, and if the errors on
the individual values were also unchanged, this error would have been
0.023, or 1/3 of the error we quote.

The new D/H is $0.6 \sigma $ lower than the value we gave in O'Meara
\etal\ (2001), log D/H $= -4.52 \pm 0.06$, because the two new values
since that paper are both lower.  However, the error, which is the
error on the mean in both cases, has not changed
significantly. Although the dispersion in the D/H values is now
larger, the error on the mean is not larger because we have two more
measurements, \qpettini\ and \object .

When we take the primordial D/H from the weighted mean, we 
assumed that the quoted errors on each D/H value are too small by the
same factor.  However, if this assumption is not true, other ways of
combining the measurements of D/H will give a better estimate of the
primordial D/H.  As an example, we could speculate that \object\ and
\qfour\ give the two most reliable D/H values because \object\ has the
most thorough treatment of the errors, and \qfour\ is the simplest
measurement with the most supporting evidence.  A similar line of
argument was explored by Pettini \& Bowen (2001) when they derived a
D/H value using three DLA systems alone, although the value they gave
is no longer acceptable because the D/H to Q0347--3819 has since increased
and now seems the least secure (\S \ref{qlev}).  The best estimate of
the primordial D/H might now be the weighted mean of the D/H values
from \qfour\ and \object\ alone.  We might then reject the other three
measurements because they have less data and they were less thoroughly
evaluated (\S \ref{newmethods}). 
This alternative, two-QSO, mean D/H is not very different:
$\log D/H = -4.604 \pm 0.032$ (7.6\% error), or D/H $=2.49 \times $\mf
, and the corresponding parameters from SBBN are \ETA $= 6.30 \pm 0.30
\times 10^{-10}$ and \obh $= 0.0230 \pm 0.0011$, all $0.8 \sigma $
different from the values that we quote in \S \ref{cospar}, in units
of the errors in that section.

\section{BBN RELATED COSMOLOGICAL PARAMETERS 
\label{cospar}
\NOTE{section\_g.tex}}

We use SBBN calculations to obtain the \ETA\ and \obh\ values that
correspond to the primordial D/H. We use this \ETA\ value to predict the 
abundance of the other light nuclei, and we compare with measurements.
There are differences that may be caused by systematic errors. We also
compare with other estimates of the \obh\ and find good agreement.

Using the SBBN calculations of Burles, Nollett \& Turner (2001), our
best estimate for primordial D/H leads to the following predictions:
$\eta = 5.9 \pm 0.5 \times 10^{-10}$, \obh $= 0.0214 \pm 0.0020$
(9.3\%), $Y_p = 0.2476 \pm 0.0010$ (predicted mass fraction of
$^4$He), $\rm ^3He/H = 1.04 \pm 0.06 \times 10^{-5}$ and $\rm ^7Li/H =
4.5^{+0.9}_{-0.8} \times 10^{-10}$ .  In the above, the error on $Y_p$
is the quadratic sum of 0.0009 from the error on the D/H measurement
and 0.0004 from the uncertainty in the $Y_p$ for a given \ETA\ (Lopez
\& Turner 1999).  We obtain slightly different central values if we
use values from Esposito \etal\ (2000a,b). The differences are 10\%
or less of the error from D/H alone, except for $^7$Li/H (Esposito
\etal\ equations give $4.05 \times 10^{-10}$) and $^3$He/H ($1.06
\times $\mf).  In Fig. \ref{p11} we compare the predicted abundances
with some recent measurements.  The vertical band shows the range of
\ETA\ and \obh\ values that SBBN specifies for our primordial D/H
value. Measurements of primordial \het\ are consistent, but all \lisv\
and most \hef\ measurements prefer lower \ETA.

\subsection{Comparison with the Abundances of other light nuclei}
\label{nuclei}

Here we discuss the other light nuclei produced during BBN -- \het,
\hef, and \lisv -- and why we believe that D/H is preferred over these
elements to determine \ob.

For several years it appeared that the primordial \het\ had not been
measured, because of chemical evolution.  Now Bania \etal\ (2002)
report a limit on the primordial \het /H ratio from their detailed
long term study of 60 Galactic H~II regions and 6 planetary nebulae.
They argue that low mass stars have neither destroyed nor released
significant \het, because \het /H changes little with metal abundance.
In 17 Galactic H~II regions for which the ionization corrections were
relatively simple, they find a mean \het /H $= 1.9 \pm 0.6 \times
$\mf, which will be an upper limit on the primordial ratio if stars
have not on average destroyed \het.  They propose that the best value
for the upper limit on the primordial \het /H is the value they
measured for one H~II region that has the lowest metal abundance in
their sample, the third lowest \het /H ratio, excellent data and a
small ionization correction of 22\%.  They then quote \het /H $ < 1.1
\pm 0.2 \times $\mf\ that is consistent with the value predicted by
D/H and SBBN.  Given the potential complexity of the chemical
evolution of \het, the relatively small range and high mean
metal abundance in the gas where they have made measurements ($0.1 < $
[O/H] $< -0.5$), and the other possible ways of extracting the
primordial abundance from the data, we suspect that the errors are
larger than quoted, as with D and \hef.

The main isotope of He, \hef, is measured in many tens of
extragalactic H~II regions to much higher accuracy than either D or
\het.  But \hef\ is fairly insensitive to $\eta $ and the differences
between the measurements allow a large range of $\eta $, probably
including the value indicated by D/H.  It is likely (Skillman \etal\
1994; Olive \& Skillman 2001) that the systematic errors were
underestimated for many measurements.  Pagel (2000) stated
``systematic errors up to about 0.005 are still not excluded", and
``\yp\ is very probably between 0.24 and 0.25'', while Fields \etal\
(2001) state: ``the systematic uncertainties in the \hef\ abundance
make it difficult to exclude \yp\ as high as 0.25".  The upper end of
this range includes the \yp\ expected from D/H and SBBN.  However,
Peimbert \etal\ (2002) argue
that the systematic errors can be ten times smaller.  Compared to
other measurements, Izotov \& Thuan (1998) report relatively high
values for \hef /H that are closest to the D/H prediction, but at \yp
$=0.2443 \pm 0.0013$ they are still approximately $2 \sigma $ below
our prediction from D/H (in units of their error).  Thuan \& Izotov
(2002) estimate that their 1998 \yp\ value may be to small by 0.0005
-- 0.0010, which still leaves \yp\ $1.5 \sigma$ below the prediction
from D/H.  Ballantyne, Ferland \& Martin (2000) reject nebulae from
the Izotov \& Thuan (1998) sample that may have significant ionization
corrections and find a very high \yp $= 0.2489 \pm 0.0030$, which is
consistent with D/H, but they use an unacceptable decrease in \hef /H
as O/H increases.  Although Izotov \& Thuan found absorption lines
that might explain why Olive, Steigman \& Skillman (1997) found less
\hef\ in some objects, it is uncertain whether the \yp\ is as high as
indicated by D/H.

\lisv\ also prefers a lower $\eta$ than the value given by SBBN and
D/H.  There are many tens of high accuracy measurements of \lisv\ in
the atmospheres of metal poor halo stars. There is very little
scatter, and hence the abundance ratio in these stars is well
determined at 1 -- 2 $\times 10^{-10}$ (Ryan \etal\ 2000; Bonifacio \&
Molaro 1997; Thorburn 1994). With the recent measurement of D/H, \het\
and the \obh\ from the CMB, it is looking increasingly likely that the
\lisv\ in these stars is not the primordial value, but rather a factor
of 3 -- 4 less (Tytler, Fan \& Burles 1996; Charbonnel 2002).  Authors
differ on whether this much depletion is reasonable.  Pinsonneault
\etal\ (2002) quote a primordial \lisv /H $=2.51^{+1.74}_{-0.93}
\times 10^{-10}$ from models of stellar rotational mixing and the
measurements of Ryan \etal\ (1999). This value is $1\sigma $ below the
D/H prediction.  However, Ryan \etal\ (2000) claim that the depletion
correction is only $0.02^{+0.08}_{-0.02}$ in the log, and that D/H is
not consistent with \lisv /H. In Fig. \ref{p11} we show their
primordial \lisv /H value that includes all their corrections,
including their depletion correction.  Vangioni-Flam, Coc \& Cass{$\rm
\acute e$} (2000) also believe that the \lisv\ depletion is small, and
hence that \lisv /H rather than D/H, gives the best \ETA .

We prefer the \ETA\ and \obh\ values from D/H rather than the lower
values from \hef\ and \lisv\ for several reasons.  The deuterium
abundance is more sensitive to \ETA\ that the abundances of the other
light nuclei (Fig. \ref{p11}).  The D/H that we see is likely the
primordial value (\S \ref{pridh}) and does not need corrections D
creation or destruction. No ionization corrections are needed. Over
the last eight years we have established that deuterium is seen in the
spectra of a few QSOs, and D/H can be measured to approximately 10\%
accuracy in favorable cases. The excess dispersion in the D/H values
can be explained if some of the measurement errors have been
underestimated.  We account for this dispersion in our estimates of
the error on the primordial D/H value, the \ETA\ value and the other
deduced parameters. The dispersion is not enough to include the lower
\ETA\ values preferred by \hef\ and \lisv.  The D/H value from
\object\ is similar to that from \qfour\ and 
the mean D/H towards the QSOs is very similar to
the primordial D/H we expect using a simple model of with Galactic
chemical evolution to estimate the depletion in the LISM (\S \ref{astration}).

Our confidence in the \ETA\ and \obh\ from D/H is increased by the
agreement with the \obh\ from the anisotropy of the CMB, but this does
not lessen the need for improved primordial abundance
measurements. Systematic errors are important, or dominant, in most
measurement methods, and many error terms are poorly known because
they are hard to estimate.

\subsection{Comparison with other measurements of the baryon density}

Aside from using the light element abundances, the cosmological baryon
density has been measured in a variety of other ways. These methods
include the amount of H in the intergalactic medium at redshift $z
\simeq 3$, the fraction of baryons in clusters of galaxies, and most
recently and with the most accuracy, the variation of the anisotropy
of the cosmic microwave background on angles of under one degree.

The estimates of \obh\ from different CMB experiments, listed in Table
\ref{table_h}, are consistent with each other and with the \obh\ from
D/H and SBBN.  The first results from BOOMERANG (de Bernardis \etal\
2000) indicated a much larger value of \obh\ $= 0.036 \pm 0.005$,
however, Netterfield \etal\ (2002) have since used revised the
pointing solutions and include data on smaller angular scales. They
find \obh\ $= 0.022^{+0.004}_{-0.003}$ using ``weak priors" which
constrain the age of the universe to $>10$ Gyr and the Hubble constant
to $0.45 < h < 0.9$.  For the same instrument, de Bernardis \etal\
(2002) find \obh\ $=0.022^{+0.004}_{-0.003}$,
$0.020^{+0.004}_{-0.004}$ and $0.019^{+0.005}_{-0.004}$ in three
complementary calculations. Analyzing the measurements from the first
season of observations with the Degree Angular Scale Interferometer
(DASI), Pryke \etal\ (2002) report a very similar value, with similar
precision: \obh\ $=0.022^{+0.004}_{-0.003}$.  The results from a third
experiment, MAXIMA-I (Lee \etal\ 2001; Stompor \etal\ 2001) are
consistent.  Early measurement on the CMB anisotropy in smaller
angular scales by the Cosmic Background Imager (CBI) had a maximum
likelihood with a lower \obh $= 0.009$, and this likelihood dropped by
a factor of two for \obh\ $= 0.019$, and a factor of 3 for \obh\ =
0.03.  Later measurements covering 40 square degrees on the sky give
much more precision. Sievers \etal\ (2002) include the COBE-DMR
results and find \obh\ $=0.022^{+0.15}_{-0.009}$.

Other methods of measuring the \ob\ have lower accuracy.  The \lyaf\
at redshifts $z \simeq 3$ typically indicates higher \obh\ values
(e.g. Hue \etal\ 2002), while the baryon fraction in clusters of
galaxies gives consistent values. For example, Steigman (2002)
multiplied the \om\ derived from SnIa, assuming the universe is flat,
by the baryon fraction in clusters of galaxies to obtain \obh $= 0.019
^{+0.007}_{-0.005}$.

The relevance of D/H measurements is changing.  Today SBBN and D/H
gives the best estimate of \obh. However, as CMB measurements
improve, the \obh\ from the CMB will be as accurate as that from the
SBBN. When we use \obh\ from the CMB in SBBN, the D/H is predicted
with no free parameters, and hence the main value of D/H will become
to test the physics in SBBN (Kaplinghat \& Turner 2001; Cyburt, Fields
\& Olive 2001; Steigman, Kneller \& Zentner 2002; Abazajian 2002).  
Such tests can be
made now, comparing the abundances of the light nuclei, but the
measurement errors are not well established, and hence the precision
will improve when \obh\ comes from the CMB and we can use D/H alone to
test the physics in SBBN.

\section{SUMMARY \NOTE{section\_i.tex}}
\label{summary}

In this paper, we have presented the detection of D towards \qfive.  We
measured D/H and obtained an accuracy nearly as good as
the best previous measurements. The value is slightly lower than the
previous mean and hence our best estimate for the 
cosmological baryon density from SBBN is slightly higher.

The most conspicuous absorption system in the optical spectrum of
\object\ makes a strong Lyman break, and we found D in the main
component of that system. The absorption system has a total column
density \lnhi $= 19.73 \pm 0.04$~\cmm\ in two main components,
separated by approximately 13~\kms . The separation is not well
determined, because the second component contains only approximately
12\% of the H~I, D~I and O~I, and hence in no case does it make a
distinct line. It makes the O~I slightly asymmetric, and it accounts
for a portion of the absorption at velocities between the main H and
the main D, but its only effect on the H~I lines is to make them
slightly wider on the red side. In total the absorption system has at
least seven components, but only components 1 and 2 show O~I, and we
have seen that nearly all of the H~I is confined to $-40 < v <
40$~\kms .  The other components have \lnhi $\simeq 16$~\cmm \ and
with one exception, they have no effect on the D/H measurement.

The column density of the D~I in the main component, D-1, is well
determined, since we see five transitions from this absorber, and
several of them are well separated from other lines, giving accurate
$b$-values, redshift and \lndi .  The column density in the second
component, D-2, is less well determined because its velocity and $b$-value
are not well known.  We explored a variety of models for the relevant
portions of the spectrum, and found the range of parameters that gave
the lowest \chisq\ values.  Some acceptable models have 7\% of their
\ndi\ in the D-2 component, while others have 28\%, the maximum
allowed by the spectrum.  The range for the column density of D-2 gives the
range in the total D column density.  This range arises because the
absorption near D-2 can be either D, or H~I from component H-3 that
has \lnhi $=15.90 \pm 0.03$~\cmm , $b = 17.0 \pm 1.0$~\kms\ and
$v=-44.3 \pm 7$~\kms . The $b$-value and $v$ are not very well known 
for H-3. As the 
$b$-value increases, or the $v$ decreases, H-3 absorbs more near
D-2, and the column in D-2 decreases.

The line near D-1 is clearly H or D because it makes a Lyman series,
and with $b = 9.2 \pm 0.2$~\kms , it is much narrower than typical H
lines. Both the velocity of the D-1 lines and their width indicate
that the line is D and not H. Comparison with the width of the O~I
line gives T $= 5500 \pm 400$~K which is cooler than we expect, but
there is more uncertainty here than indicated by the quoted error
because the O~I and D~I are slightly separated in velocity. We are not
as certain as we were for \qone , \qtwo , \qthree , and \qfour\ that
we have detected D, because for \object\ all the H~I lines are
saturated, and there is extra absorption in other components on either
side of the main H~I, and hence we know little about the velocity
distribution of the H~I. We can not make a detailed comparison the H~I
and D~I velocity structures that would prove that all of D-1 is D.

While the main error in the D column density comes from the uncertain
velocity structure, the main error  on the H column density comes from the
uncertain continuum shape. The \lya\ line alone gives the best
constraints on the \lnhi\ and it is in the short wavelength side of
the \lya\ emission line, where the continuum level is especially hard
to reconstruct. Had the \lya\ line been well removed from the emission
line, the error on the \lnhi\ might have been much less.

We determined the \lnhi\ by fitting the continuum, \lyaf\ and the main
\lya\ line simultaneously.  We model the continuum with smooth curves
represented by B-splines, and we use an optimizing code to vary the
hundreds of parameters in the model. We re-started the optimization
thousands of times, using different initial values for the
parameters. All models with the lowest \chisq\ values have \lnhi $=
19.73$~\cmm , which we use as the best fit value. The error on \lnhi\ is
harder to estimate.  The change in \chisq\ values suggests that the $1
\sigma $ error is near 0.005, but we reject this error as too small, because
we are unsure if we have adequately explored the parameter space or range
of models that might be consistent with the spectrum. Instead, we find
the range of \lnhi\ values that give models with acceptable \chisq\
values, and we represent the \lnhi\ error with a normal distribution
chosen to give a 20\% chance that the \lnhi\ value is outside the
range of acceptable models.  The errors on the H~I and D~I are then
comparable.

There are now measurements or limits on D/H towards seven QSOs, all listed
in Table \ref{table_f}. In the appendix, we explain why measurements
towards two other QSOs are no longer considered useful, and why we
will not use measurements towards Q0347--3819, because the D line is
not seen, and the velocity structure is too uncertain. Ignoring the one
consistent limit, there are then measurements to five QSOs, four from
our group.

These five measurements have a larger dispersion than we expect given
their quoted measurement errors. There is only a 1.5\% chance of a
larger \chisq\ value by chance. We suspect that the dispersion is
large because one or more of the D/H values is inaccurate,
or has errors that have been underestimated. The work we
present here on \object\ has reinforced this belief, because we have found that
it is surprisingly hard to obtain a reliable D/H measurement. Many 
details are relevant (\S \ref{measerrors}) and can change the value or the 
error estimate, and we must attempt to model all D/H values that might be 
consistent with the spectrum, and not stop with the simplest models.

The five D/H measurements do not correlate with metal abundance, but
instead there is a correlation with \lnhi\ that is not expected from
cosmology or astrophysics. We suspect that this correlation is an artifact, a
chance ordering of the D/H values, that themselves show excess
dispersion.  Even though we suspect that some of the D/H measurements
have larger errors than have been quoted, if the relative
size of the errors are indicative of the real errors, we can use
the weighted mean D/H from the five QSOs as the best cosmological
D/H. For the error, we use the error on the mean, and not the weighted
mean, because the former better represents the dispersion. The best
cosmological D/H is $ 0.6 \sigma $ lower than the value that we quoted
in O'Meara \etal\ (2001) because the two new measurements are both
below that value. The error is unchanged, because the larger number of
measurements compensates for the increased dispersion.

The lower value for the cosmological D/H increases the tension with
the primordial abundances of \hef\ and \lisv , both of which prefer
lower values for \ETA\ and \obh . However, considering the
difficulties in obtaining these primordial abundances, we consider
that the relative abundances of the elements are consistent with SBBN.

For several decades now most measurements have given less $^4$He than
we predict, although the dispersion of those measurements suggests that
this difference is because of systematic errors. A recent estimate of
the primordial $^3$He abundance by Bania \etal\ (2002) agrees exactly
with D/H, but they rely on one measurement and we still do not
understand the change in $^3$He caused by Galactic chemical
evolution.  The abundance of $^7$Li in halo stars is a factor of 3 --
4 less than predicted.  Significant $^7$Li may have been destroyed in
these stars, although the amount of destruction implied by SBBN and
D/H is near the maximum allowed.  There is a clear need for more work on 
all four nuclei.

We believe that Deuterium is the best baryometer (Schramm and Turner 1998)
among the light nuclei for the reason given at the end of \S \ref{nuclei}.
The agreement between the \obh\ from D/H and that from the CMB adds
confidence.  These two \obh\ measurements use very different physics,
at very different epochs in the universe. The agreement implies that
we understand the main physical processes that effect the observations of 
both SBBN and the CMB anisotropies.

\subsection{New methods used in this paper}
\label{newmethods}

For \object , we have attempted to improve the methods that we use to
measure D/H. We employed a more accurate relative flux calibration, a
more thorough explorations of the models that can represent the D and
H absorption, more realistic continuum fitting and better
representation of the \lyaf\ and other blended absorption. The D/H
value and the errors should be more reliable.

We have paid much more attention to the accuracy of the relative flux
calibration of the HIRES spectra of \object\ than we did for any of
our past publications.  We have several low resolution spectra from
the Lick Kast spectrograph, and also a spectrum from ESI.  These
spectra have allowed us to compare several methods of flux calibrating
our HIRES spectra, and to verify that we get similar results if we
start with different calibration spectra.

Our continuum fitting procedure is more general than the methods we
used for our measurement of D/H in other QSOs. For \qone\ and \qtwo ,
we fit the continuum twice; once as a smooth curve passing through the
highest flux levels, and again as one smooth curve for each portion of
the spectrum which was adjusted to give a good fit.  For \qfour , the
continuum was a single smooth curve which passed through the high
points in the flux which appeared to be least absorbed by the \lyaf .
We estimated the errors by moving this curve up and down by a factor,
such that it was below the flux in many places, or above the flux in
all places.
The B-spline continua that we fit to \object\ should give make our
\nhi\ estimate more reliable.  The B-spline continuum model allows for
a wide variety of smooth shapes, and we are able to fit the continuum
parameters simultaneously with the absorption line parameters.  In
addition, the B-spline gives us the ability localize continuum degrees
of freedom in wavelength, which we have had difficulty doing in past
work.  These continuum can be adjusted iteratively to allow for the effects
of line absorption, and hence we no longer have to require that the
flux is at the continuum level in some pixels.

For \object, we fit the continuum, \lyaf\ and the DLA all
simultaneously.  For \qfour , we fitted the continuum before and
independently of the \lyaf\ and the DLA.  
When we used methods like those that we used for \qfour\ on \object , we
found a significantly lower \nhi\ value, probably because that method
required that the flux in at least a few pixels is at the continuum level.
We have seen here that the
continuum level, the \lyaf\ and the DLA are connected in a complex
way. In contrast to \qfour , where we may have use too few degrees of freedom
to model the continuum, for \qone\ and \qtwo , we may have allowed too many 
degrees of freedom.

For \object , we have also attempted a more thorough exploration of the
models that might represent the spectrum. For \qone\ and \qtwo , we
used a minimization method to find the optimum parameters.  Here, for the 
\lndi , we use a full grid search, where each parameter takes on every allowed 
value, ensuring a thorough search.

For \object , we have also used different ways of estimating
the measurement errors. For \qone\ and \qtwo , we used the $\delta $\chisq\ 
method that we used here for the \lndi\ alone.
For \qfour , we quoted the error on \lnhi\ of 0.009~\cmm ,       
compared to 0.04~\cmm\  for \object . The smaller error for
\qfour \ is the weighted mean of three estimates of \lnhi\ from    
different parts of the \lya\ line. It is reasonable that the value is smaller
than for \object\ because the DLA suffers less contamination near its core. 
To estimate the error from the continuum, we shift the
continuum up or down by some factor, but we did not change its shape.

\subsection{Five Outstanding Issues}

We end with six outstanding issues.
Some are specific to our D/H measurement from \object\ and others 
are long standing.

{\bf Velocity offsets.}
The O~I absorption is centered $2.8$~\kms\ away from the center of the D 
absorption, not enough to change the identification of the D, but enough to
raise concerns that we may have misunderstood the velocity structure of the 
absorption system (\S \ref{velagree}).
The deepest components of C~II and S~II are also offset from O-1
by 2~\kms , in the other direction from the D-1 component.  These offsets
make it to estimate the column densities of different ions in the same gas, 
and hence the ionization is uncertain.

{\bf Missing D components.} 
The main error on the \lndi\ comes from the amount of D in the component
near 13~\kms . This component is hard to recognize and fit 
because it is weak and blended with the main component (Fig. \ref{fig_d3}).
There might be D in other components that we have mistakenly modeled as H
(\S \ref{missingd}).
We have not included this possibility in the \lndi\ error because 
we do not know how to estimate the chance that it has occurred. 
This additional D could be $>$10\% of the total if these
components have [O/H] $<-3$. 

{\bf Accuracy of the \lnhi\ error.}  We spent considerable time
investigating and debating the error associated with \lnhi\ (\S
\ref{nhi}). We mentioned three such error estimates in this paper: the
covariance matrix from the optimized fits, the $\delta $\chisq\
method, and the range of \lnhi\ that gave fits with acceptable
\chisq .  They gave very different errors: 0.0002, 0.005 and 0.04~\cmm\
respectively.  We know that our estimate from the covariance
matrix is unreliable, and we mention it only because it is the standard
error quoted in most QSO absorption line studies.

The error that we use, from the range of acceptable fits, is
unconventional.  It could be in error, perhaps by a factor of two, for
several reasons.  The largest acceptable \chisq\ value might be
incorrect (\S \ref{chisqscale}), we may have been over generous in
allowing \lyaf\ lines with $b = 150$~\kms\ (\S \ref{fitting}), or we
may have considered too narrow or too wide a range of \lyaf\ (\S \ref{fitting})
or continuum shapes (\S \ref{contshape}).

{\bf Dispersion in the D/H values.}
While we suspect that the excess dispersion arises from underestimated
measurement errors, and we discuss 
circumstantial evidence for this (\S \ref{measerrors}, 
\ref{dispersioniserrors}, \& \ref{newmethods})
we will not know whether this is correct until more reliable 
measurements are made.

We are struck that the D/H values and their errors depend more on the  choice 
of the models that we use to estimate D/H, than upon the details of the 
fit to the data and the \chisq\ value (\S \ref{measerrors}). 
A vital role of high S/N spectra is to guide us to examine an appropriate 
set of models.  The errors on D/H values are hard to estimate because it is 
hard to know whether we have explored a full range of models.

{\bf Correlation of D/H with \lnhi .}
The correlation (Fig. \ref{p9}) looks rather impressive, because all five 
measurements, and the limit, allow a monotonic decrease of D/H with 
increasing \lnhi .
We suspect the correlation is spurious, and a random accident of the
measurement errors. 
We do not know of any single systematic error that could explain all of the 
effect (\S \ref{dhvnhi}). Rather, a variety of errors may be involved.

{\bf The Lack of Precise Agreement of the Measured Abundances in SBBN.}
The measurements of the light nuclei abundances have preferred slightly 
different \ETA\ values since our first measurement of D/H (Tytler, Fan \& 
Burles 1996). Although Izotov \& Thuan (1998) reported higher \yp\ values
that were closer to the predictions from SBBN and D/H, our D/H values have
decreased over the years, exacerbating the difference in the favored \ETA\
values. There are excellent reasons to believe that measurement
problems account for the lack of precise agreement.

\section{Acknowledgments \NOTE{section\_j.tex}}

This work was funded in part by NSF grant AST-9900842. The flux
calibration was funded in part by NASA grant NAG5-9224.

The spectra were obtained from the Lick Observatory, and the W.M. Keck
observatory, which is managed by a partnership among the University of
California, Caltech and NASA.  The Observatory was made possible by
the generous financial support of the W.M. Keck foundation.  We are
grateful to Steve Vogt, the PI for the Keck HIRES instrument which
enabled our work on D/H, to the W.M. Keck Observatory staff, to the
Lick Observatory staff, to
Tom Barlow, who wrote the software used to extract HIRES spectra, and to
Ann Boesgaard, 
George Fuller, Toshitaka Kajino, Sergei Levshakov, Mike Norman, Nikos Prantzos, 
Jason Prochaska, Gary Steigman 
and Art Wolfe
for helpful discussions and comments.

% appendix aaaaaaaaaaaaaaaaaaaaaaaaaaaaaaaaaaaaaaaaaaaaaaaaaaaaaaaaaaaa

\section{APPENDIX I: B-spline continuum model}
\label{splinesection}

We model the continuum level with a B-spline curve.  A B-spline is
defined by a set of control points, $P_0, P_1, \ldots , P_n$, $n>3$
where $n+1$ is the number of control points defining the curve.  The
B-spline passes near to the control points, but does not generally
pass through them.  A B-spline is defined by the control points as a
series of $n-2$ piecewise parametric polynomial segments.  The $i$'th
segment is given by
\begin{equation}
C_i(t) = \frac{1}{6} [t^3\ t^2\ t\ 1]
    \left[ \begin{array}{cccc}
       -1 &  3 & -3 & 1 \\
        3 & -6 &  3 & 0 \\
       -3 &  0 &  3 & 0 \\
        1 &  4 &  1 & 0
       \end{array}
    \right]
    \left[ \begin{array}{c}
	P_{i-1} \\
        P_{i}   \\
        P_{i+1} \\
        P_{i+2}
        \end{array}
    \right]
\end{equation}
where $t$ is the parametric variable and ranges from 0 to 1.  The full
B-spline curve, including the joints between segments, has continuity
of order two.  That is the curve itself, and it's first and second
derivatives are continuous.

In the past we have defined continuum levels with Chebyshev
polynomials and with cubic splines, which are the options available in
the {\tt IRAF} task {\tt continuum}.  We decided against using such
representations of the continuum in this work because they both define
global curves -- if you change one control point in a cubic spline or
a single coefficient in a polynomial the {\it entire} curve is
affected.  In contrast, when $P_i$ is changed, only the B-spline
segments between $C_{i-2}(t)$ and $C_{i+1}(t)$ are affected.  This
locality has two advantages.  First, it is easy to define the
continuum over a large wavelength range in a piecewise fashion
e.g. modifications to the emission line continuum do not mess up the
continuum level defined near the Lyman limit.  Secondly, it greatly
simplifies the task of efficiently computing the continuum changes
that occur as a result of changing a single control point, as often
occurs during non-linear optimizations.

\section{APPENDIX II: OPTIMIZER}
\label{optimizersection}

Here we describe the code we used to vary the parameters of a model
for the spectrum.
In our previous absorption line work, we have usually fit absorption
line parameters using the \chisq\ optimize {\tt VPFIT}, which was
kindly made available to us by R.F. Carswell and J. Webb and collaborators. 
We were unable to use {\tt VPFIT} for this work because we desired to
optimize both our continuum and absorption line models simultaneously.
In addition, we wanted to be able to optimize many hundreds of
parameters simultaneously, which we had difficulty doing with standard
tools.  The optimizer we developed to address these issues implements
the Levenberg-Marquardt algorithm where we regularize the normal
equations by inverting the curvature matrix by means of a truncated
singular value decomposition (SVD).

Our algorithm is based on the Press \etal\ (1992) routine {\tt mrqmin}, and
we refer the reader to Press \etal\ for a general description of the 
Levenberg-Marquardt (LM) algorithm, which we will not repeat here.  We only
describe the differences between our algorithm and the algorithm implemented
by {\tt mrqmin}.

Using the Press \etal\ notation, a key step in the LM algorithm is to
solve the following set of linear equations for the changes $\delta a_l$
that need to be applied to the parameter vector $a_l$ during the next
iteration of the algorithm
\begin{equation}
\label{normalequation}
\sum_{l=1}^M \alpha_{kl} \delta a_l = \beta_k
\end{equation}
where M is size of the parameter vector.  The curvature matrix
$\alpha$ is defined as
\begin{equation}
\alpha_{kl} = \frac{1}{2} \frac{\partial^2\chi^2}{\partial a_k \partial a_l}
\end{equation}
and $\beta$ is given by 
\begin{equation}
\beta_k = - \frac{1}{2} \frac{\partial \chi^2}{\partial a_k}
\end{equation}
In the most implementations of the LM algorithm, Equation
\ref{normalequation} is solved by direct inversion of the curvature
matrix.  For example, {\tt mrqmin} does this by Gauss-Jordan
elimination.

The problem with a direct solution of Equation \ref{normalequation} is
that when fitting absorption lines the curvature matrix is frequently
ill-conditioned.  When $\alpha$ has high condition number, a direct
inversion gives a solution vector $\delta a_l$ that is usually
nonsense. As a result, the LM algorithm will fail to find acceptable
parameter updates and will stop.

Unfortunately, it is very easy for to get an ill-conditioned $\alpha$
during an absorption line optimization -- all that is required is a
pair of nearly degenerate parameters.  In practice this seems to occur
when fitting several heavily blended lines, or when both broad lines
and continuum degrees of freedom are in the parameter set.  There are
likely many other scenarios which result in an ill-conditioned
$\alpha$, as we have observed that the LM algorithm will almost always
have a few iterations with an ill-conditioned $\alpha$ if the number
of optimized parameters is large enough.  The critical number of
parameters appears to be something in the range of 30-50 though we
have not investigated this carefully.  We just note that in the course
of this work an ill-conditioned $\alpha$ occurred frequently enough to
make the standard LM algorithm nearly useless. 

In most situations when the curvature matrix is ill-conditioned, there
are still parameter updates $\delta a_l$ available that would
significantly improve the \chisq\ between the model being optimized
and the underlying data.  To find these, and to thus improve the performance
of our optimizer, our algorithm regularizes $\alpha_{kl}$ before solving
Equation \ref{normalequation}.  We do this by inverting $\alpha_{kl}$
via a truncated singular value decomposition.  Briefly, we compute
the SVD of the $\alpha$ using the {\tt LAPACK} routine {\tt dgesdd},
which computes orthogonal matrixes $U, V$ and the diagonal matrix
$\Sigma = {\rm Diag}(\sigma_1, \ldots, \sigma_n)$.  $U, V$ and $\Sigma$
are defined such that
\begin{equation}
\label{svdequation}
  \alpha = U\Sigma V^*
\end{equation}
Because $U$ and $V$ are orthogonal and $\Sigma$ is diagonal it is trivial
to invert $\alpha$ once it's SVD is known
\begin{equation}
\label{piequation}
  \alpha^{-1} = V\Sigma^{-1}U^*
\end{equation}
which works as long as all of the $\sigma_k$ are non-zero.  The SVD
(Equation \ref{svdequation}) always exists, even if $\alpha$ is
ill-conditioned or even singular.  Unfortunately, if $\alpha$ is
ill-conditioned, Equation \ref{piequation} does no better than direct
methods for computing $\alpha^{-1}$.  However, the magnitude of the
$\sigma_k$ tell us which of the orthogonal vectors in $U$ are
responsible for the near singular behavior.  The standard trick is to
compute $\alpha^{-1}$ via
\begin{equation}
  \alpha^{-1} = V\Sigma^+U^*
\end{equation}
where 
\begin{equation}
  \Sigma^+ = {\rm Diag}(\sigma_k^+), \ \ \ \ \
  \sigma_k^+ = \left\{
               \begin{array}{ll}
		   1/\sigma_k  &   {\rm if} \sigma_k > h \\
                   0           &   {\rm otherwise.}
               \end{array}
               \right.
\end{equation}
In this scheme $h$ is called the regularization parameter.  Setting
it is something of a black art.  If $h$ is set too small, the near
singular components of $U$ will corrupt $\alpha^{-1}$.  If $h$ is set
too large, otherwise good component vectors of $U$ will not be
updated, and the optimizer will perform poorly.  We could find no
guidance from the literature on an appropriate value for $h$ -- the
general advice given is too experiment and see what works for the
application.  This is exactly what we did, and in our algorithm
$h=10^{-8}$.  Our entire algorithm is implemented in double precision.

This is the algorithm that was used for all of the optimizations
performed in this work.  This optimizer works very well in practice,
and we have routinely performed successful optimizations of over 500
parameters, many of which were highly blended \lyaf\ lines.  

\subsection{Error estimates}
\label{apperror}

The standard way to estimate the errors in the fitted parameters is to
estimate the covariance matrix $C$ from the curvature matrix $\alpha$
\begin{equation}
\label{ceequation}
  C = \alpha^{-1}
\end{equation}
The parameter errors are then just given by the square root of the 
diagonals of $C$.

This is exactly the error estimate that we use in our optimizer.  We
directly invert $\alpha$ using Gauss-Jordan elimination instead of
using the truncated SVD regularization described above.  This is
because the validity of an approximate inverse is easily tested when
updating parameter values: if the solution to Equation
\ref{normalequation} results in a better \chisq, the solution has some
value.  However, we are not sure how an approximate inverse affects
the errors inferred from the covariance matrix, so we just use a
direct method.

Note that the estimated errors are not likely to be valid when
$\alpha$ is ill-conditioned.  In practice, they generally appear to be
much to small when the curvature matrix is ill-conditioned.  We did a
small amount of Monte-Carlo testing which indicated that the estimated
covariance matrix appeared to be accurate when optimizing a small
number of lines with a well conditioned covariance matrix.  However,
it appeared from the same testing that the estimated covariance matrix
begins to go bad long before $\alpha$ becomes ill-conditioned enough
to jeopardize the solution of Equation \ref{normalequation}.  Because
we are sure that Equation \ref{ceequation} is a valid estimate in only
very simple optimizations, we use error estimates from Equation
\ref{ceequation} only cautiously in this work.

\section{APPENDIX III: DISCUSSION OF OTHER D/H MEASUREMENTS}
% how number appedix?
\label{otherqsos}

In this appendix we discuss D/H measurements and limits to other QSOs.
In some cases the absorption systems are too complex, or existing
spectra are inadequate to give convincing measurements or constraints.
Related issues are discussed in reviews by Tytler \& Burles (1997),
Tytler \etal\ (2000) and Lemoine \etal\ (1999).

Towards Q0014+813 Songaila \etal\ (1994) reported an upper limit of
D/H $< 25 \times 10^{-5}$ in the $z_{abs} = 3.32$ LLS.  Using
different spectra, Carswell \etal\ (1994) reported $<60 \times
10^{-5}$ in the same object, and they found no reason to think that
the deuterium abundance might be as high as their limit. Improved
spectra (Burles, Kirkman \& Tytler 1999) showed that D/H $< 35 \times
10^{-5}$. A high D/H value is allowed, but is highly unlikely because
the absorption near D is at the wrong velocity, by $17 \pm 2$ \kms, it
is too wide, and it does not have the expected distribution of
absorption in velocity, which is given by the H absorption.  Instead
this absorption is readily explained entirely by H (D/H $\simeq 0$) at
a different redshift.

Towards \qhst\ the LLS at $z_{abs}=0.701$ was also considered as a
possible site showing a large D/H value (Webb \etal\ 1997), but others
argued that this was not the most convincing interpretation of the HST
spectra (Levshakov, Kegel, \& Takahara 1998; Tytler \etal\ 1999). New
higher resolution spectra from HST (Kirkman \etal\ 2001) show the
velocity structure of the H for the first time. We found that it is
unlikely that this QSO gives any useful information on D/H, because
the absorption near D is at the wrong velocity and has the wrong line
width to be all D.  Hence it is probably mostly contaminating H.

Molaro \etal\ (1999) claimed that the z=3.514 absorption system
towards AMP 08279+5255 showed low D/H, but they and Levshakov
\etal\ (2000) note that since only the \lya\ line has been observed,
the hydrogen velocity structure and the H~I column density are poorly
known, and the D feature can be fit using H alone.

\subsection{Q2206--0199}

Pettini and Bowen (2001, hereafter PB) report D/H $=1.65 \pm 0.35
\times $\mf \ in the $z=2.076$ DLA towards 
Q2206--199.  Since the D is measured in a very low S/N HST spectrum it is
hard to determine whether the errors are reasonable, and 
whether D is seen.  The reported D/H value is very low 
compared to the D/H we expect from Galactic chemical evolution and the D/H 
in the LISM (\S \ref{astration}).
We now discuss some ways in which the errors might have been underestimated.

We concentrate on the \ndi\ which seems to be less reliable than the
\lnhi .  The authors solve for the \ndi\ by fitting features at the
position of D in three Lyman series transitions: Ly-7, Ly-9, and
Ly-12.  The \ndi\ value depends on the $b$-value chosen 
for the D.  The D lines are too noisy to give this $b$-value, so the
authors interpolate between the widths they measure for the $b$-values
of the H~I and the metal lines, giving 10.6~\kms .  We question
whether this $b$-value is accurate, perhaps because the $b$-value of
the H~I is itself uncertain.

Although the low ion transitions such as Fe~II and Al~II indicate a
single component nature for the gas, Prochaska \& Wolfe (1997) also
fit multi-component CIV and SiIV gas on either side of the $v = 0$
position of the H which shows D.  These additional components may have
enough H absorption to widen the Lyman lines that were used to get the
$b$-value of the main H~I.  As such, we explore the implications of a
lower $b$(D) on the range of D/H allowed by the data.  If we assume
the gas is cold, in the range of 300--400 K, and motions are dominated
by turbulent velocities, then values of \lndi\
giving D/H $\simeq 2.9\times $\mf\ are consistent with the 1 $\sigma$ errors 
obtained by PB's fit to Ly-9, but not to Ly-12.

We consider the other sources of error relevant to the determination
of \lndi .  In the Lyman limit region presented by PB, very few pixels
reach the continuum in their normalized data.  In particular, for Ly-9
and Ly-12, no pixels reach the continuum within $v \simeq 150$~\kms\
of the position of D.  Due to the high level of contamination
throughout the Lyman limit, the continuum is very poorly constrained,
and could contribute significant errors to the \lndi .

Continuum placement aside, contamination also affects the
determination of \lndi\ in this system.  In particular we note that the
fit to the contamination on the large wavelength side of the D feature in 
Ly-9 seems
incompatible with the size of the 1~$\sigma$ errors given to the D
line.  Since this additional absorption adds significant optical depth
at the position of D, it might change the \lndi\ by more than $2\sigma
$.

Our main concern with this D/H measurement is that the HST spectra 
have low S/N and provide very little supporting evidence. However, the
ions do show a simple velocity structure, and hence the feature may 
be D with the measured D/H.

\subsection{Q0347--3819}
\label{qlev}

D'Odorico, Dessauges-Zavadsky \& Molaro (DDM, 2001) report D/H $= 2.24
\pm 0.67 \times $\mf\ in the z=3.025 absorption system to Q0347-3819,
while Levshakov, Dessauges-Zavadsky, D'Odorico \& Molaro (LDDM, 2002)
report a very different value, D/H $= 3.75 \pm 0.25 \times $\mf ,
from a sophisticated analysis of the same spectrum.  
We are not convinced that D has been detected in the spectrum of this QSO. 
The D lines are completely blended with the H lines and in no case can we see
D clearly separated from the H. Hence we do not know the central
velocity, velocity structure or $b$-value for the absorption near D.
This absorption could be partly or predominantly H.
It is hard to obtain either a secure value or a limit on D/H, because the 
velocity structure for this absorption system is critical and not well enough 
known. We will discuss this system in detail here because many of the
issues are very familiar from our discussion of \object . 

The total H~I column should be well determined from the damping wings
of the \lya\ line. The fits shown in Fig. 11 of LDDM seem to
over-absorb in \lya\ near $-2800$ ~\kms , and from +450 to +750 \kms\
for \lyb\ which suggests that the model has excess \lnhi\ or that 
there might be errors in the flux calibration or continuum level.

The metal lines show many components that can contribute to the Lyman
series H~I lines. Several components may have enough \lnhi\ to also
show D.  The low ions, especially H$_2$, indicate the velocity for
much of the H~I, and the expected D. There is absorption on the blue
side of \ly\ lines 8, 10 and 12 at the position expected for D, which
resembles a Lyman series, and hence it may be caused by D~I or H~I or
a combination of the two.

To determine the total \lndi\ that is associated with the total
\lnhi\ we need to know the velocity structure of the D and the H.  DDM
assumed that the H~I is at the velocities of three of the components
seen in the metal lines, and they assumed that the H~I columns are
distributed approximately like the metals. They fit the H~I lines
by adjusting the \lnhi\ and $b$-values of these components.
LDDM stated that their model may not
be unique, which seems likely because their model seems to have more
degrees of freedom than are needed to fit the data.
We expect that two components might suffice, one to 
fit each side of the H Lyman series lines.

It is difficult to measure the amount of H~I in the different
components because their lines are all blended. This is especially
true for the main component ($v=0$, component 3 in DDM) because in no
case do we see H~I absorption from this component alone. The other
components, which are on either side of it, account for both sides
of the higher order Lyman lines.

DDM find that additional absorption is needed near the velocity
expected for D.  It is difficult to measure a column density for this
extra absorption, whether of not it is D, because we do not know its
velocity structure.  LDDM demonstrate that the fit tabulated by DDM,
and hence the D/H value, is not unique, and that the differences are
highly significant because some of the H~I, and we expect a similar
proportion of the D~I, is in a saturated component with an unusually
low $b \simeq 2.9$~\kms .  The \lndi\ is sensitive to the $b$-value of
the D, and the proportion of the H~I and D~I that is in this
component, neither of which can be measured.

The \lndi\ also depends on the amount of H~I absorption on the red of
the D position. The more flux absorbed by the H~I, the less can be
absorbed by the D, and hence the lower the D/H. The total \lnhi\ is
not changed because it is fixed by the damped \lya\ line that is
insensitive to the velocity structure.  If we adjust the relative
amount of H~I in the two main components, both the \lnhi\ and the 
$b$-values, which are not known, then the amount of absorption that could
be D also changes.  There is also a component near $-65$~\kms ,
seen in C~IV and Si~IV (Prochaska \& Wolfe 1999), which might also
have enough H~I to effect the absorption on the red side of where D is
expected in Ly-8, 10 and 12.  The uncertainty in continuum near Ly-10
and 12, two of the three regions used to get \lndi , and the lines
blended on the blue side of the expected positions of the D lines
further increase the uncertainty in the \lndi .

DDM and LDDM get very different \lndi\ and hence D/H because they make
different assumptions about the velocity structure. They both 
assume that the H~I, and the D~I, have the components at the
velocities that show low ionization metal lines, and they assume that
the proportion of the H~I in these components is similar to that of
the metals, which will not work if the metal abundances vary.  DDM
find that the D can be fit with two main components with $b$-values of
$14.1 \pm 0.5$ and $16.2 \pm 3.0$~\kms . LDDM assume that most of the
D has $b \simeq 2.8$~\kms\ which produces saturated D lines that
require much larger \lndi\ to explain a given amount of absorption.
It is reasonable that much of the H~I and hence the expected D, has
$b$ closer to 2.8 than to $15$~\kms , and hence the D/H will be larger
than reported by DDM. However we do not know how much of the H~I has
this low $b$-value, and hence we do not know whether the D/H is lower,
similar to, or high than proposed by LDDM.

LDDM tie the unknown velocity structure of the H~I and D~I to that of
the H$_2$ and various metal lines.  This is not a unique procedure
because in this system the metal lines themselves differ: Fe~II
(1143.2, 1133.7, 1125.4) has different velocity structure from Si~II
1808 and other ions. LDDM note that the Fe~II lines may lack column in
the main component at $v=0$ because the Fe there is locked up in dust.

The metal abundance in the gas where D would be seen is uncertain
because this gas has a very low $b$-value and 
there may be dust.  LDDM find [Zn/H] $=-1.51 \pm 0.05$, but elements C, P
and Ar are higher by 0.3 dex, while O and Si are higher by 0.7
dex. These enhancements are unexpected because Zn should be the least
depleted onto dust, and we do not expect such a large enhancement of O
and Si. Although LDDM quote [Si/H] $=-0.95 \pm 0.02$, we expect that
the uncertainty is much larger than 0.02 dex, and we anticipate that
further analysis will show that the total abundances including metals
in dust, is lower than this.

We prefer not to take the DDM D/H value as a lower limit, because we
do not know whether D has been seen. Before we could take the LDDM
value as an upper limit we would need to know that there
is at least as much H absorption near the D as they assume, and that
the proportion of the D in the low $b$-value component is at least as
much as they assume.  This absorption system is not ideal for the
measurement of D/H.

\clearpage

%---------------------------------------------------
\begin{deluxetable}{llccccl}
\tablecaption{\label{table_a}OBSERVATIONS OF \qfive\ \NOTE{table\_a.tex}}
\tablewidth{0pt}
\tablehead{
\colhead{Instrument} &
\colhead{Date\tablenotemark{a} }&
\colhead{Integration Time} &
\colhead{Slit} &
\colhead{Wavelengths Observed}
\\
\colhead{} &  \colhead{} & \colhead{(seconds)} &
\colhead{(arcsec)} &
\colhead{(\AA)}
}
\startdata
Kast  & February 13, 1997 & 3300 & 3 & 3120 -- 5800 \\
Kast  & May 10, 1999      & 6982 & 2 & 3137 -- 7188 \\
Kast  & May 11, 1999      & 7200 & 2 & 3137 -- 7188 \\
Kast  & May 17, 2001 & 5400 & 2 & 3191 -- 5881 \\ 
Kast  & May 17, 2001 & 5400 & 2 & 3191 -- 5881 \\ 
ESI   & January 11, 2000  & 1364 & 1 & 4000 - 10,000 \\
HIRES & April 15, 1999    & 3600 & 1.14 & 3494 -- 5842 \\
HIRES & April 16, 1999    & 7200 & 1.14 & 3168 -- 4705 \\
HIRES & April 17, 1999    & 8100 & 1.14 & 3168 -- 4705 \\
HIRES & April 17, 1999    & 8100 & 1.14 & 3168 -- 4705 \\
HIRES & March 12, 2000    & 9000 & 1.14 & 3214 -- 4705 \\
HIRES & March 12, 2000    & 9000 & 1.14 & 3214 -- 4705 \\
HIRES & March 13, 2000    & 7200 & 1.14 & 3214 -- 4705 \\
HIRES & March 13, 2000    & 7200 & 1.14 & 3214 -- 4705 \\
\enddata 
\tablenotetext{a}{\footnotesize{We list the local calendar date at
sunset at the start of the night.}}
\end{deluxetable}

%---------------------------------------------------
\begin{deluxetable}{lccccc}
\tablewidth{0pt}
\tablecaption{\label{resolution} \label{table_b} RESOLUTION AND S/N OF SPECTRA 
   \NOTE{table\_b.tex}}
\tablehead{
\colhead{Spectrograph} &
\colhead{Slit} &
\colhead{Pixel Width} &
\colhead{S/N\tablenotemark{a} }&
\colhead{S/N\tablenotemark{a} }&
\colhead{FWHM}
\\
\colhead{} & \colhead{(arcsec)} & \colhead{(kms$^{-1}$)}&
\colhead{(3250\AA )} & \colhead{(4250\AA )} & \colhead{(kms$^{-1}$)}
}

\startdata
Kast & 2 & 105\tablenotemark{b}  & 20 & 60 & $283 \pm 25$ \\
ESI  & 1  & 11.5 & -- & 40 & $63.2 \pm 3.0$\\
HIRES & 1.14 & 2.1 & 10 & 90 & $8.0 \pm 0.2$ \\
\enddata
\tablenotetext{a}{\footnotesize{S/N per pixel.}}
\tablenotetext{b}{\footnotesize{Mean value. 
We measure variation with wavelength and from spectrum to spectrum.}}

\end{deluxetable}

%---------------------------------------------------

\begin{deluxetable}{llllr}
\tablewidth{0pt}
\tablecaption{\label{dhlinetab} \label{table_c} IONS 
IN THE $z \simeq 2.526$ ABSORPTION SYSTEM TOWARDS \object
\NOTE{table\_c.tex}}
\tablehead{
\colhead{Ion} & \colhead{log N} & \colhead{$b$} &
\colhead{$z$} & \colhead{$v$} 
\\ 
\colhead{} & \colhead{(\cmm)} & \colhead{(\kms)} &
\colhead{} & \colhead{(\kms)}
 }
\startdata 
H~I total\tablenotemark{a} & $19.73 \pm 0.04$ & ... & ... & ... \\
H~I (H-3) & $15.90 \pm 0.03$ & $17.0 \pm 1.0$ & 2.525171 & $-44.3 \pm 7.0$ \\
H~I (H-1) & 19.63 & $14.8 \pm 2.9$ & 2.525659 & $-$2.8\\
H~I (H-2) & 19.05 & $10.9 \pm 3.3$ & 2.525804 & 9.5 \\
H~I (H-4) & $16.25 \pm 0.02$ & $25.8 \pm 0.9$ & 2.526939 & $106.0 \pm 0.7$ \\
H~I (H-5) & $16.35 \pm 0.02$ & $26.6 \pm 0.5$ & 2.528108 & $205.4 \pm 0.3$ \\
D~I total & $15.113^{+0.042}_{-0.026} $ & ... &...  &...  \\
D~I (D-1) &
$15.058 \pm 0.03$ &
$9.2 \pm 0.2$ &
2.525659 &
$-2.8 \pm 0.6$\\
D~I (D-2) &
$14.191 \pm 0.10$ &
$<12$ &
2.525804 &
$9.2^{+2}_{-3.5}$ \\
O~I (O-1)\tablenotemark{b} & 
13.570 & 
6.766  &
2.525692 & 
0.0\tablenotemark{c} \\
O~I (O-2)\tablenotemark{b} & 
12.755 & 
6.766  &
2.525848 & 
13.3\tablenotemark{c} \\ 
Si~II & 
11.884 & 
3.44  &
2.525209 & 
$-$41.0 \\ 
Si~II & 
11.783 & 
5.89  &
2.525342 & 
$-$29.8 \\ 
Si~II & 
12.020 & 
7.85  &
2.525568 & 
$-$10.5 \\ 
Si~II & 
11.830 & 
3.76  &
2.525635 & 
$-$4.8\tablenotemark{c} \\ 
Si~II & 
12.766 & 
4.46  &
2.525728 & 
3.1\tablenotemark{c} \\ 
Si~II & 
12.473 & 
5.99  &
2.525885 & 
16.4\tablenotemark{c} \\ 
Si~II & 
11.508 & 
3.27  &
2.525984 & 
24.8 \\ 
Si~II & 
12.120 & 
7.41  &
2.526083 & 
33.3 \\ 
Si~II & 
11.751 & 
5.97  &
2.526856 & 
99.0 \\ 

C~II & 
12.521 & 
3.03  &
2.525215 & 
$-$40.5 \\ 
C~II & 
12.987 & 
11.06  &
2.525324 & 
$-$31.2 \\ 
C~II & 
12.693 & 
9.97  &
2.525552 & 
$-$11.9 \\ 
C~II & 
12.956 & 
7.58  &
2.525626 & 
$-$5.6 \\ 
C~II & 
13.344 & 
5.21  &
2.525708 & 
1.4\tablenotemark{c} \\ 
C~II & 
13.260 & 
7.17  &
2.525872 & 
15.4\tablenotemark{c} \\ 
C~II & 
12.921 & 
12.73  &
2.525959 & 
22.8 \\ 
C~II & 
13.150 & 
10.40  &
2.526076 & 
32.7 \\ 
C~II & 
12.916 & 
10.59  &
2.526834 & 
97.1 \\ 
Al~II\tablenotemark{d} &
$<12.2$ &
7 &
... & 0\tablenotemark{c} \\
Si~III\tablenotemark{e} &
$<13.0$ &
7 &
... & 0\tablenotemark{c} \\
C~IV\tablenotemark{f} &
$<12.75$ &
 7 &
... & 0\tablenotemark{c} \\
C~IV & 
13.190 & 
38.81  &
2.525750 & 
4.9 \\ 
C~IV & 
13.256 & 
20.98  &
2.526171 & 
40.8 \\ 
C~IV & 
13.248 & 
29.53  &
2.526985 & 
110.0 \\ 
C~IV & 
12.514 & 
14.11  &
2.528017 & 
197.7 \\ 
Si~IV\tablenotemark{f} &
$<12.4$ &
 7 &
... & 0\tablenotemark{c} \\
Si~IV & 
12.905 & 
37.64  &
2.525560 & 
$-$11.2 \\ 
Si~IV & 
12.730 & 
17.21  &
2.526145 & 
38.6 \\ 
Si~IV & 
12.709 & 
24.33  &
2.526944 & 
106.5 \\ 
Si~IV & 
11.920 & 
10.71  &
2.527978 & 
194.4 \\ 
\enddata 

\tablenotetext{a}{\footnotesize{Includes components H-1 and H-2 only.
}}

\tablenotetext{b}{\footnotesize{Values correspond to model where the $b$
of O-1 and O-2 have been set to equal each other.}}

\tablenotetext{c}{\footnotesize{
The column density for this component was used to constrain the ionization 
in components 1 \& 2.
}}

\tablenotetext{d}{\footnotesize{This ion was observed in the ESI spectrum 
only.  The quoted upper limit on the column density
is obtained when we fix $b=7.0$ \kms , 
$v=0$~\kms\ and fit the entire region from --60 \kms\ to +60 \kms .}}

\tablenotetext{e}{\footnotesize{
Not fit because strongly blended. We quote the maximum column near $v=0$
for $b=7$~\kms\ that is consistent with the residual flux.
}}

\tablenotetext{f}{\footnotesize{
The upper limit comes from a fit with fixed $b=7$~\kms\ and $v=0$~\kms .
}}
\end{deluxetable}
%---------------------------------------------------
\begin{deluxetable}{lllccc}
\tablecaption{\label{dhinferred} \label{table_j} INFERRED PHYSICAL CONDITIONS WHERE D/H IS MEASURED
   \NOTE{table\_j.tex}}
\tablewidth{0pt}
\tablehead{
\colhead{QSO} &
\colhead{$\log H~I/H$} &
\colhead{$\log H$} &
\colhead{Size} &
\colhead{Hydrogen Gas Mass} 
\\
\colhead{} & \colhead{} & \colhead{(\cmm )} & \colhead{(kpc)} & 
\colhead{($M_{\odot} $)} 
}
% --table1
\startdata
PKS 1937--1009\tablenotemark{a,b}
& $-2.35,-2.29$ & 20.05,19.74 
& 0.9, 0.4 & $3.9 \times 10^{5}$,$2.9 \times 10^{4}$\\
Q1009+299\tablenotemark{b,c}
& $-2.97,-2.84$ & 19.90, 19.93 &
1.8, 1.5 & $1.1 \times 10^{6}$,$7.5 \times 10^{6}$\\
HS 0105+1619\tablenotemark{b,d}
& $>-0.1$ & $< 19.52$ & $<1.1$ & $< 1.6 \times 10^{5}$\\
Q1243+3047\tablenotemark{e} 
& $-0.69$ & 20.42 &
2.7 & $8.0 \times 10^{6}$\\
Q2206--199\tablenotemark{f}
& unknown\tablenotemark{g}  & unknown & unknown & unknown\\
Q0347--3819\tablenotemark{h} & $>-0.3$ & $< 20.93$ & 0.014 & 211\\
Q0130--403\tablenotemark{i}
& $-3.4$ & 20.06 & 30\tablenotemark{j} & $4.5 \times 10^{8}$\\
\enddata
\tablenotetext{a}{\footnotesize{
We list the parameters for each of the two components, where
available,
from Tytler, Fan \&\ Burles (1996); Burles \&\ Tytler (1998a).}
}
\tablenotetext{b}{\footnotesize{Scaled from previous references to 
$J_{912} = 10^{-21}$ ergs \cmm\ s$^{-1}$ Hz$^{-1}$ sr$^{-1}$ }.}
\tablenotetext{c}{\footnotesize{
We list the parameters for each of the two components, where
available, from Tytler \& Burles (1997)
and Burles \&\ Tytler (1998b).} }
\tablenotetext{d}{\footnotesize{O'Meara \etal\ 2001.} }
\tablenotetext{e}{\footnotesize{This paper.}}
\tablenotetext{f}{\footnotesize{Pettini \& Bowen 2001.} }
\tablenotetext{g}{\footnotesize{Currently, there is not enough diagnostic 
information to measure the ionization state in this absorber, yet it is 
presumably neutral, given the \lnhi .} }
\tablenotetext{h}{\footnotesize{Levshakov \etal\ 2002.}}
\tablenotetext{i}{\footnotesize{Kirkman \etal\ 1999.} }
\tablenotetext{j}{\footnotesize{Calculated from $log~ n_H = -2.9$ \cmmm\ from
Kirkman \etal\ 2000 and $log~ n_{HI}/n_H = -3.4$ from O'Meara \etal\ 2001 .
}}
\end{deluxetable}

%---------------------------------------------------
\begin{deluxetable}{llllccc}
\tabletypesize{\small}
\tablewidth{0pt}
\tablecaption{\label{fourqsos} \label{table_f} D/H MEASUREMENTS TOWARDS QSOS 
   \NOTE{table\_f.tex}}
\tablehead{
\colhead{QSO} &
\colhead{$z_{DH}$} &
\colhead{D/H $\pm 1\sigma $} &
\colhead{$\log $D/H} &
\colhead{$X_i $\tablenotemark{a} }&
\multicolumn{2}{c}{$b$(D) (\kms )} 
\\
\colhead{} & \colhead{} & \colhead{($\times 10^{-5}$)} & \colhead{} &
\colhead{} &
\colhead{predicted} & 
\colhead{observed}
}
% --table1
\startdata
PKS 1937--1009\tablenotemark{b}
& 3.572 & $3.25 \pm 0.3 $ & $-4.49 \pm 0.04$ & +1.65 & $12.5 \pm 2.1$\tablenotemark{c} 
&$14.0 \pm 1.0$ \\
Q1009+299\tablenotemark{d}
& 2.504 & $3.98 ^{+0.59}_{-0.67}$ & $-4.40^{+0.06}_{-0.08}$ & +1.95 &
$13.5 \pm 0.5 $\tablenotemark{c} & $15.7 \pm 2.1$ \\
HS 0105+1619\tablenotemark{e}
& 2.536 & $2.54 \pm 0.23$ & $-4.596 \pm 0.040$ & $-1.00$ & 
$10.1 \pm 0.3$\tablenotemark{c} & 
$9.85 \pm 0.42$ \\
Q1243+3047\tablenotemark{f} 
& 2.525675 & $ 2.42^{+0.35}_{-0.25}$ & $-4.617 ^{+0.058}_{-0.048}$ & $-1.05$ & 
$11.3 \pm 1.8$ & $9.2 \pm 0.2$ \\
Q2206--199\tablenotemark{g}
& 2.0762 & $1.65 \pm 0.35$\tablenotemark{h} & $-4.78^{+0.08}_{-0.10}$ & $-2.17$ &
10.6 & -- \\
Q0347--3819\tablenotemark{i} & 3.024855 & $3.75 \pm 0.25$\tablenotemark{h} & $-4.43 \pm 0.03$ 
& +3.35 & 3,14.1,16.2 & -- \\
Q0130--403\tablenotemark{j}
& 2.799 & $< 6.8$ & $< -4.17$ & -- & $16.2 \pm 0.3$\tablenotemark{c} & --\\
\enddata
\tablenotetext{a}{\footnotesize{
$X_i = (Y_i -mean)/\sigma(Y_i)$, where $Y_i = log (D/H)_i$ and we use the 
weighted mean of the first five QSOs,
log D/H = $-4.556 \pm 0.064$.}}
\tablenotetext{b}{\footnotesize{
We list combined results for the two components,
from Tytler, Fan \&\ Burles (1996); Burles \&\ Tytler (1998a).}
}
\tablenotetext{c}{\footnotesize{
Calculated from the published data and first presented here.}}
\tablenotetext{d}{\footnotesize{
We list combined results for the two components,
from Tytler \& Burles (1997)
and Burles \&\ Tytler (1998b).} }
\tablenotetext{e}{\footnotesize{O'Meara \etal\ 2001.} }
\tablenotetext{f}{\footnotesize{This paper.}}
\tablenotetext{g}{\footnotesize{Pettini \& Bowen 2001.} }
\tablenotetext{h}{\footnotesize{Discussed in the the appendix of this paper.}}
\tablenotetext{i}{\footnotesize{First analyzed by D'Odorico \etal\ 2001. 
We quote results from Levshakov \etal\  2002.}}
\tablenotetext{j}{\footnotesize{From Kirkman \etal\ 2000.} }
\end{deluxetable}

%---------------------------------------------------
\begin{deluxetable}{llclcc}
\tablecaption{\label{dhphysical} \label{table_g} COLUMN DENSITIES AND 
METAL ABUNDANCES WHERE D/H IS MEASURED
   \NOTE{table\_g.tex}}
\tablewidth{0pt}
\tablehead{
\colhead{QSO} &
\colhead{\lnhi\ } &
\colhead{Element} &
\colhead{Abundance} 
\\
\colhead{} & \colhead{(\cmm )} & 
\colhead{$\alpha $} & \colhead{[$\alpha $/H]} 
}
% --table1
\startdata
PKS 1937--1009\tablenotemark{a}
& $17.86 \pm 0.02$ & 
Si & $-2.7,-1.9$\\
Q1009+299\tablenotemark{b}
& $17.39 \pm 0.06$ & 
Si & $-2.4,-2.7$\\
HS 0105+1619\tablenotemark{c}
& $19.422 \pm 0.009$ &
O\tablenotemark{d}& $-1.73$\\
Q1243+3047\tablenotemark{e} 
& $19.73 \pm 0.04$ & 
O\tablenotemark{d}& $-2.79 \pm 0.05$\\
Q2206--199\tablenotemark{f}
& $20.436 \pm 0.008$ &  
Si & $-2.23$\tablenotemark{g} \\
Q0347--3819\tablenotemark{h} & $20.626 \pm 0.005$ & Si
& $-1$\tablenotemark{i}\\
Q0130--403\tablenotemark{j}
& $16.66 \pm 0.02$ & Si & $-2.6$\\
\enddata
\tablenotetext{a}{\footnotesize{
We list the parameters for each of the two components, where
available,
from Tytler, Fan \&\ Burles (1996); Burles \&\ Tytler (1998a).}
}
\tablenotetext{b}{\footnotesize{
We list the parameters for each of the two components, where
available, from Tytler \& Burles (1997)
and Burles \&\ Tytler (1998b).} }
\tablenotetext{c}{\footnotesize{O'Meara \etal\ 2001.} }
\tablenotetext{d}{\footnotesize{Using log~O/H $=-3.31$ from  Allende Prieto, Lambert \& Asplund  (2001).}}
\tablenotetext{e}{\footnotesize{This paper.}}
\tablenotetext{f}{\footnotesize{Pettini \& Bowen 2001.} }
\tablenotetext{g}{\footnotesize{Prochaska \& Wolfe 1997.}}
\tablenotetext{h}{\footnotesize{Levshakov \etal\ 2002.}}
\tablenotetext{i}{\footnotesize{Discussed in this paper.}}
\tablenotetext{j}{\footnotesize{Kirkman \etal\ 1999.} }
\end{deluxetable}

%---------------------------------------------------
% sbbn predictions
% to include theory errors, esposito and bnt
\begin{deluxetable}{lll}
\tablewidth{0pt}
\tablecaption{\label{baryondensity1} \label{table_h} Recent Estimates 
   of the Baryon Density
   \NOTE{table\_h.tex}}
\tablehead{
\colhead{Method} &
\colhead{$\Omega _b h^{2}$} &
\colhead{Ref. } 
}
\startdata
BBN + D/H & $0.0214 \pm 0.0020$ & this paper \\
CMB:- BOOMERANG & $0.021^{+0.003}_{-0.003}$ & Netterfield \etal\ 2002 \\
CMB:- DASI & $0.022^{+0.004}_{-0.003}$ & Pryke \etal\ 2002 \\
CMB:- MAXIMA-I & $0.033 \pm 0.013$ (95\%) & Stompor \etal\ 2001 \\
CMB:- CBI  & $0.022^{+0.15}_{-0.009}$ & Sievers \etal\ 2002 \\ 
%CMB:- CBI  & 0.009 & Padin \etal\ 2001 \\ 
Clusters + SNIa & $0.019^{+0.007}_{-0.005}$ & Steigman 2002 \\
\enddata
%\tablenotetext{a}{\footnotesize{Approximate values}}
\end{deluxetable}
%%%%%%%%%%%%%%%%%%%%%%%%%%%%%%%%%%%%%%%%%%%%%%%%%%%%%%%%%%%%%%%%%%%%%%%%%%%%%%
%%%%%%%%%%%%%%%%%%%%%%%%%%%%%%%%%%%%%%%%%%%%%%%%%%%%%%%%%%%%%%%%%%%%%%%%%%%%%%
%%% --fig FIGURES   FIGURES   FIGURES   FIGURES   FIGURES   %%%%
%%%%%%%%%%%%%%%%%%%%%%%%%%%%%%%%%%%%%%%%%%%%%%%%%%%%%%%%%%%%%%%%%%%%%%%%%%%%%%
%%%%%%%%%%%%%%%%%%%%%%%%%%%%%%%%%%%%%%%%%%%%%%%%%%%%%%%%%%%%%%%%%%%%%%%%%%%%%%

\clearpage

\begin{figure*}
\begin{center}
\includegraphics[angle=-90, scale=0.60]{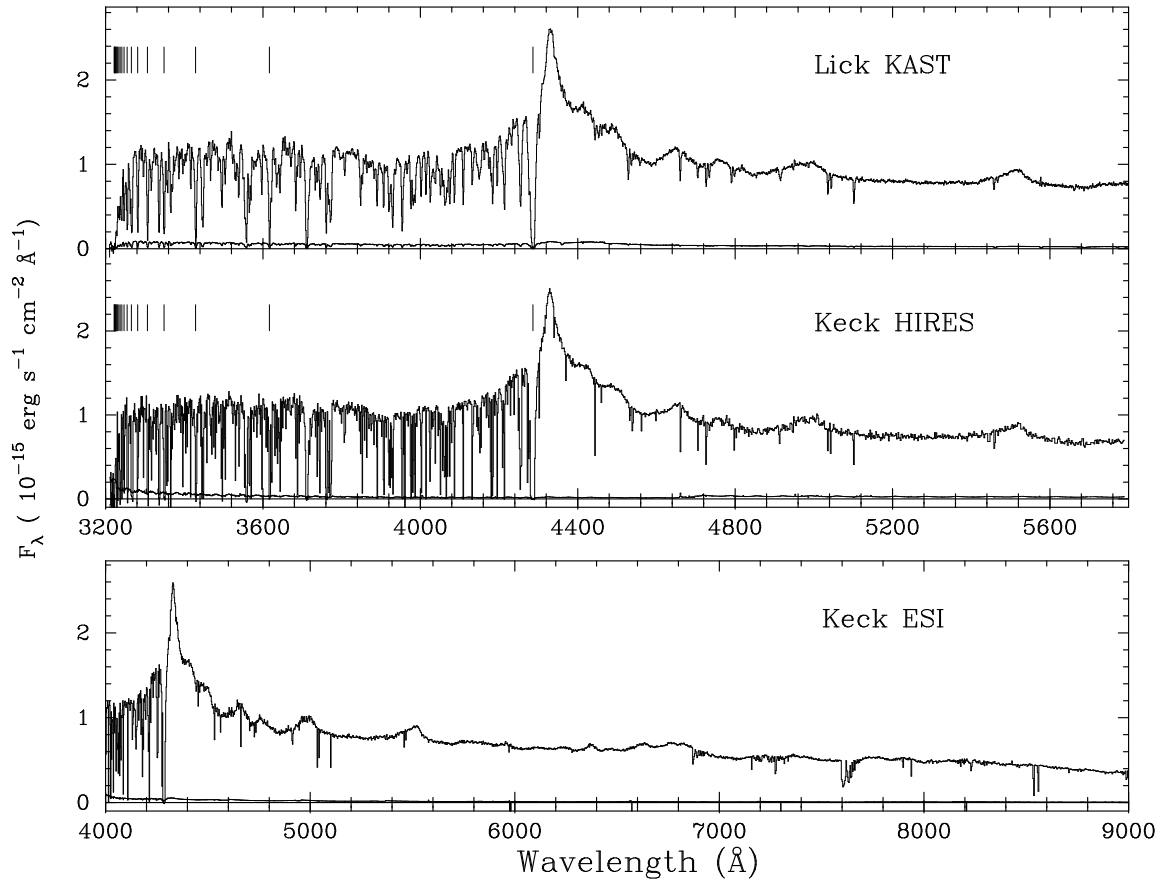}
\end{center}
\caption{\label{lickhiresesi} \NOTE{p1v1.ps} The spectra of
\object\ from the KAST spectrograph (top), HIRES (middle) and ESI
(bottom). We show the complete wavelength coverage for the Kast and
HIRES spectra, but not for the ESI, which extends to 10,000~\AA .  We
have applied relative flux calibration to all three spectra.  
The emission lines blend
to give a continuously undulating continuum level from 4400 --
5000~\AA .  The vertical marks above the Kast and HIRES spectra show
the positions of the \ly\ series lines in the absorption system at
\zdh\ that gives the D/H value. The \Lya\ absorption line of
this system, from which we get the H~I column density, is near
4285~\AA , just to the left of the peak of the \lya\ emission line. We
do not plot most pixels, to reduce the file size.}
\end{figure*}

%
% Too big for astro-ph, Nao not around to make new one on Friday.
%
%\begin{figure*}
%\begin{center}
%\includegraphics[angle=-90, scale=0.6]{lyseries.ps}
%\end{center}
%\caption{\label{lyseries} \NOTE{lyseries.ps} Expansion near the Lyman
%limit of the Kast (top half of each panel) and HIRES (lower half of
%each panel) spectra from Figure 1. In the top panel, the 
%Ly-4 is near 3348.5~\AA , 
%while the lower panel shows Ly-3 on the left and Ly-2
%on the right.}
%\end{figure*}

\begin{figure*}
\begin{center}
\includegraphics[angle=-90, scale=0.6]{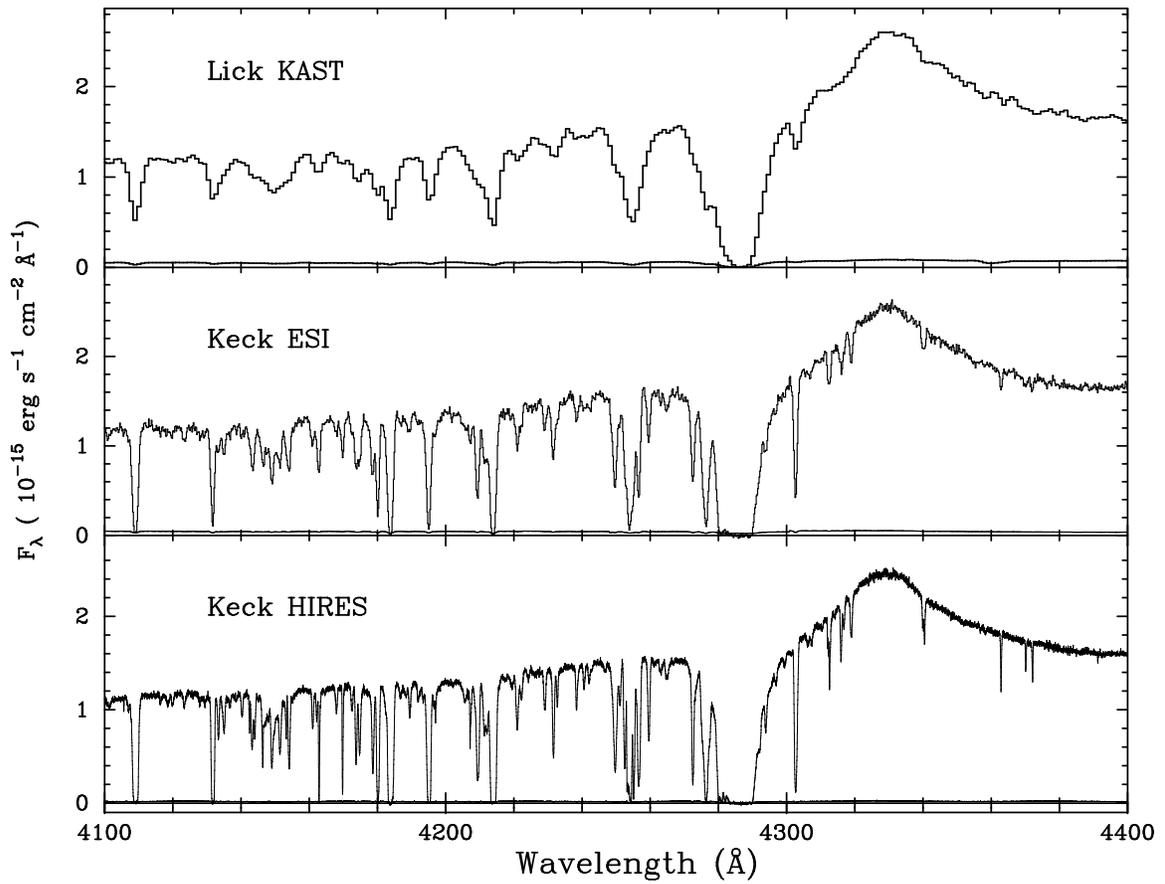}
\end{center}
\caption{\label{3instoriginal} \NOTE{p2v1.ps} Expansion of
the Kast, ESI and HIRES spectra from Figure 1. The \lya\ absorption
near 4285~\AA\ is from the system in which we measure D/H.}
\end{figure*}

\begin{figure*}
\begin{center}
\includegraphics[angle=-90, scale=0.6]{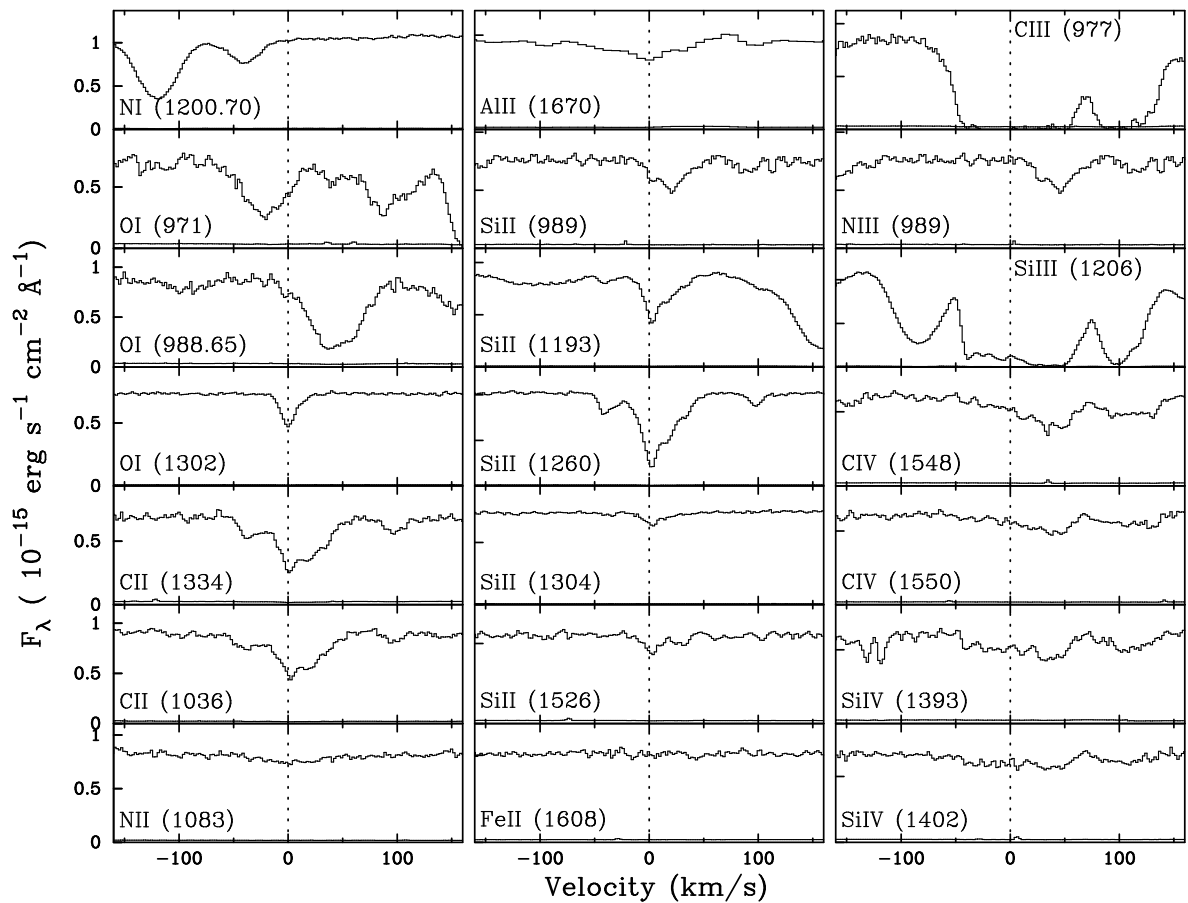}
\end{center}
\caption{\label{p5} \NOTE{p5v1.ps} Most of the metal absorption
lines near $z = 2.526 $. We shifted the Al~II
1670 spectrum, which is the only one from ESI, by $-16.5$~\kms\ to
correct a likely error.  
We see three types of components, grouped by
ionization; the low ionization lines represented by O~I alone,
intermediate ions C~II, Si~II and high ionization C~IV and Si~IV.  }
\end{figure*}

\begin{figure*}
\epsscale{0.7}
\plotone{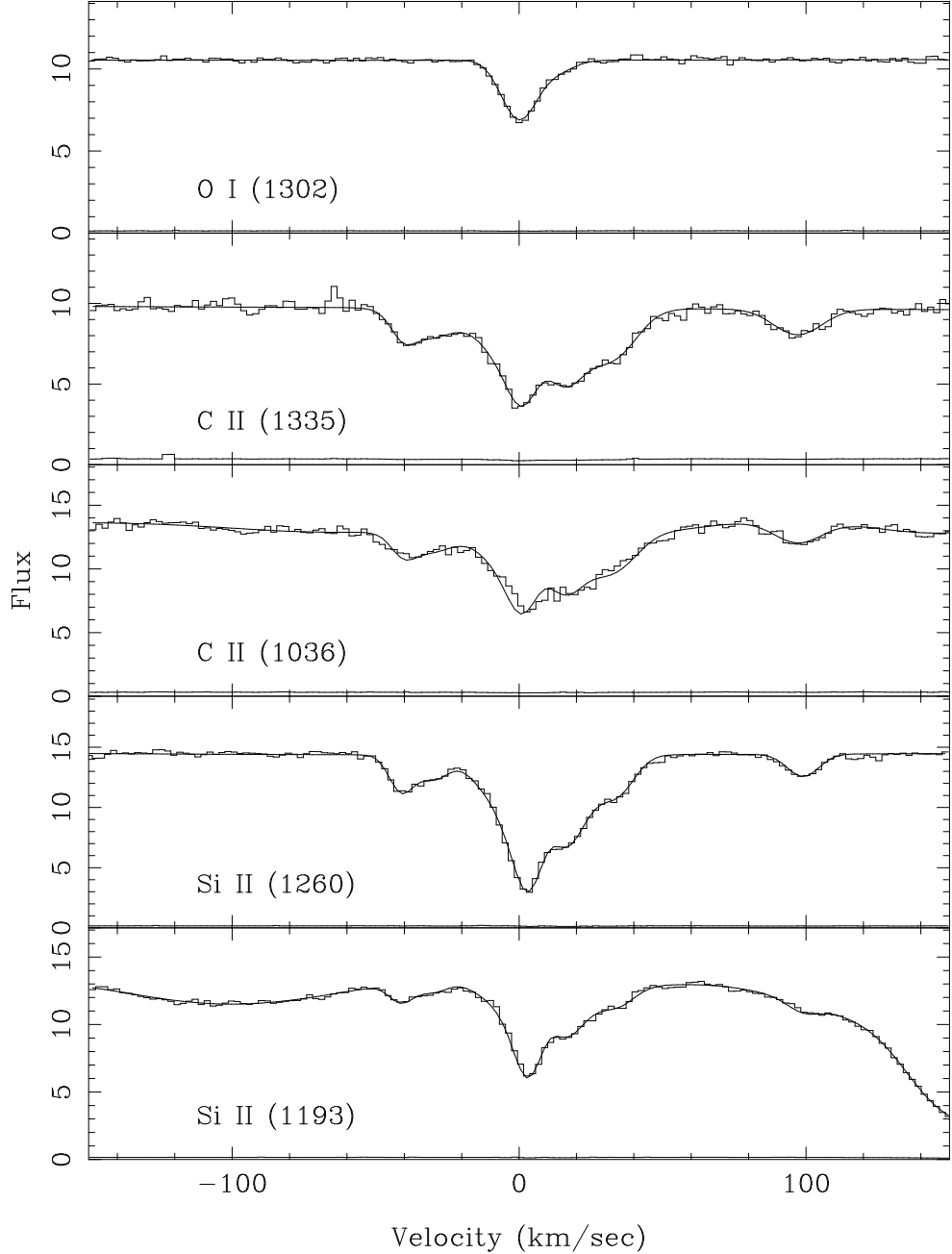}
\caption{\label{lowions1} \label{lowions} \NOTE{lowions1.ps} Voigt
profile fits to the lines of the low ionization ions. The line
parameters used to generate these profiles are given in Table
\ref{dhlinetab}.  The data is our combined HIRES spectrum, with
$1\sigma $ errors shown just above the zero flux level.  Flux is in
units of $10^{-16}$~erg~s$^{-1}$~cm$^{-2}$~\AA\ $^{-1}$.  The O~I
absorption shows a well defined component 1 at $v =0.0$~\kms, and
additional absorption in component 2 near $+13$~\kms.  
C~II and Si~II show several
components, including a main component near $v = 0$~\kms\ for C~II (1335) or
3~\kms\ for C~II (1036) and Si~II, and a second component near 16~\kms . 
We see no absorption near --82~\kms. }
\end{figure*}

\begin{figure*}
\epsscale{0.7}
\plotone{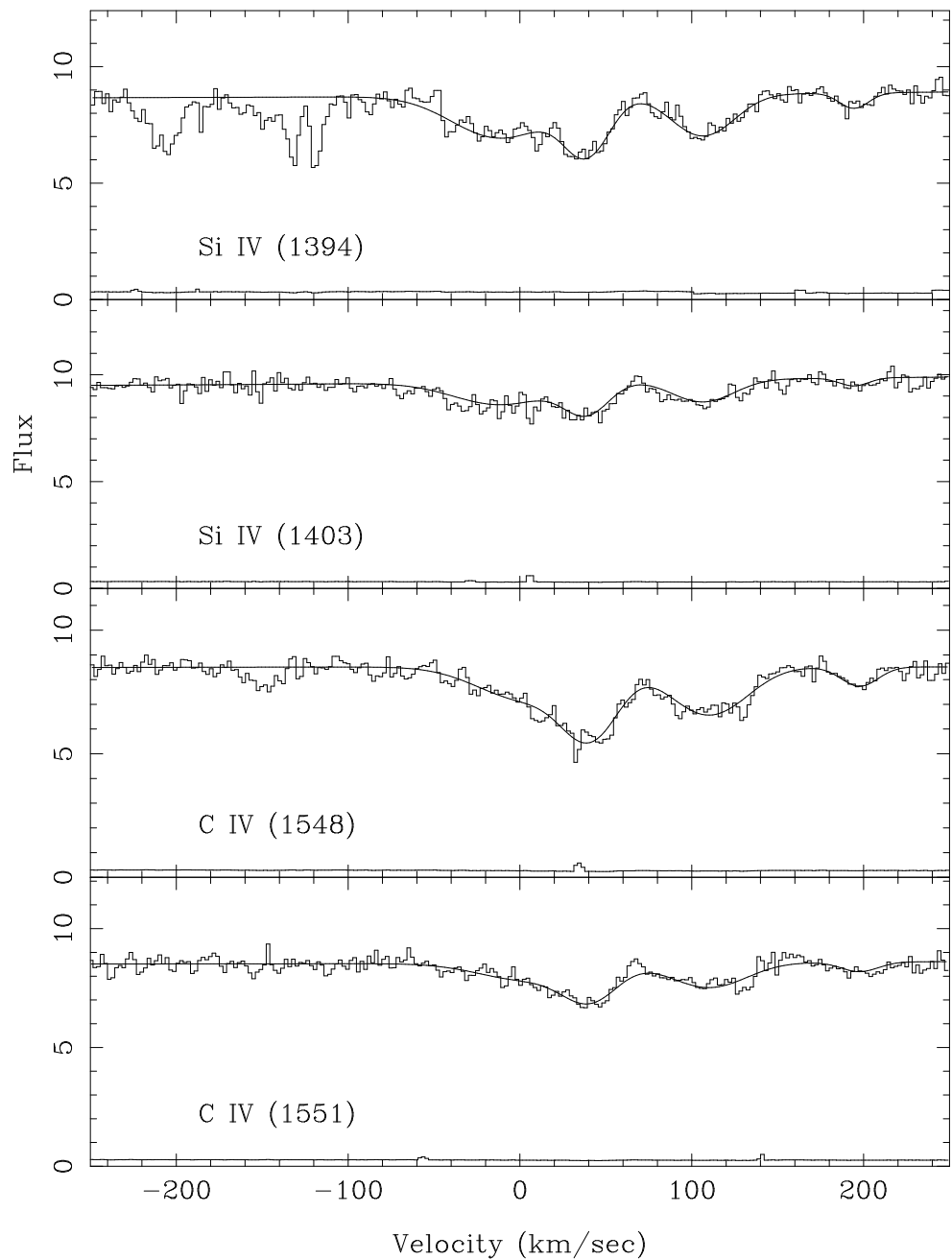}
\caption{\label{highions1} \NOTE{highions1.ps} As Figure \ref{lowions},
but here we show the lines of the high ionization ions.  Again, there is
no evidence of metal line absorption near --82 \kms.}
\end{figure*}

\clearpage

\begin{figure*}
\epsscale{0.7}
\plotone{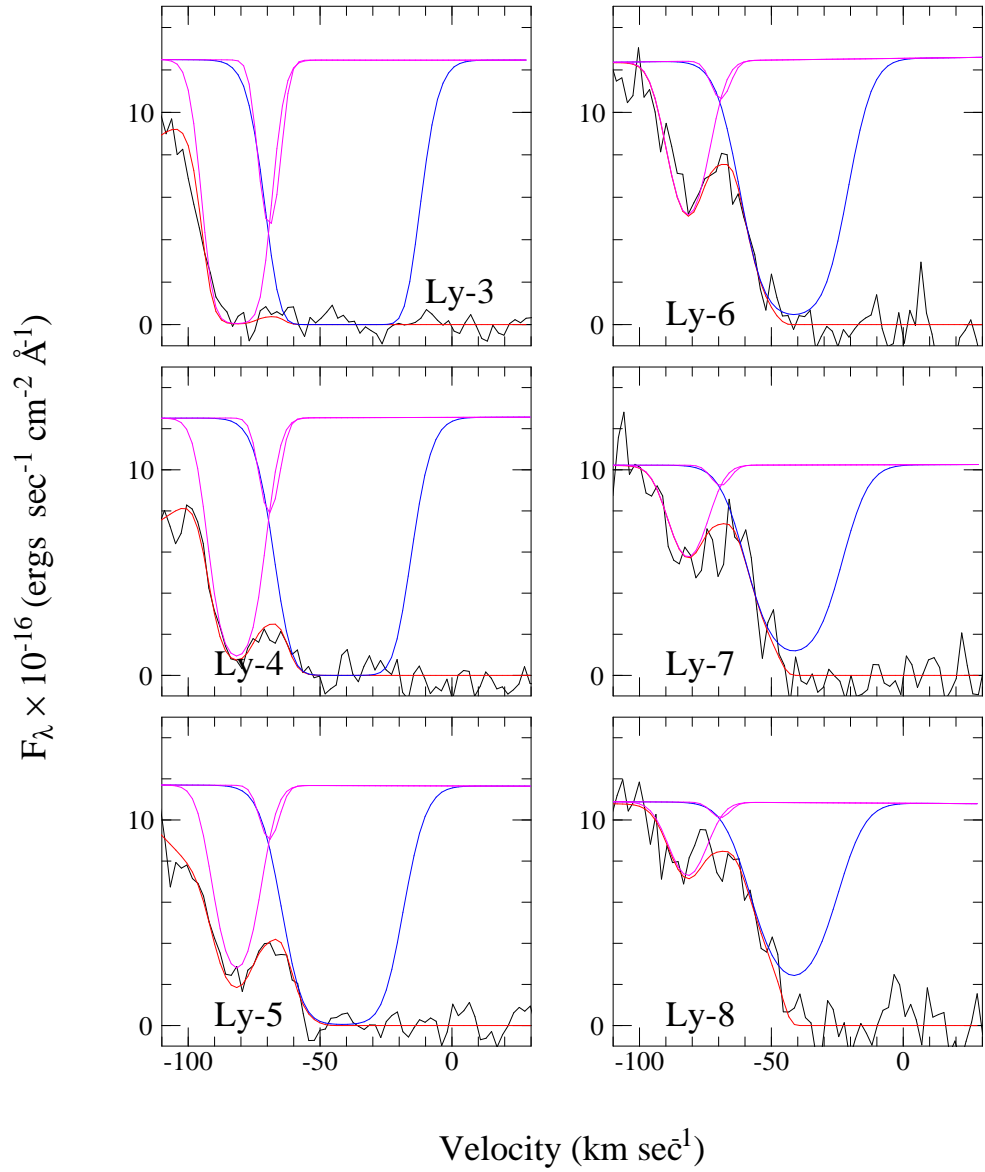}
\caption{\label{fig_d3} \NOTE{fig\_d3.ps} The blending of the
absorption line components D-1, D-2 and H-3 in Lyman series lines 4 --
9.  The two components of the D~I absorption are centered at $-82\
\kms$ and $-70\ \kms$.  H-3 is the deep broad line centered near $-40\
\kms$.  The data is our summed HIRES spectrum.  Also shown is the full
fit (D-1, D-2, H-3, and all other absorption associated with the
system).  Note that D-2 is strongly blended with both D-1 and H-3.}
\end{figure*}

\begin{figure*}
\epsscale{0.7}
\plotone{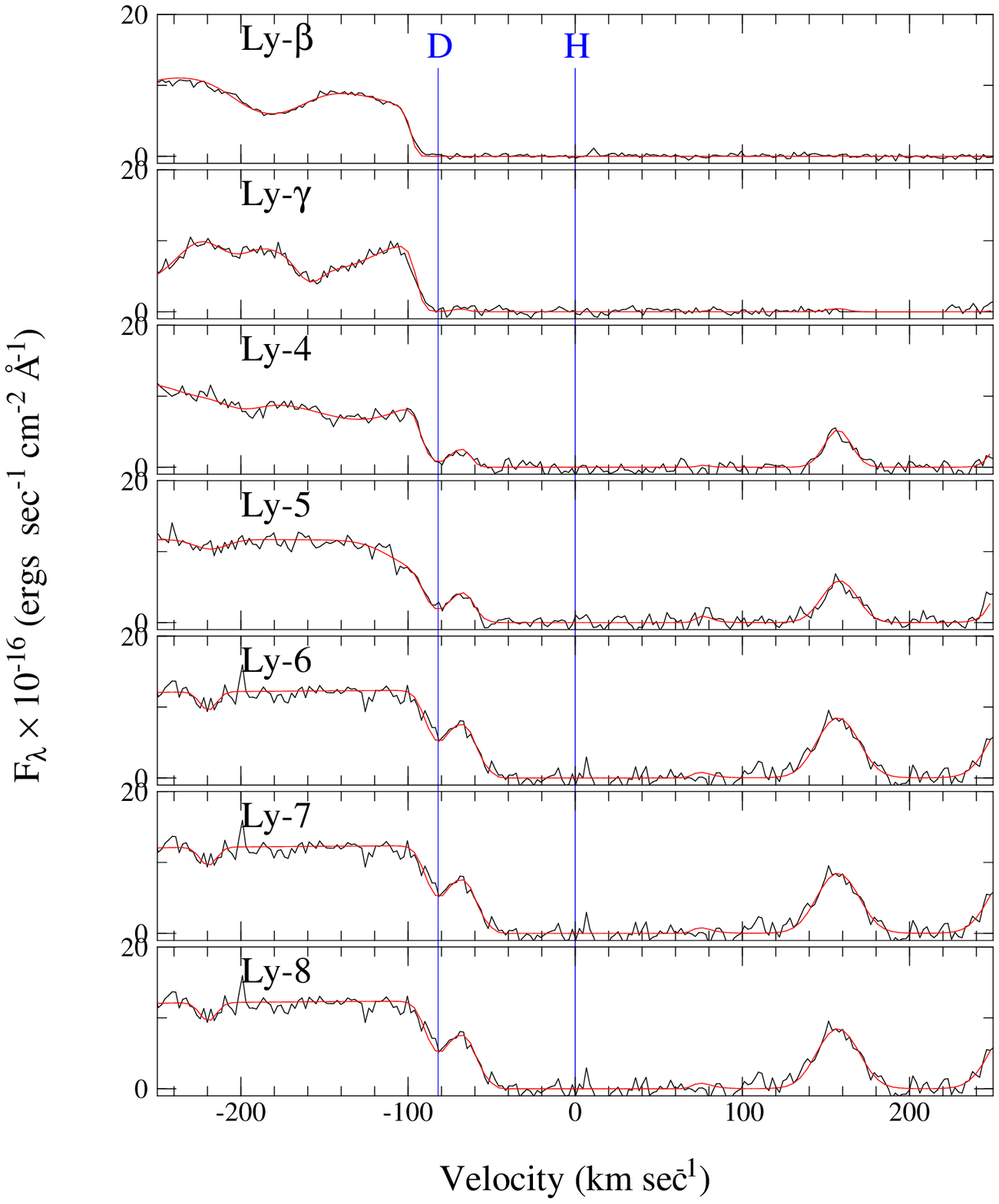}
\caption{\label{p4v0} \label{p4} \NOTE{p4v1.ps} The HIRES spectrum of
Ly-2 to 8, together with our model of the system, as given in Table
\ref{table_c}.}
\end{figure*}

\begin{figure*}
\epsscale{0.7}
\plotone{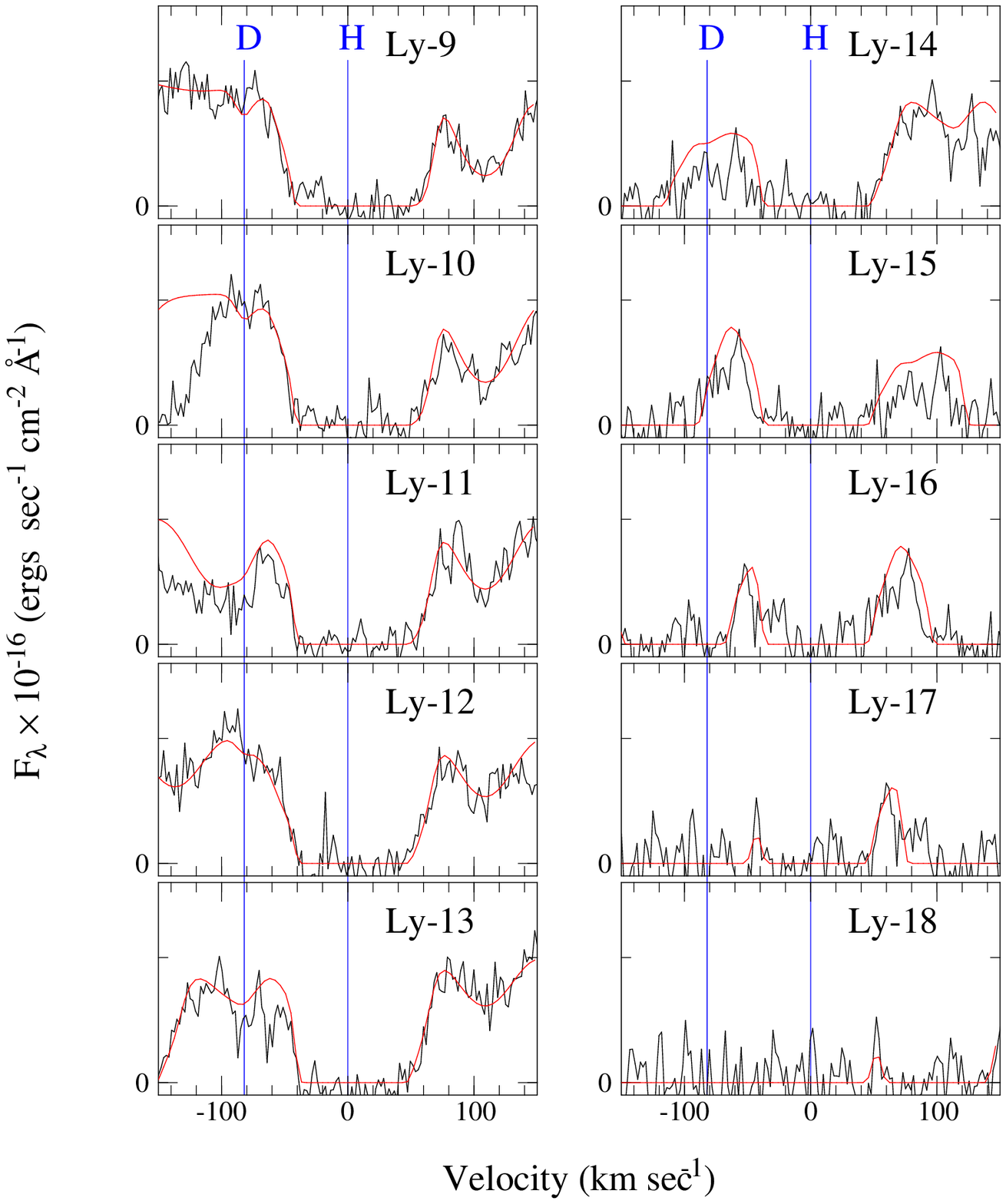}
\caption{\label{p15} \NOTE{p15v1.ps} The HIRES spectrum of Ly-9 to 18,
together with our model of the system, as given in Table \ref{table_c}.
}
\end{figure*}

\begin{figure*}
\epsscale{0.7}
\plotone{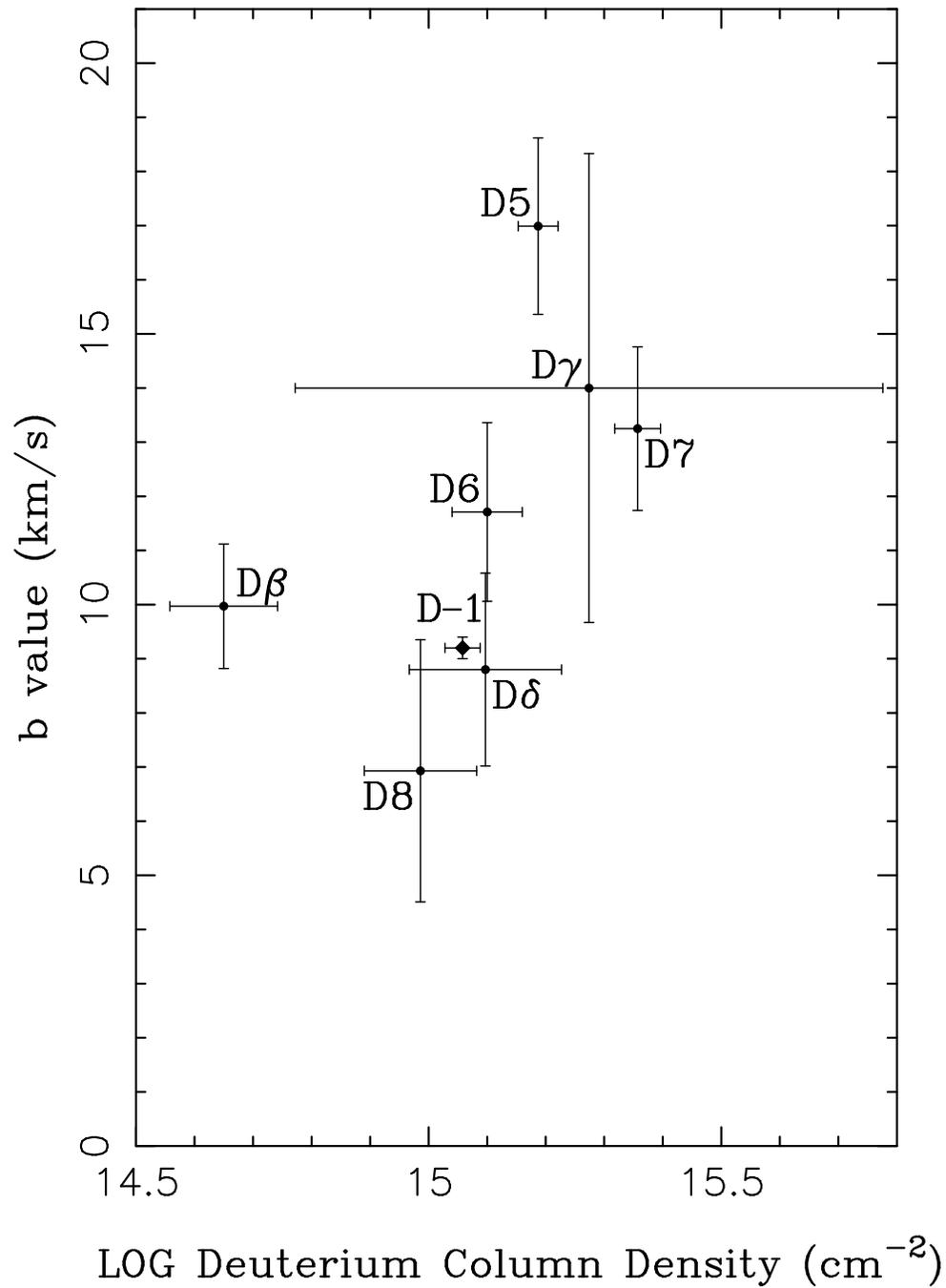}
\caption{\label{p8v0} \label{p8} \NOTE{p8v3.ps} The $b$ and \ndi\
values from single component fits to the individual D lines.  The only
free parameters in the optimizations were $b$(D) and \ndi(D).  All
other parameters, including $v$(D)=0, and $b$(H-3)=16.3\kms, were
fixed during the fitting process.  The displayed errors com from the
estimated covariance matrix we produced during the optimization
process.}
\end{figure*}

\begin{figure*}
\epsscale{0.9}
\plotone{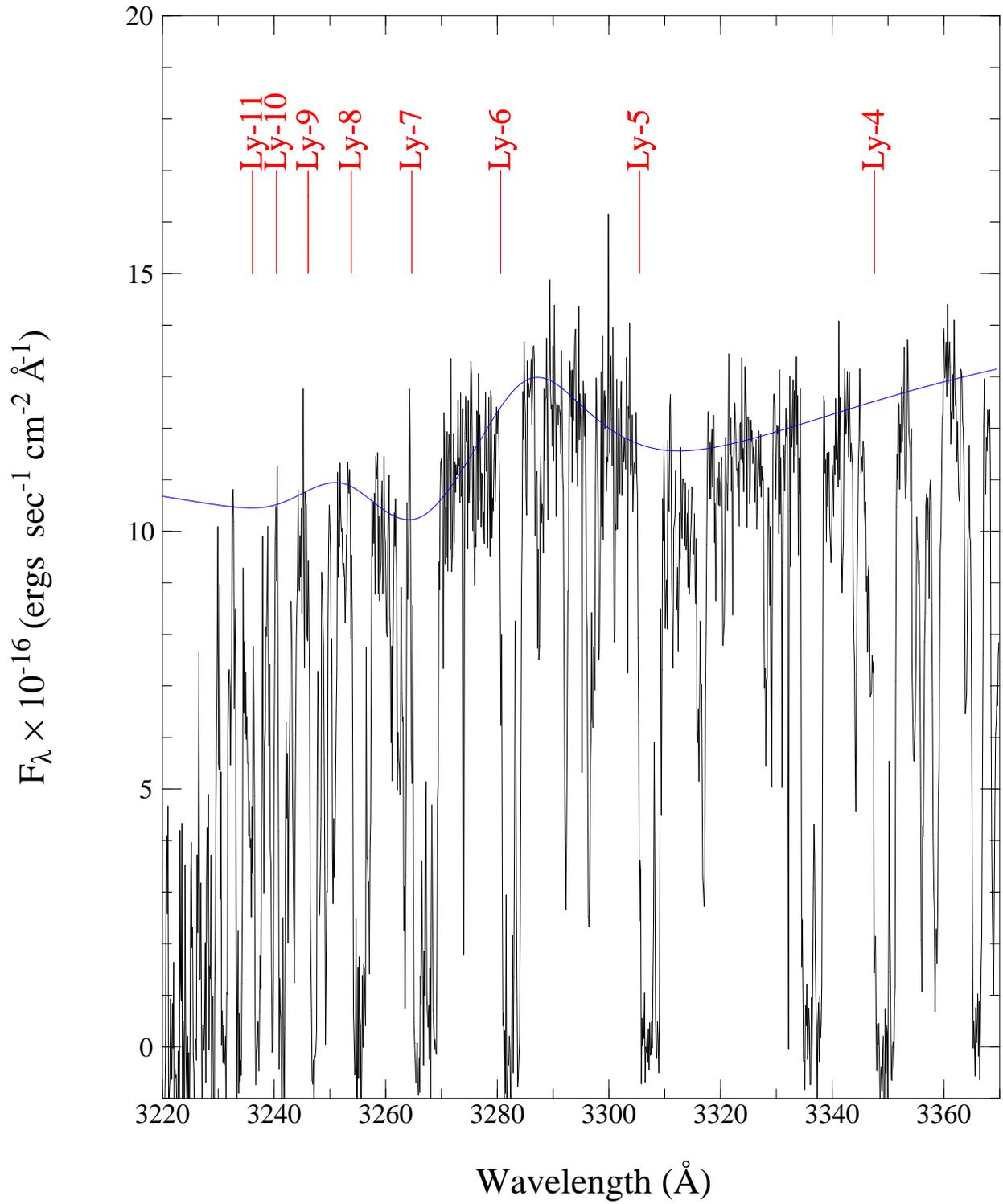}
\caption{\label{fig_d4} \NOTE{fig\_d4.ps} The continuum level used to construct
our models of the D absorption.  The positions of the Lyman series lines are
marked at the top of the panel.}
\end{figure*}

\begin{figure*}
\epsscale{0.9}
\plotone{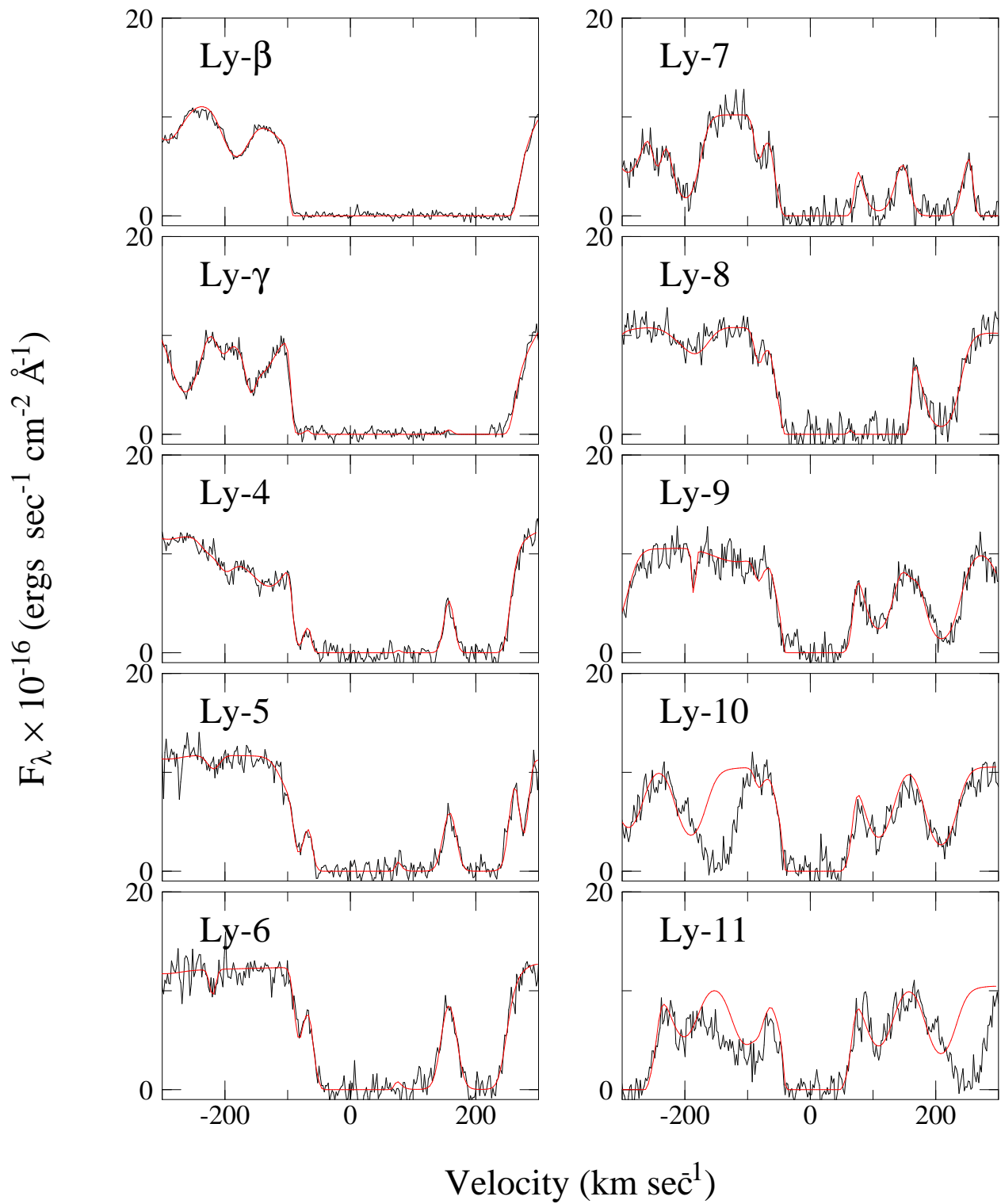}
\caption{\label{fig_d1} \NOTE{fig\_d1.ps} The complex continuum shown in
Fig. \ref{fig_d4} is required to obtain the fit shown here using relatively few
\lyaf\ lines.
}
\end{figure*}

\begin{figure*}
\epsscale{1.0}
\plotone{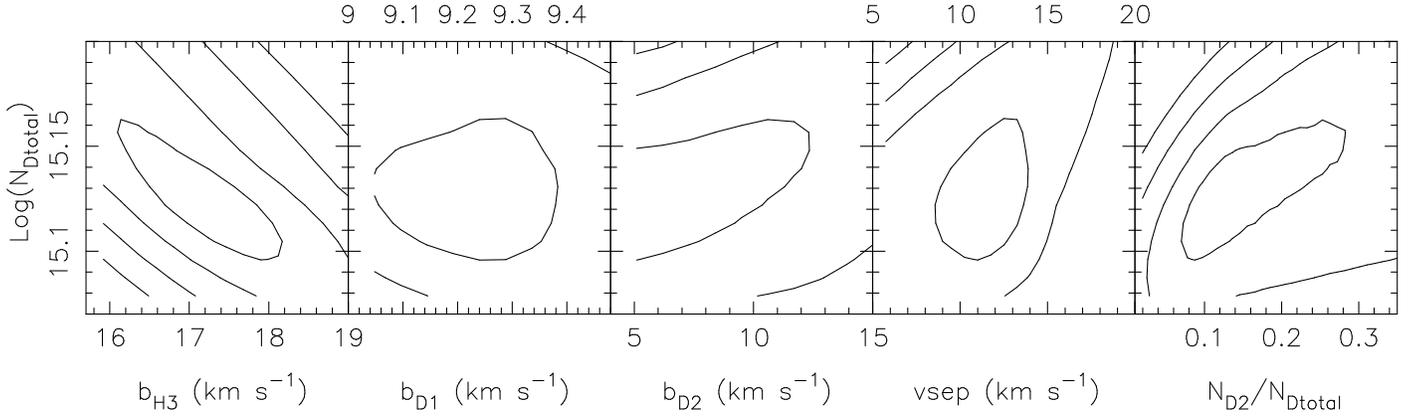}
\caption{\label{dgrid} \NOTE{dgrid.ps} The results of our grid search to 
measure \ndi.  We plot contours of constant \chisq\ for \ndi\ against each
of the parameters we varied in the search.  The contours are at the 1,2,3, 
and 4 sigma levels.}
\end{figure*}

\begin{figure*}
\epsscale{0.5}
\plotone{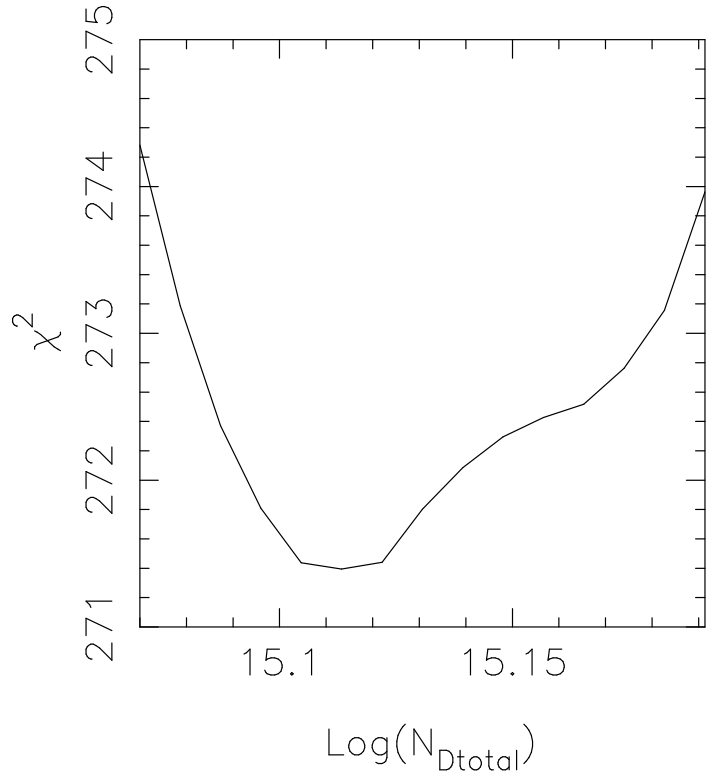}
\caption{\label{dchi} \NOTE{dchi.ps} The minimum \chisq\ value that we
found in our grid search, as a function of the total \ndi\ (\cmm ) in
components 1 and 2}
\end{figure*}

\begin{figure*}
\epsscale{0.6}
\plotone{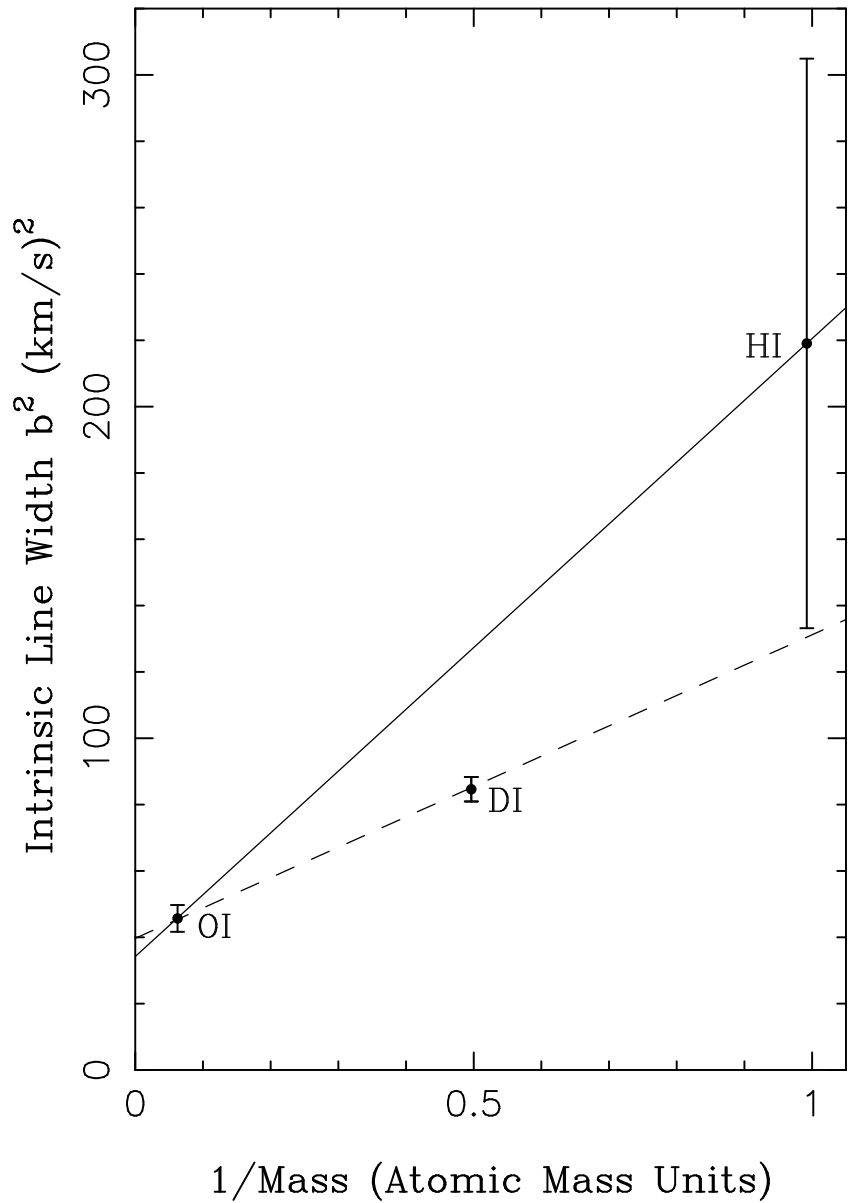}
\caption{\label{p7v3} \label{p7} \NOTE{p7v3.ps} Square of the
intrinsic width of the lines of ions as a function of 1/mass of the
ion in atomic mass units.  We plot the $b_{int}^2$ for the main
components of the O~I, D~I and H~I (components O-1, D-1 and H-1). The
solid line connects the O~I and H~I points and ignores the D~I.  The
slope of this line gives a temperature $T=1.1 \pm 0.6 \times 10^4$~K
and the intercept gives the turbulent velocity of $b_{turb} = 5.8 \pm
0.6$~\kms .  The line predicts $b(D-1) = 11.3 \pm 1.8$~\kms . The
observed $b(D-1) = 9.2 \pm 0.2$~\kms\ is $1.2\sigma $ below, and consistent
with this prediction.  The dashed line that is the best fit to O, H and D.
The data are also consistent with this fit, which we prefer.  Although
the $b$-value of the H-1 component is not well known, the data shown
on this plot provide evidence that D-1 is D rather than H.  }
\end{figure*}

%\clearpage

\begin{figure*}
\plotone{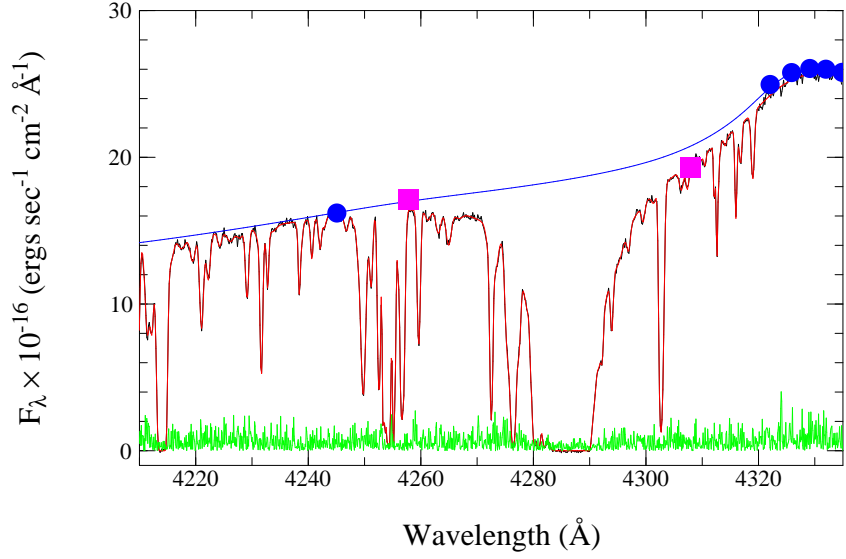}
\caption{\label{f1008_v1} \NOTE{f1008\_v1.ps} Our initial model of the
spectrum 4284 \AA.  We show the observed data, our model and the
continuum level of our model.  The jagged horizontal array below 2
units of flux shows five times the absolute value of the difference
between our model and the data. The squares and circles are the
b-spline control points that define the continuum model.  The circles
are fixed in wavelength and free in flux, while the squares are free
in both wavelength and flux.  The \chisq\ for this model is 5911, with
4759 degrees of freedom.}
\end{figure*}

\begin{figure*} 
\epsscale{0.7}
\plotone{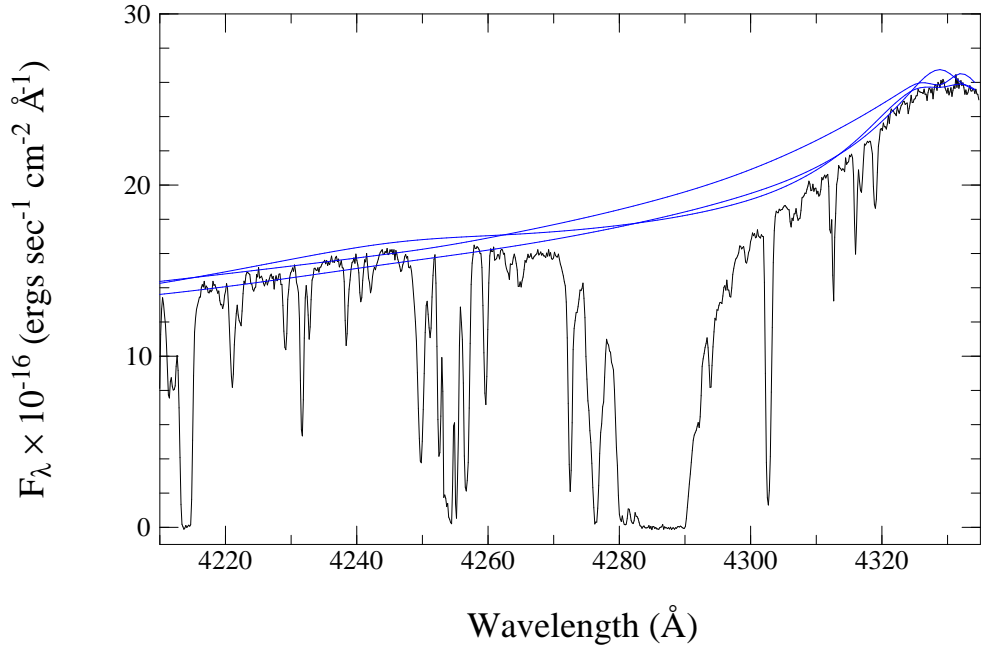}
\caption{\label{f1009_v2} \NOTE{f1009\_v2.ps} Three of the randomly
generated continua used at the start of the optimization restarts that
we used
to estimate \nhi.}
\end{figure*}

\clearpage

\begin{figure*}
\epsscale{0.6}
\plotone{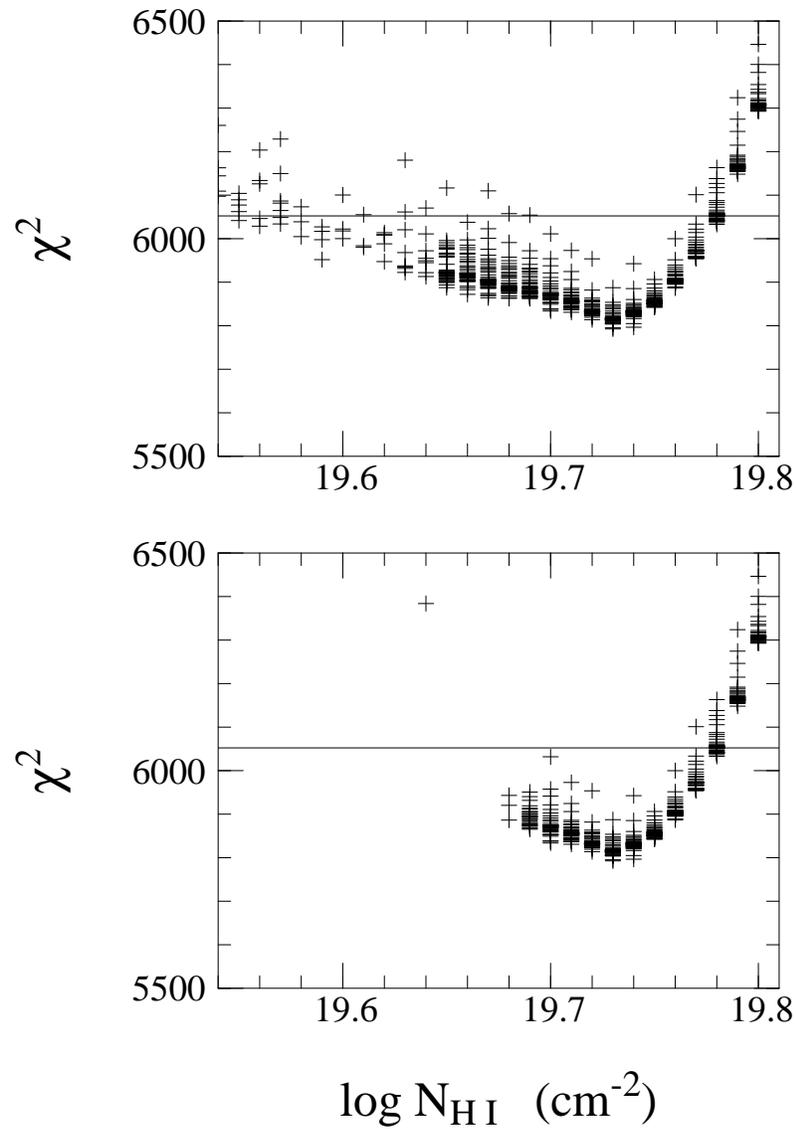}
\caption{\label{f1014_v1} \NOTE{f1014\_v1.ps} \chisq\ values as a
function of the \lnhi\ in components 1 and 2. The \chisq\ values are
those returned by the optimization code. The results are shown for
3195 restarts. Models containing \lyaf\ absorbers with $b > 150\ \kms$
are excluded from the lower panel. We accept \lnhi\ values that gave
\chisq\ values below the horizontal lines at 6052.3.}
\end{figure*}

\begin{figure*}
\plotone{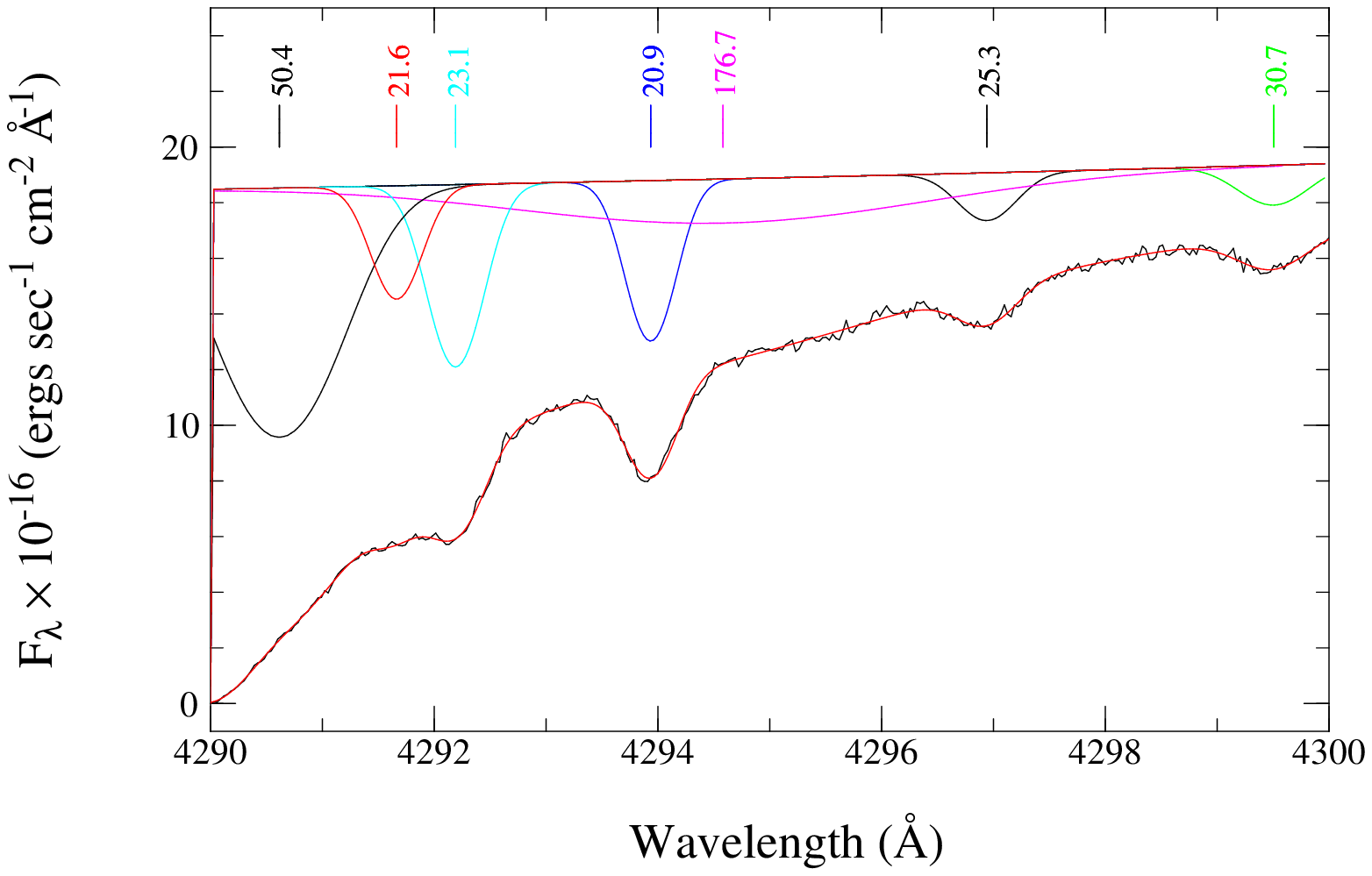}
\caption{\label{f1012_v3} \NOTE{f1012\_v3.ps} Example of a model with
a wide \lyaf\ absorption line.  This model has $\lnhi = 19.67\ \cmm$
and uses the \lyaf\ lines that we show, including the
wide, shallow line centered near 4294.5~\AA\ with $b = 176.7$~\kms .
The line running through the spectrum is the complete model that
includes the continuum level, the DLA and the \lyaf\ lines.
The optimizer restarts found a number of models, typically with $\lnhi
< 19.69$~\cmm , which used similar wide \lyaf\ lines.  In these cases, 
it appears
that the optimizer is using wide \lyaf\ absorption to provide opacity
that should be provided by the DLA or the continuum. Because such wide
\lyaf\ lines are not observed in other studies of the \lyaf, we rejected
models with wide \lyaf\ lines.}
\end{figure*}

\begin{figure*}
\plotone{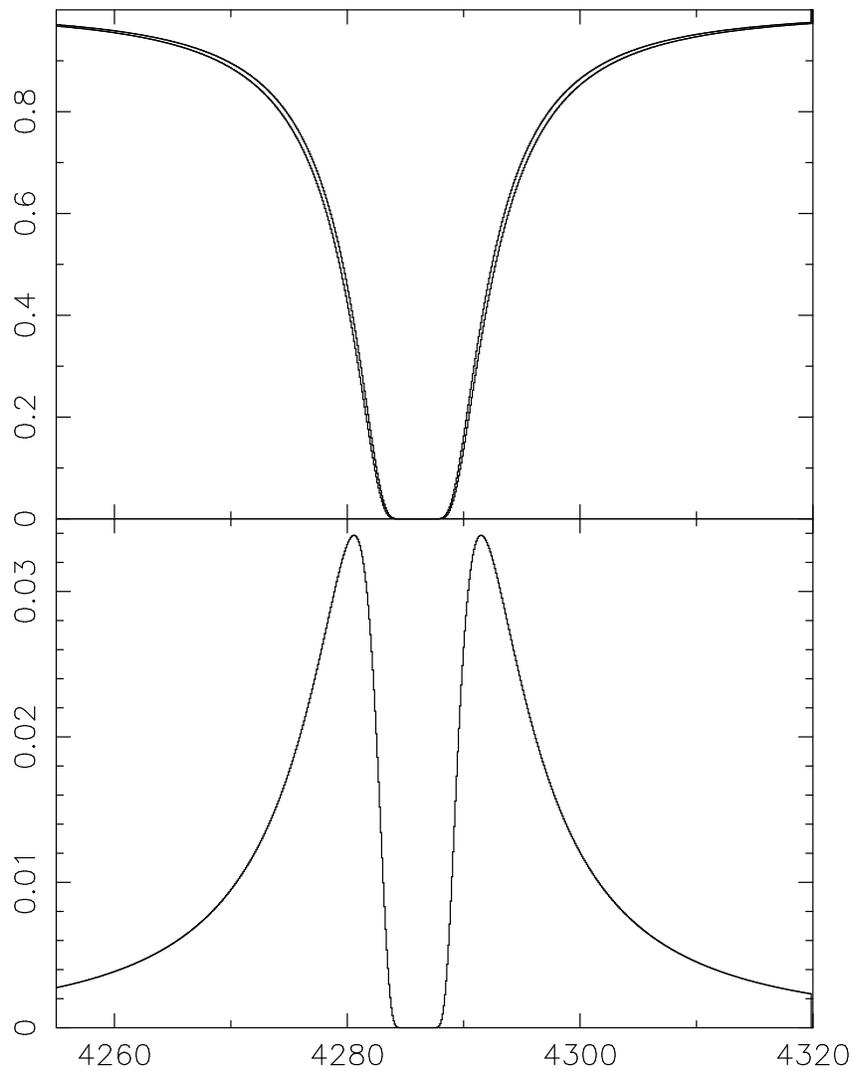}
\caption{\label{deltaflux} \NOTE{deltafluxv1.ps} The upper panel shows
the normalized in flux profiles of two absorption lines whose column
densities differ by \lnhi\ $=0.04$ \cmm , the error we quote on the
\lnhi\ value for the system that shows D.  The upper line has a column
density of \lnhi\ $= 19.73$ \cmm.  The lower panel displays the
difference in flux between these two absorption line profiles.  }
\end{figure*}

\begin{figure*}
\epsscale{0.6}
\plotone{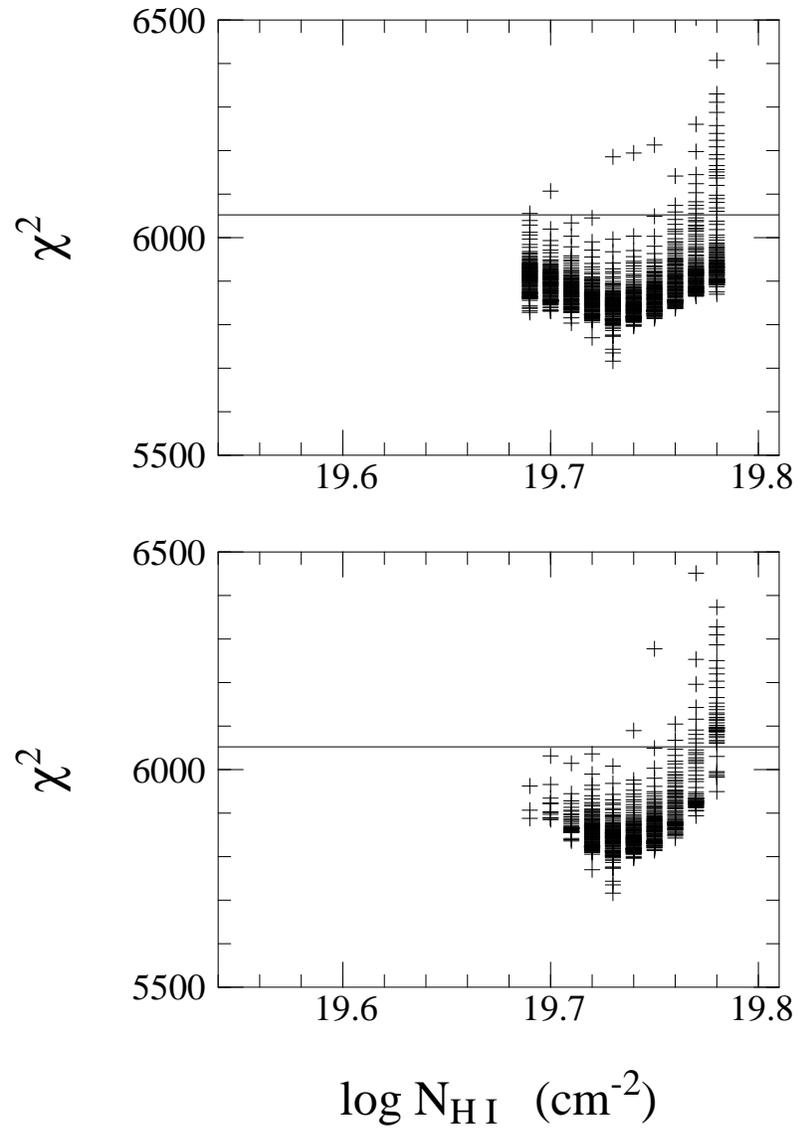}
\caption{\label{f1013_v2} \NOTE{f1013\_v2.ps} As \ref{f1014_v1}, only
for the 7898 optimization restarts with three continuum control points
in the region between 4260~\AA\ and 4325~\AA.}
\end{figure*}

\begin{figure*}
\epsscale{0.7}
\plotone{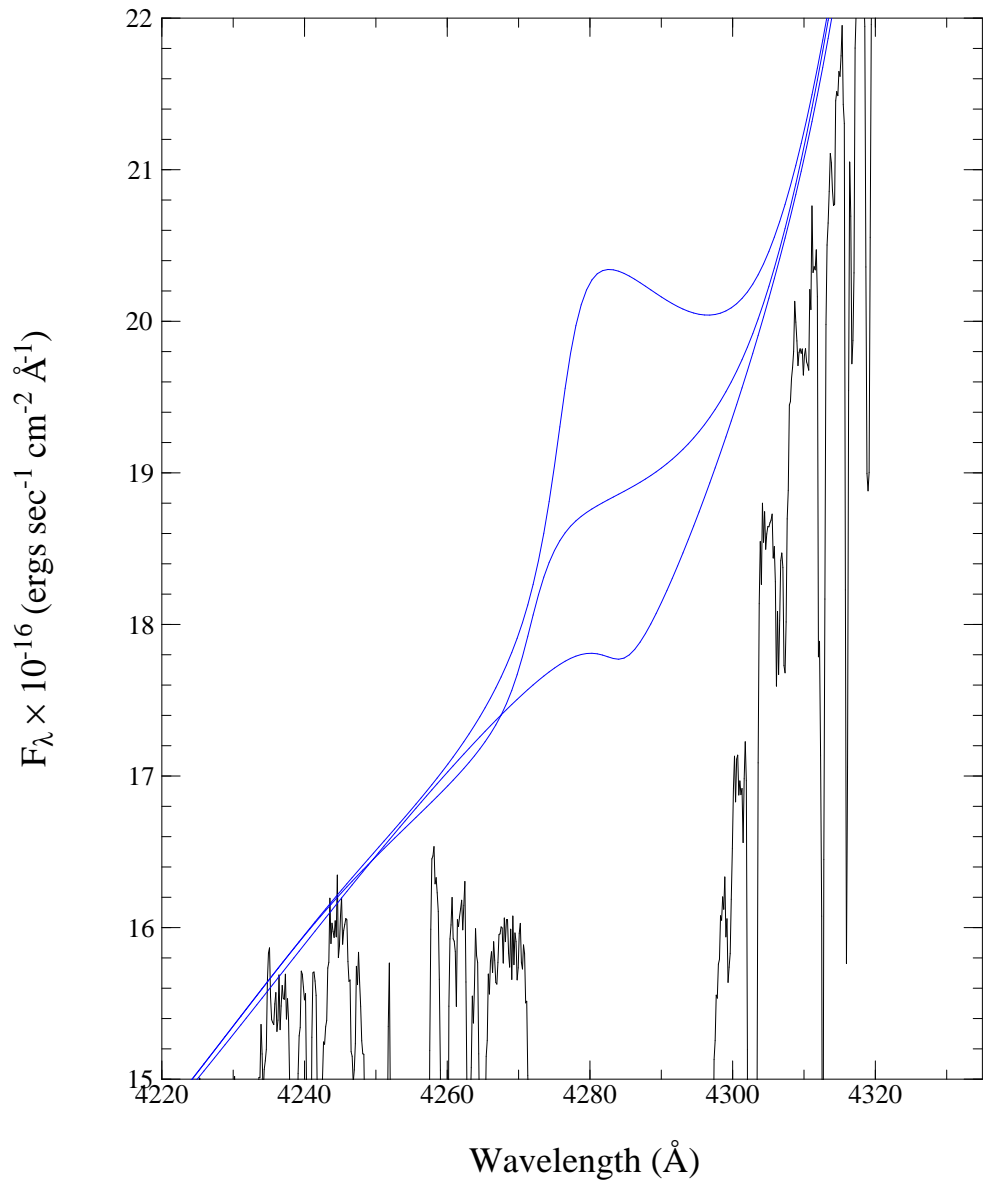}
\caption{\label{f1002ab_v1} \NOTE{f1002ab\_v1.ps} Three examples of
continuum shapes that were rejected by the filter we applied to the
three continuum control point models.}
\end{figure*}

\begin{figure*}
\plotone{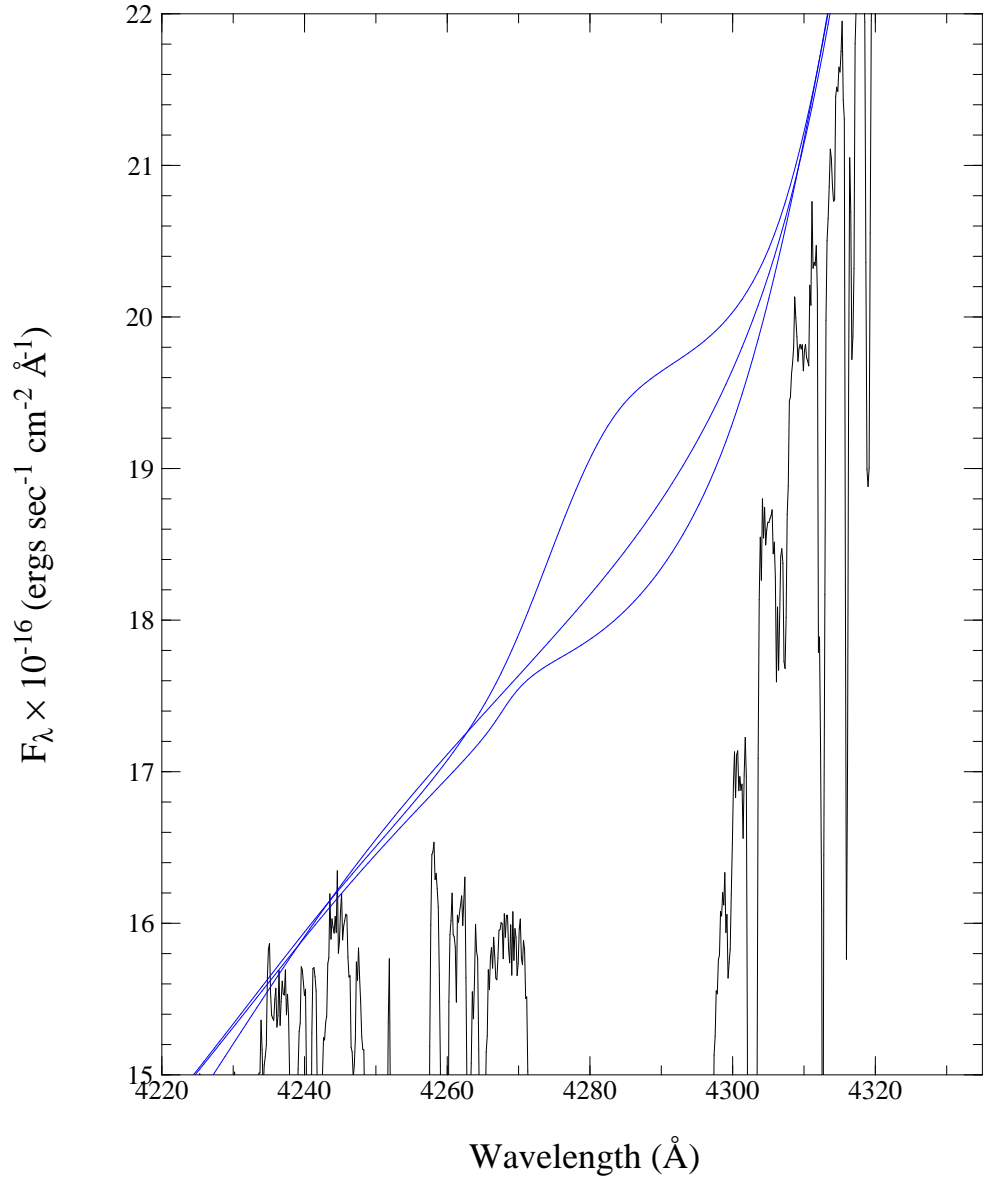}
\caption{\label{f1002bb_v1} \NOTE{f1002bb\_v1.ps} Three examples of
continuum shapes that were found acceptable by the filter we applied to
the three continuum control point models.  The top curve is from a
model with $\lnhi = 19.78\ \cmm$, the middle curve is from a model
with $\lnhi = 19.73\ \cmm$, and the bottom curve is from a model with
$\lnhi = 19.69\ \cmm$.}
\end{figure*}

\begin{figure*}
\epsscale{0.7}  %1.4 is way too big, and does not print well
\plotone{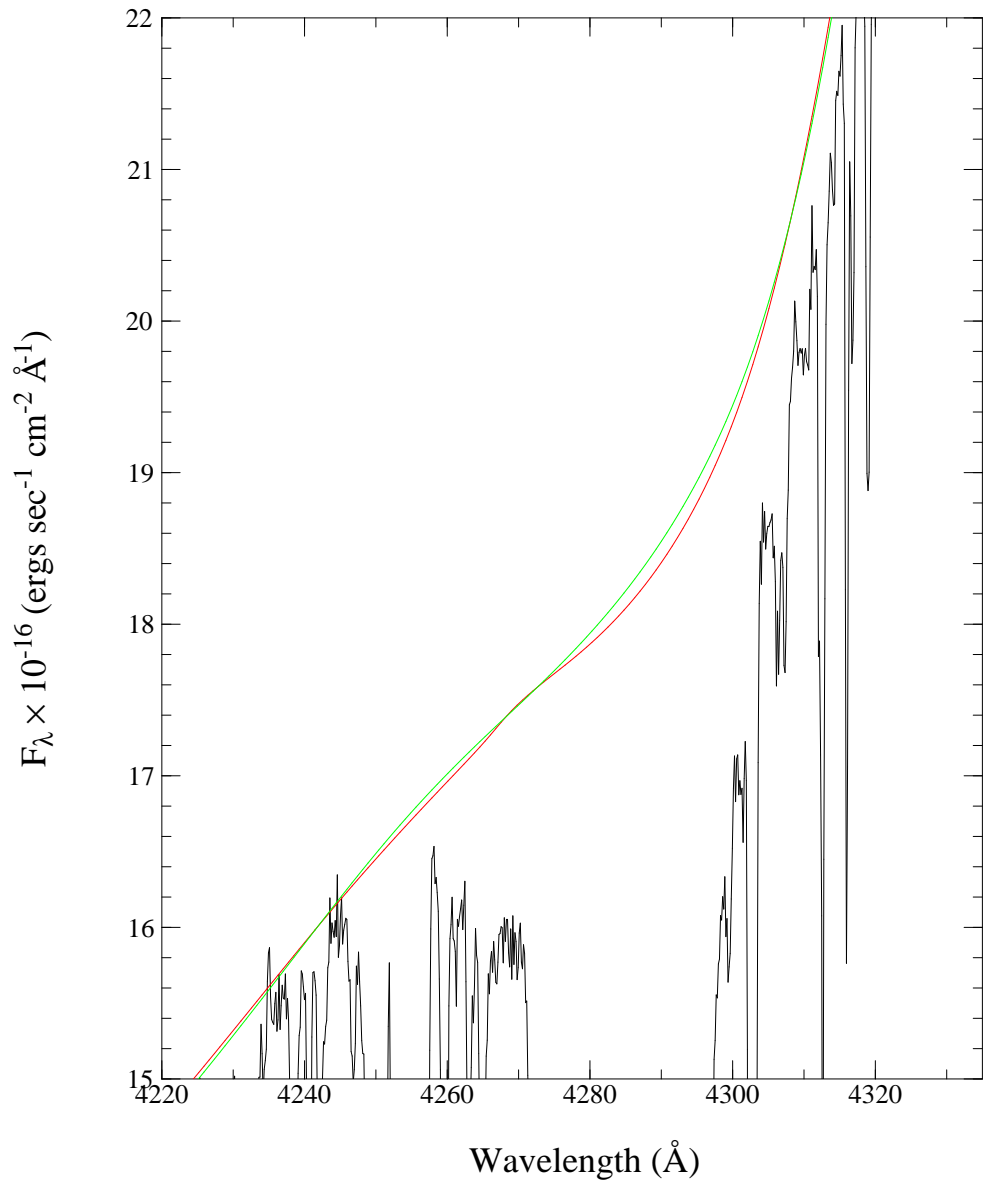}
\caption{\label{f1022} \NOTE{f1022\_v1.ps} The continua for the best
models we were able to find with $\lnhi = 19.69\ \cmm$.  The continuum
that is low at 4290\AA\ is from the best model with three continuum
control points in the region between 4260\AA\ and 4325\AA . The other
curve is from the best model with two continuum control points.}
\end{figure*}

\begin{figure*}
\epsscale{0.7}
\plotone{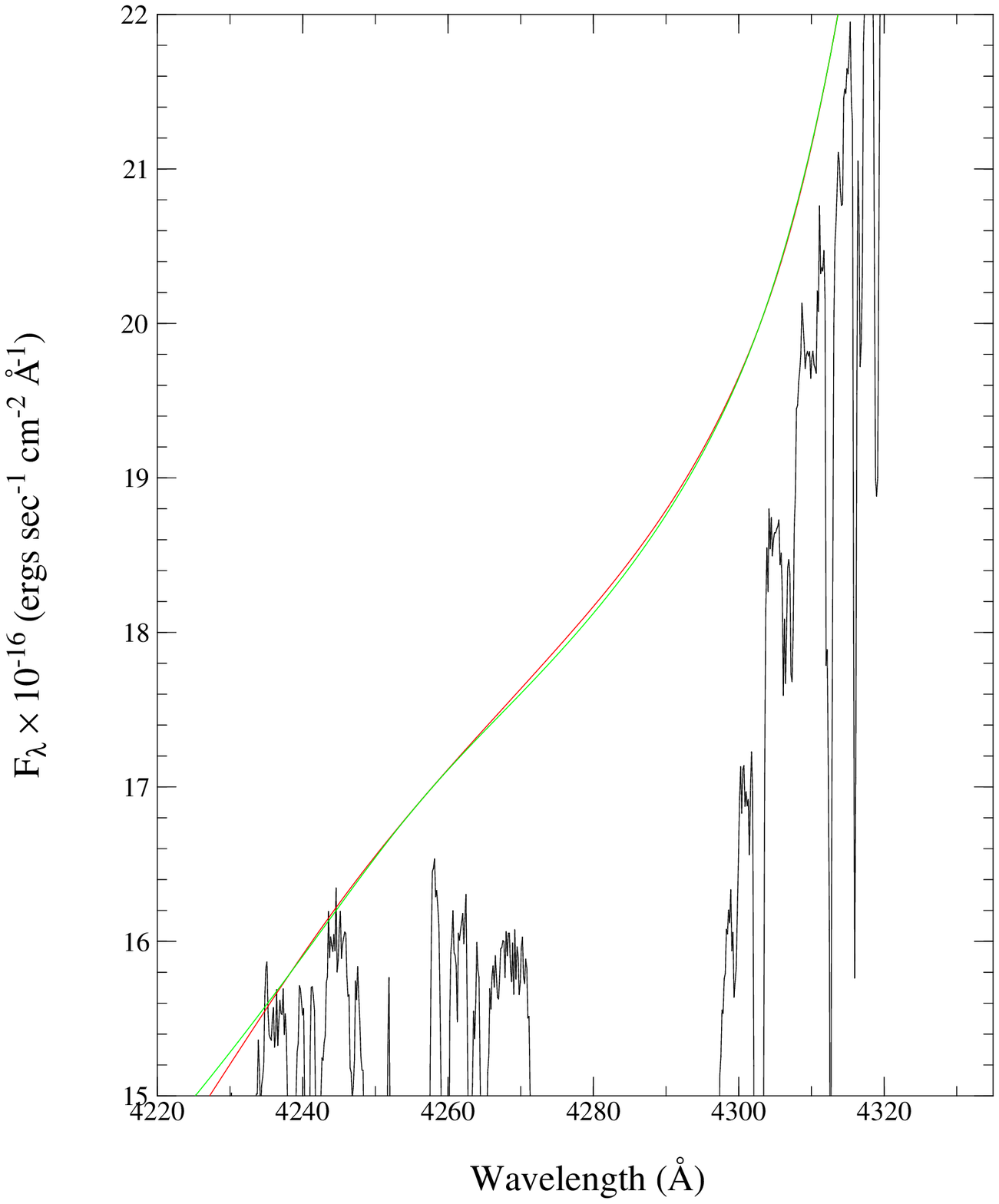}
\caption{\label{f1023} \NOTE{f1023\_v1.ps} As \ref{f1022}, but for
models with $\lnhi = 19.73\ \cmm$. }
\end{figure*}

\begin{figure*}
\epsscale{0.7}
\plotone{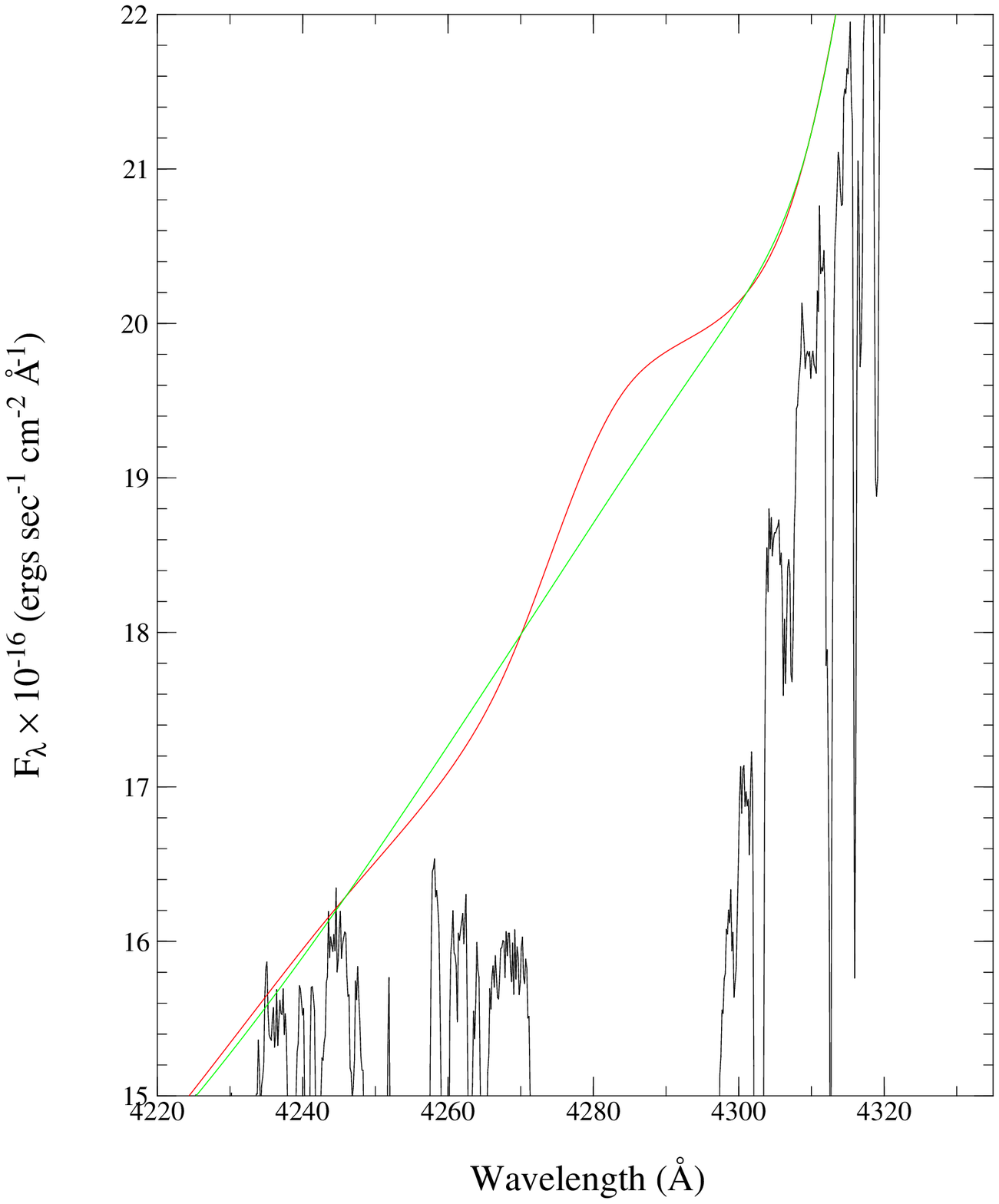}
\caption{\label{f1024} \NOTE{f1024\_v1.ps} As \ref{f1022}, but for 
models with $\lnhi = 19.78\ \cmm$.  In this case the three continuum
control point model is high at 4290\AA.}
\end{figure*}

\begin{figure*}
\epsscale{0.7}
\plotone{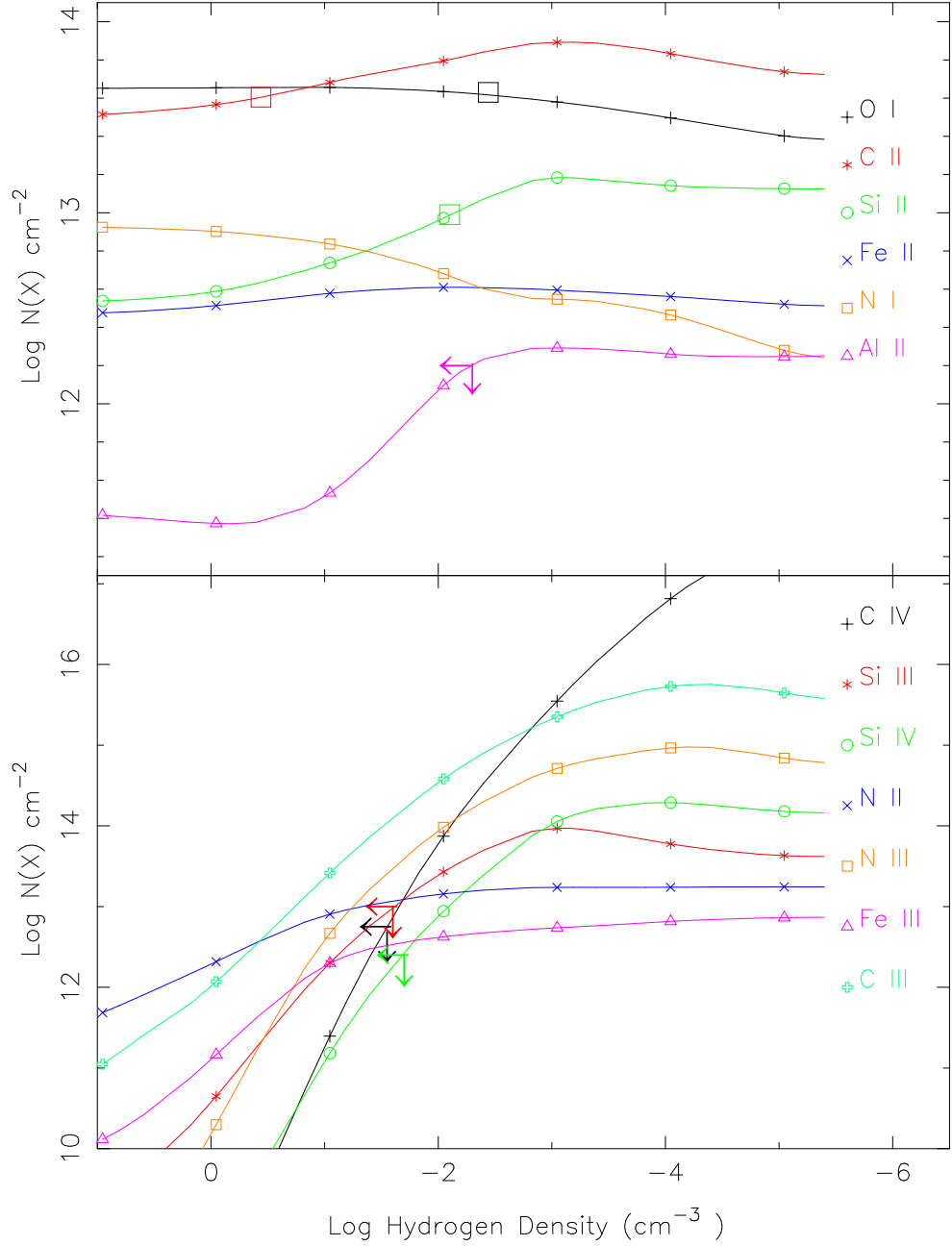}
\caption{\label{cloudy1975} \NOTE{result\_19.75} Column densities for
various metal ions as a function of the hydrogen density $n_{H}$
for a photoionized cloud with \lnhi\ $=19.73$~\cmm , a metal abundance 
$[X/H] = -2.77$, and a $J_{912} = 10^{-21}$ ergs \cmm\ s$^{-1}$
Hz$^{-1}$ sr$^{-1}$.
Ionization increases to the right.  We show solar abundance ratios.
The O~I column density is insensitive to the ionization.
Measured column densities in components 1 \& 2 are
shown by the three large boxes in the upper panel, while four
allowed upper limits are shown by arrows.
The preferred density is $\log n_H \simeq -1.5$~\cmmm . }

\end{figure*}

\begin{figure*}
\epsscale{0.7}
\plotone{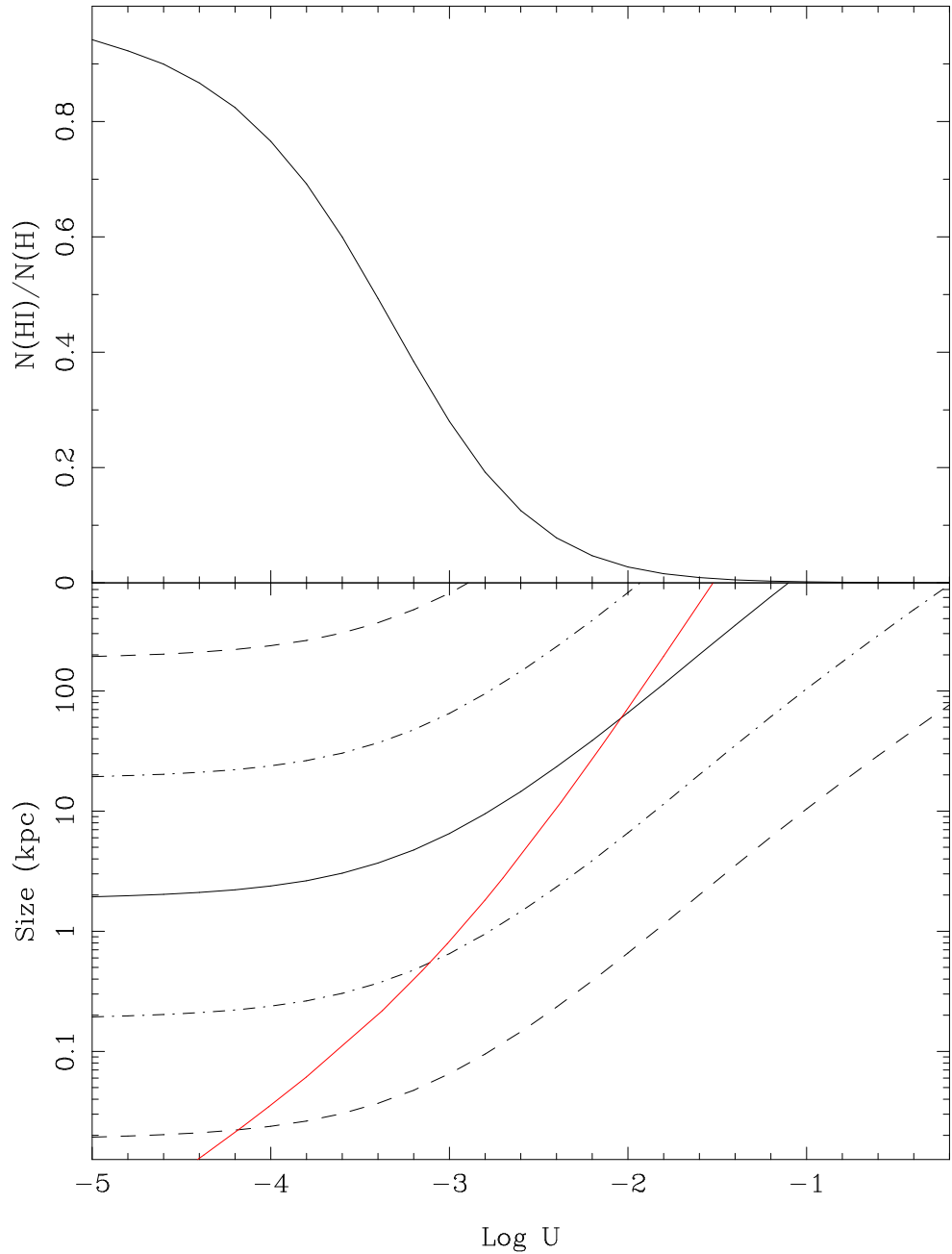}
\caption{\label{abs_size} The upper panel shows the neutral fraction
of the gas with \lnhi $=19.73$~\cmm\ as a function of ionization
parameter $\log U$ as returned by Cloudy, assuming a constant Hydrogen
density of $n_{H} = 0.01$ \cmmm.  The lower panel shows the size of
the absorber along the line of sight in kpc, again as a function of
$\log U$.  In the lower panel, the solid curve that begins at 2 kpc
displays the absorber size when $n_{H} = 0.01$~\cmmm . The $J_{912}$ increases
to the right along this and other curves with the same shape.
We also show curves for $n_{H}$
increased (lower on the plot) or decreased 
by a factor of 10 (dot-dashed lines) or 100 
(dashed lines). The steeper curve that begins on the horizontal
axis at $\log U = -4.4$ shows the size when the
$J_{912} = 10^{-21}$ ergs \cmm\ s$^{-1}$ Hz$^{-1}$ sr$^{-1}$. The density
increases to the right along this curve.
The ion column densities indicate $\log U \simeq -2.84$. }
\end{figure*}

\begin{figure*}
\epsscale{0.7}
\plotone{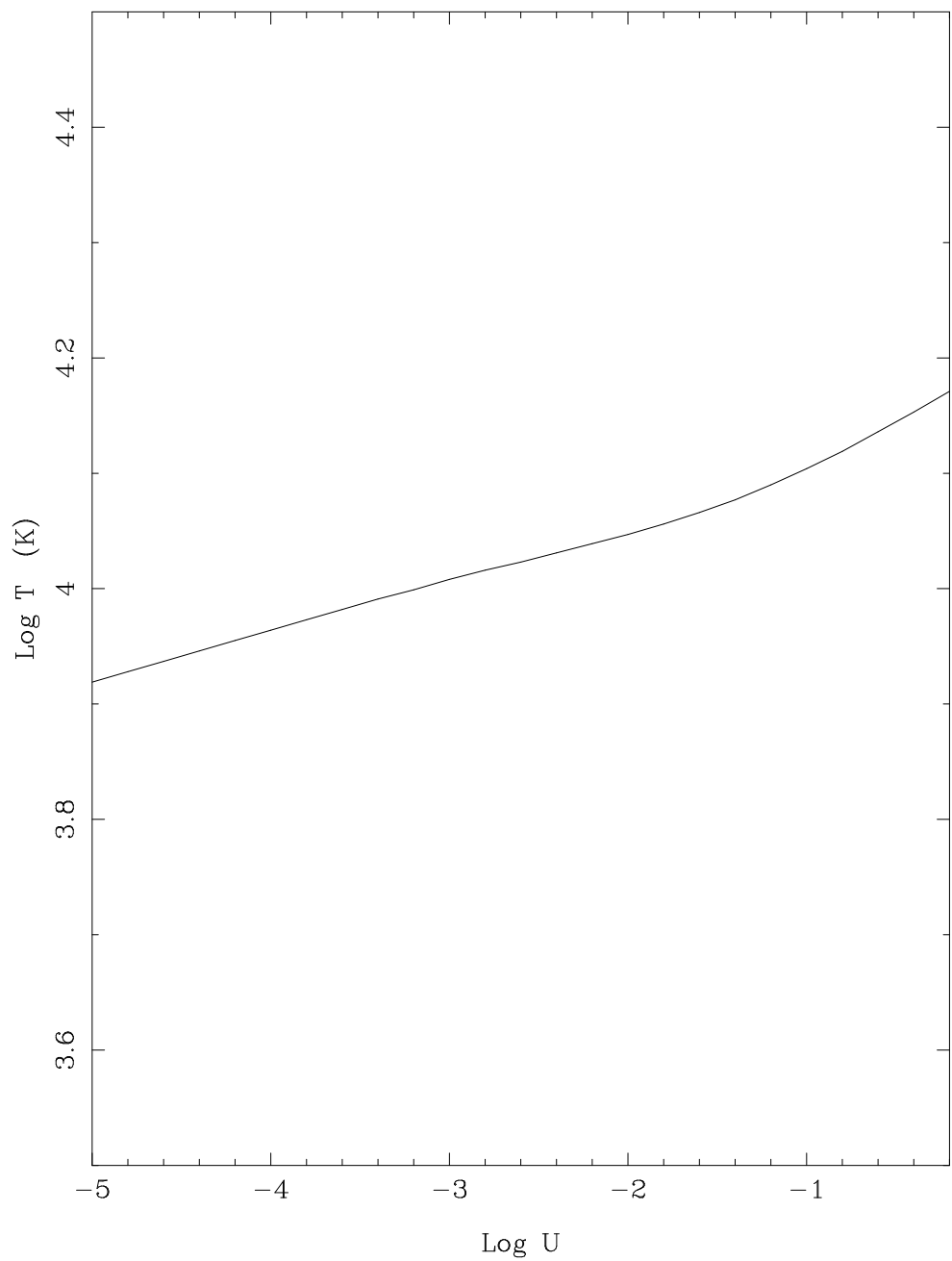}
\caption{\label{gas_temp}
\NOTE{temperature.ps}
Temperature for the gas returned by Cloudy for an absorber with \lnhi $=19.73$
\cmm as a function of the ionization parameter $\log U$.
}
\end{figure*}

\begin{figure*}
\plottwo{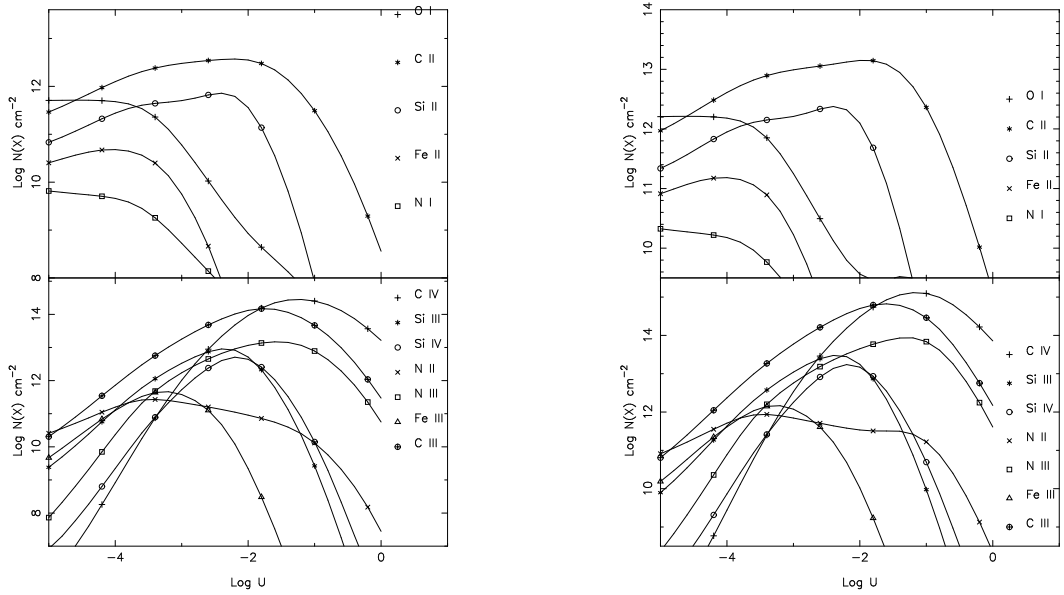}{result_16.5.ps}
\caption{\label{cloudylowcol} 
\NOTE{result\_16.5.ps}
Cloudy simulations for \lnhi $=
16.0$\cmm\ (left panel), and \lnhi $=16.5$\cmm\ (right) assuming a
Hydrogen density $n_{H} = 0.001$ \cmm\ and a metal abundance of [X/H] =
$-1.5$ as a function of ionization parameter $\log U$. 
For this figure alone, we enhanced the O and Si abundances 
by 0.3 dex, and lowered N by 
0.45 dex to match [N/O] in DLAs and elsewhere (Prochaska \etal\ 2002;
Edmunds, Henry \& Koppen 2001). The results on
the left are applicable to the component near --40~\kms\ that we call
component 3, while those on the right are for the component near
+100~\kms .}
\end{figure*}

\begin{figure*}
\plotone{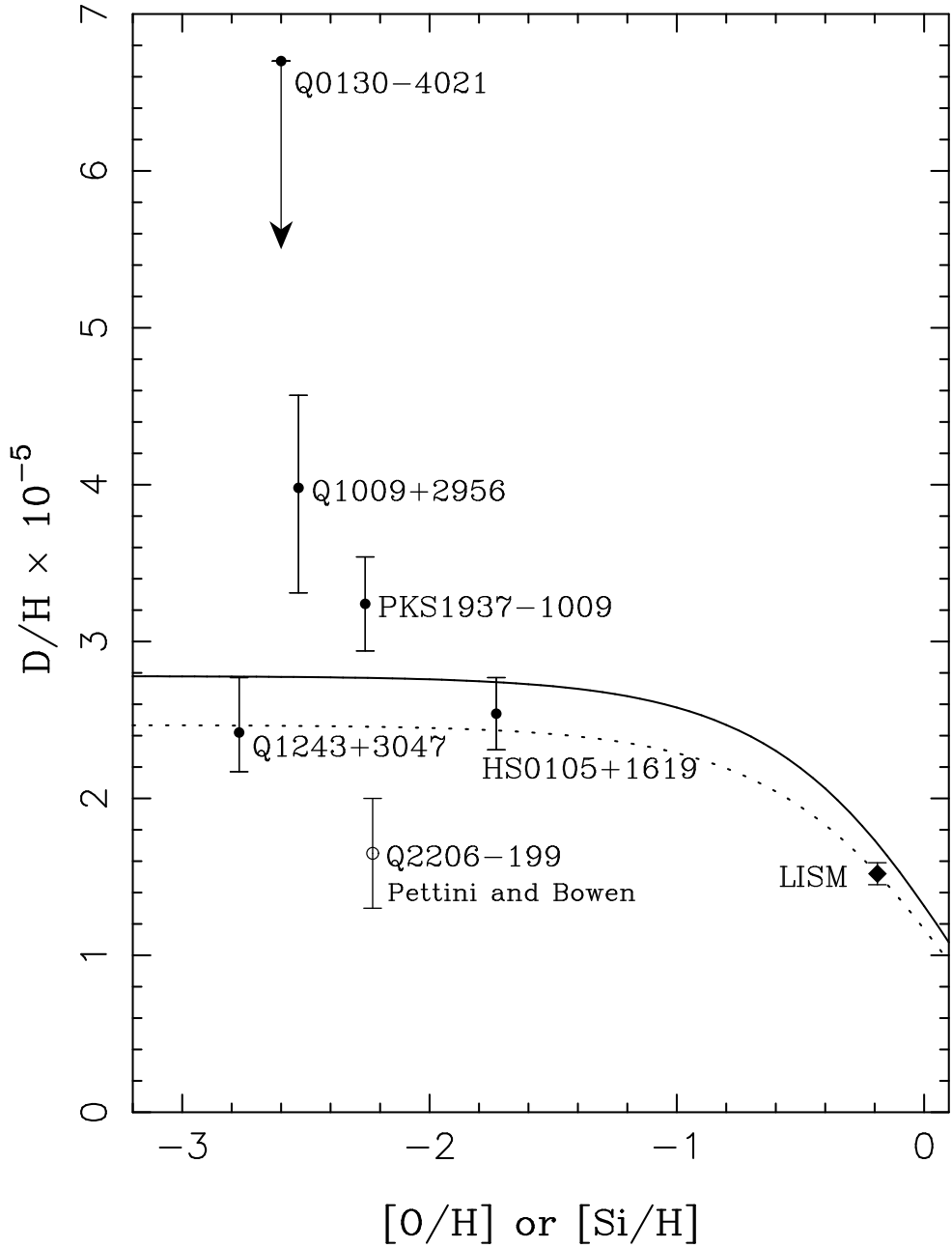}
\caption{\label{p10} \label{p10v3} \label{p10v1} \NOTE{p10}
Measurements of D/H as a function of the Silicon or Oxygen abundance in the 
gas. The solid circles are from our
group, Q2206--199 is from Pettini \& Bowen (2001) and the diamond is
the local interstellar medium (LISM) measurement (Oliveira \etal\
2002).  The error bars are intended to be 1$\sigma $ but we suspect
that in some cases they have been underestimated.  The curves show a
closed box model for the expected D/H evolution.  The solid curve is 
normalized to the primordial D/H from five QSOs 
while the dotted curve uses the D/H value in the LISM. 
}

\end{figure*}

\begin{figure*}
\plotone{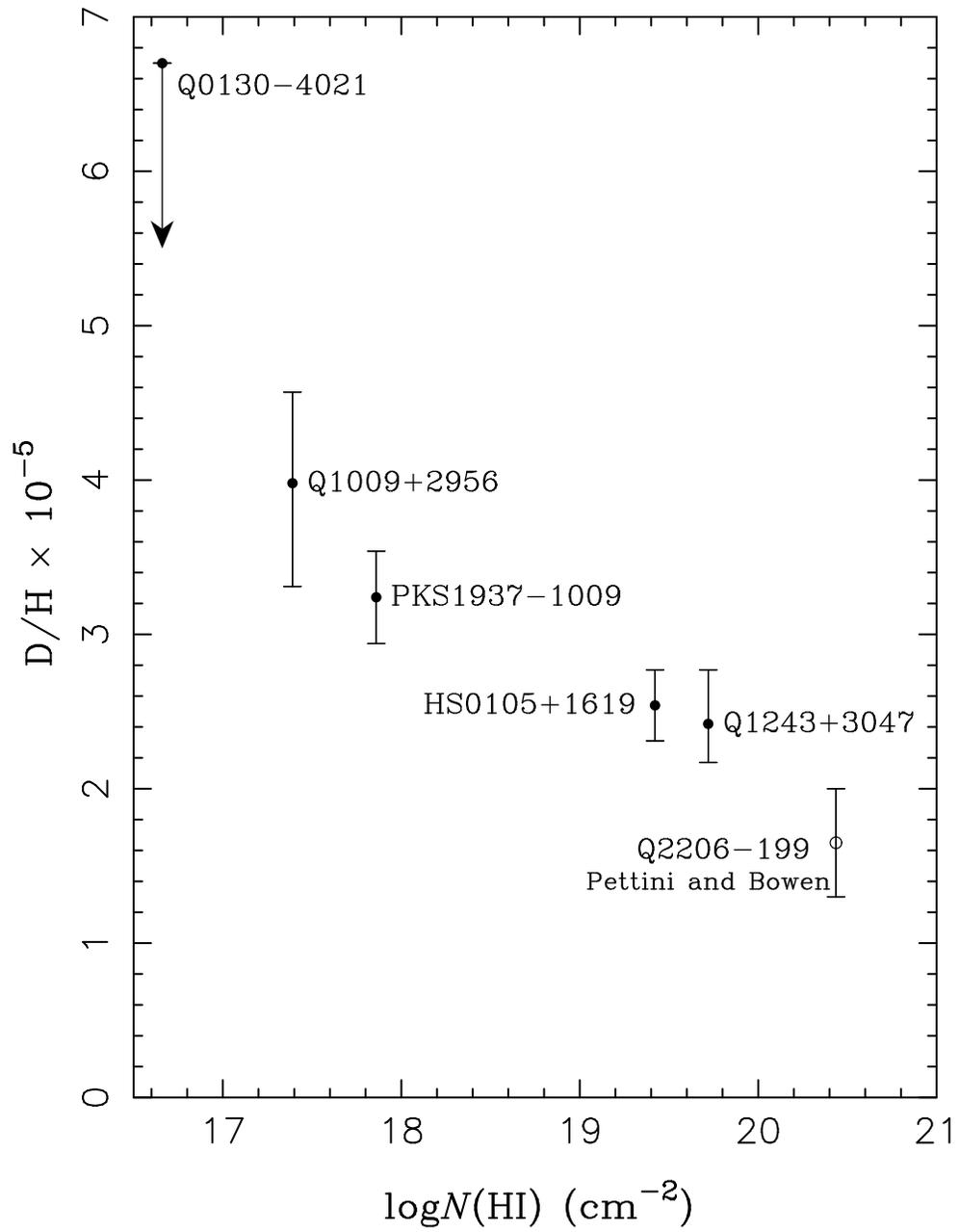}
\caption{\label{p9} \label{p9v2} \label{p9v1} \NOTE{p9} As
Fig. \ref{p10v3} but showing D/H as a function of the H~I column
density. This correlation is unexpected and we believed it is an
accident.  }
\end{figure*}

\begin{figure*}
\epsscale{0.5}
\plotone{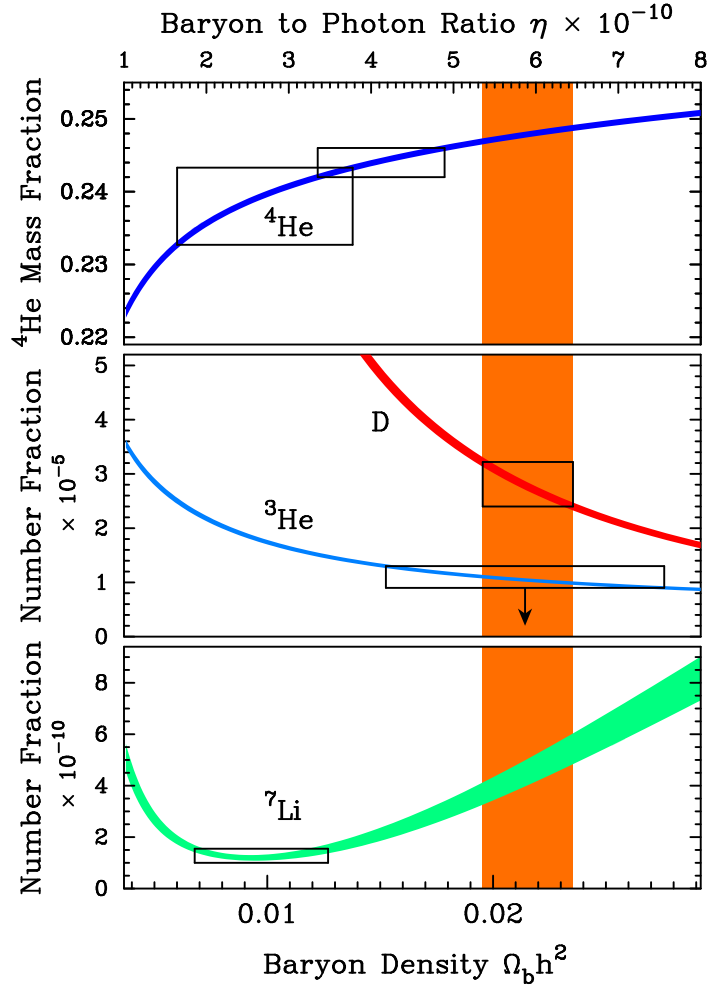}
\caption{\label{p11} \label{p11v1} \NOTE{p11v1.ps} Comparison of
predicted and measured abundances of four light nuclei as a function
of the baryon density.  The figure has three vertical panels each with
a different linear scale.  The curves show the abundance ratios
predicted for SBBN, from the calculations by Burles, Nollett and
Turner(2001).  The top curve is the \hef\ mass as a fraction of the
mass of all baryons, while the three lower curves are the number
fractions D/H, \het /H and \lisv /H. The vertical widths of the curves
show the uncertainties in the predictions. The five boxes show
measurements, where the vertical extension is the 1$\sigma $ random
error, and the horizontal range is adjusted to overlap the prediction
curves.  For \hef\ the larger box is from Olive, Steigman and Skillman
(1997), and the error includes in quadrature the systematic error from
Olive and Skillman (2001).  \NOTE{ New Astronomy 6 119} The smaller
\hef\ box is from Izotov \& Thuan (1998).  The D/H box is the mean
from five QSOs from this paper.  The \het\ from Bania, Rood \& Balser
(2002) is an upper limit.  \NOTE{ Nature 2002 415, 54} The \lisv\ is
from Ryan \etal\ (2000).  \NOTE{Ryan et al(1999) is a raw measurement,
but they do include various corrections in Ryan et al(2000)} We expect
that all the data boxes should overlap the vertical band that covers
the D/H data. They do not, probably because of systematic errors.}
\end{figure*}

\end{document}